\documentstyle[12pt,epsf]{article}
\font\tenrm=cmr10

 1
 1
 1

\begin{document}
\begin{titlepage}
\def\baselinestretch{1.2}
\topmargin     -0.25in

\vspace*{\fill}
\begin{center}
{\large
{\bf  Double Higgs Production at the Linear Colliders and the Probing of
the Higgs Self-Coupling}}

\vspace*{0.5cm}

F.~ Boudjema$^{1}$ and E.~Chopin$^{2}$\\

{\it Laboratoire de Physique Th\'eorique
EN{\large S}{\Large L}{\large A}PP}
\footnote{ URA 14-36 du CNRS, associ\'ee \`a l'E.N.S de Lyon
et \`a l'Universit\'e de Savoie.}\\
{\it B.P.110, 74941 Annecy-Le-Vieux Cedex, France} \\
{\tenrm 1. E-mail:BOUDJEMA@LAPPHP8.IN2P3.FR}\\
{\tenrm 2. E-mail:CHOPIN@LAPPHP8.IN2P3.FR}
\end{center}
\vspace*{\fill}

\centerline{ {\bf Abstract} }
\baselineskip=14pt
\noindent
 {\small  We study double Higgs production in the $e^+ e^-$ and
$\gamma \gamma$ modes of the linear collider.
It is also shown how one can probe the scalar potential in these
reactions. We discuss the effective longitudinal $W$ approximation in
$\gamma \gamma$ processes and the $W_L W_L$ luminosities in the
two modes of a high-energy linear collider. A generalised
non-linear gauge-fixing condition, which is particularly useful
for tree-level calculations of electroweak processes
for the laser induced collider, is presented. Its connection
with the background-field approach to gauge fixing is given.}
\vspace*{\fill}

\vspace*{0.1cm}
\rightline{ENSLAPP-A-534/95}
\rightline{hep-ph/9507396}
\rightline{July 1995}
\end{titlepage}
\baselineskip=18pt


\newcommand{\be}{\begin{equation}}
\newcommand{\beq}{\begin{equation}}
\newcommand{\eeq}{\end{equation}}
\newcommand{\ee}{\end{equation}}

\newcommand{\beqn}{\begin{eqnarray}}
\newcommand{\eeqn}{\end{eqnarray}}
\newcommand{\bea}{\begin{eqnarray}}
\newcommand{\ena}{\end{eqnarray}}
\newcommand{\ra}{\rightarrow}

\newcommand{\su}{$ SU(2) \times U(1)\,$}

\newcommand{\gag}{$\gamma \gamma$ }
\newcommand{\gam}{\gamma \gamma }

\newcommand{\np}{Nucl.\,Phys.\,}
\newcommand{\pl}{Phys.\,Lett.\,}
\newcommand{\pr}{Phys.\,Rev.\,}
\newcommand{\prl}{Phys.\,Rev.\,Lett.\,}
\newcommand{\prep}{Phys.\,Rep.\,}
\newcommand{\zp}{Z.\,Phys.\,}
\newcommand{\sovjnp}{{\em Sov.\ J.\ Nucl.\ Phys.\ }}
\newcommand{\nuclinst}{{\em Nucl.\ Instrum.\ Meth.\ }}
\newcommand{\annp}{{\em Ann.\ Phys.\ }}
\newcommand{\intjmp}{{\em Int.\ J.\ of Mod.\  Phys.\ }}

\newcommand{\eps}{\epsilon}
\newcommand{\mw}{M_{W}}
\newcommand{\mww}{M_{W}^{2}}
\newcommand{\mwmw}{M_{W}^{2}}
\newcommand{\mhmh}{M_{H}^2}
\newcommand{\mz}{M_{Z}}
\newcommand{\mzz}{M_{Z}^{2}}

\newcommand{\lra}{\leftrightarrow}
\newcommand{\tr}{{\rm Tr}}

\newcommand{\cms}{centre-of-mass\hspace*{.1cm}}

\newcommand{\dkg}{\Delta \kappa_{\gamma}}
\newcommand{\dkz}{\Delta \kappa_{Z}}
\newcommand{\dz}{\delta_{Z}}
\newcommand{\dgz}{\Delta g^{1}_{Z}}
\newcommand{\dgzt}{$\Delta g^{1}_{Z}\;$}
\newcommand{\la}{\lambda}
\newcommand{\lag}{\lambda_{\gamma}}
\newcommand{\laz}{\lambda_{Z}}
\newcommand{\lnl}{L_{9L}}
\newcommand{\lnr}{L_{9R}}
\newcommand{\lt}{L_{10}}
\newcommand{\lu}{L_{1}}
\newcommand{\ld}{L_{2}}

\newcommand{\epm}{$e^{+} e^{-}\;$}
\newcommand{\epemt}{$e^{+} e^{-}\;$}
\newcommand{\epem}{e^{+} e^{-}\;}
\newcommand{\eeww}{e^{+} e^{-} \ra W^+ W^- \;}
\newcommand{\eewwt}{$e^{+} e^{-} \ra W^+ W^- \;$}
\newcommand{\eennhht}{$e^{+} e^{-} \ra \nu_e \bar \nu_e HH\;$}
\newcommand{\eennhh}{e^{+} e^{-} \ra \nu_e \bar \nu_e HH\;}
\newcommand{\ppwg}{p p \ra W \gamma}
\newcommand{\wwhh}{W^+ W^- \ra HH\;}
\newcommand{\wwhht}{$W^+ W^- \ra HH\;$}
\newcommand{\ppwz}{pp \ra W Z}
\newcommand{\ppwgt}{$p p \ra W \gamma \;$}
\newcommand{\ppwzt}{$pp \ra W Z \;$}
\newcommand{\gamgamt}{$\gamma \gamma \;$}
\newcommand{\gamgam}{\gamma \gamma \;}
\newcommand{\egamt}{$e \gamma \;$}
\newcommand{\egam}{e \gamma \;}
\newcommand{\gamgamwwt}{$\gamma \gamma \ra W^+ W^- \;$}
\newcommand{\gamgamwwht}{$\gamma \gamma \ra W^+ W^- H \;$}
\newcommand{\gamgamwwh}{\gamma \gamma \ra W^+ W^- H \;}
\newcommand{\gamgamwwhht}{$\gamma \gamma \ra W^+ W^- H H\;$}
\newcommand{\gamgamwwhh}{\gamma \gamma \ra W^+ W^- H H\;}
\newcommand{\ggwwt}{$\gamma \gamma \ra W^+ W^- \;$}
\newcommand{\ggwwht}{$\gamma \gamma \ra W^+ W^- H \;$}
\newcommand{\ggwwh}{\gamma \gamma \ra W^+ W^- H \;}
\newcommand{\ggwwhht}{$\gamma \gamma \ra W^+ W^- H H\;$}
\newcommand{\ggwwhh}{\gamma \gamma \ra W^+ W^- H H\;}

\newcommand{\ptu}{p_{1\bot}}
\newcommand{\vecptu}{\vec{p}_{1\bot}}
\newcommand{\ptd}{p_{2\bot}}
\newcommand{\vecptd}{\vec{p}_{2\bot}}
\newcommand{\ie}{{\em i.e.}}
\newcommand{\cm}{{{\cal M}}}
\newcommand{\cl}{{{\cal L}}}
\newcommand{\cd}{{{\cal D}}}
\newcommand{\cv}{{{\cal V}}}
\def\slashc{c\kern -.400em {/}}
\def\slashL{L\kern -.450em {/}}
\def\slashcl{\cl\kern -.600em {/}}
\def\Ww{{\mbox{\boldmath $W$}}}
\def\B{{\mbox{\boldmath $B$}}}
\def\noi{\noindent}
\def\nn{\noindent}
\def\sm{${\cal{S}} {\cal{M}}\;$}
\def\nph{${\cal{N}} {\cal{P}}\;$}
\def\sb{$ {\cal{S}}  {\cal{B}}\;$}
\def\ssb{${\cal{S}} {\cal{S}}  {\cal{B}}\;$}
\def\ssbe{{\cal{S}} {\cal{S}}  {\cal{B}}}
\def\cviol{${\cal{C}}\;$}
\def\pviol{${\cal{P}}\;$}
\def\cpviol{${\cal{C}} {\cal{P}}\;$}

\newcommand{\lgg}{\lambda_1\lambda_2}
\newcommand{\lww}{\lambda_3\lambda_4}
\newcommand{\ppin}{ P^+_{12}}
\newcommand{\pmin}{ P^-_{12}}
\newcommand{\ppout}{ P^+_{34}}
\newcommand{\pmout}{ P^-_{34}}
\newcommand{\sinsq}{\sin^2\theta}
\newcommand{\cossq}{\cos^2\theta}
\newcommand{\yt}{y_\theta}
\newcommand{\hppll}{++;00}
\newcommand{\hpmll}{+-;00}
\newcommand{\hpplt}{++;\lambda_30}
\newcommand{\hpmlt}{+-;\lambda_30}
\newcommand{\hpptt}{++;\lambda_3\lambda_4}
\newcommand{\hpmtt}{+-;\lambda_3\lambda_4}
\newcommand{\dk}{\Delta\kappa}
\newcommand{\klam}{\Delta\kappa \lambda_\gamma }
\newcommand{\kac}{\Delta\kappa^2 }
\newcommand{\lac}{\lambda_\gamma^2 }
\def\gamgamtzz{$\gamma \gamma \ra ZZ \;$}
\def\gamgamtww{$\gamma \gamma \ra W^+ W^-\;$}
\def\gamgamtwwe{\gamma \gamma \ra W^+ W^-}

\setcounter{section}{1}

\setcounter{subsection}{0}
\setcounter{equation}{0}
\def\thesubsection {\thesection.\arabic{subsection}}
\def\theequation{\thesection.\arabic{equation}}

\setcounter{equation}{0}
\def\thequation{\thesection.\arabic{equation}}

\setcounter{section}{0}
\setcounter{subsection}{0}

\input{FEYNMAN}
\bigphotons
\section{Introduction}
Now that there has been striking direct evidence for the top quark
\cite{Topdiscovery} with a mass that
fits neatly with what one deduces from the precision measurements at LEP1 and
SLC, the matter content of the Standard Model, \sm, is complete. What is still
desperately missing is the scalar particle of the model, the Higgs. Intimately
related to the existence of this cornerstone particle is the mechanism of
symmetry breaking. The elucidation of its realisation will most probably
have to await the next generation of high energy colliders. Once the Higgs
has been
discovered it will be essential to scrutinize all its properties, like its
couplings to the other particles and its parity. These will be precision
measurements that are best conducted in a clean environment and therefore
one hopes to conduct these tests at the next linear collider in its
\epemt mode as well as the much discussed \gamgamt mode\cite{Laser}.
Among these tests
one should include the probing of the self-couplings of the Higgs. These
self-couplings have undeservedly received very little
attention\cite{Fernandhh,Bijhh,Jikiahh} and yet, in the \sm,
they directly emerge from the pure non-gauge Higgs-Goldstone potential that
realises the symmetry breaking. Let us recall that with the assumption of
one Higgs doublet $\Phi$,
which most naturally implements the well confirmed
$\rho=1$, and in order that spontaneous symmetry breaking  ensues with the
correct value of the vacuum expectation value
that gives the gauge boson masses,
the most general  potential has the form
\beqn
\label{generalpotential}
V_{\ssbe}=-\cl_{\ssbe}= \lambda
\left\{ \left[ \Phi^\dagger \Phi - \frac{v^2}{2}  \right]^{2} +
\sum_{n\geq 3} \frac{\kappa_n}{\Lambda^{2(n-2)}}
\left[ \Phi^\dagger \Phi - \frac{v^2}{2} \right]^{n} \right\}
\eeqn
Only the first term, of dimension four,  is needed in the
\sm to trigger symmetry breaking and to ensure renormalisabity.
This term introduces the one scale that feeds the masses for all the
particles in the SM, $v=246GeV$. It is also
characterised by the parameter $\lambda$ whose presence has
not been established, let alone measured since this extra parameter is direcly
related to the Higss mass
and measures its self-couplings as well as its coupling with
the would be longitudinal
weak bosons. These self-couplings are large when the Higgs
mass is large. With
 the minimal prescription the tri-linear and quadri-linear
Higgs self-couplings will be
 directly deduced when the Higgs mass is known.   \\
The higher order
terms necessarily parameterise the most general scalar potential and could
indicate a non-standard Higgs, possibly a bound state
that evades the naturality
argument without invoking supersymmetry. These ``beyond-the-\sm"
terms introduce a
new scale $\Lambda$, in a sense a new ``curvature", that may have
nothing to do with the Fermi scale, {\it i.e.} $v$.
This new potential changes the tri-linear and quadri-linear couplings.
Since these new couplings
 involve neither the matter particles nor the gauge particles a direct
unambiguous litmus test for their existence is only possible through  mutiple
Higgs production. Since one expects these multiparticle cross sections to be
small one should look for these signatures in an environment where one is not
swamped by  a large background and where there is the least theoretical
uncertainty in the standard model calculation, thus high energy \epemt machines
seem to be the ideal place to study these reactions. \\

\epemt operating in the TeV range are being very seriously considered. These
also offer the possibility of running not only in the ``classic" \epemt mode
but can also be turned into \gamgamt or \egamt colliders by converting
the electron into a highly energetic photon through Compton backscattering with
the help of a laser\cite{Laser}. Recently there has been an intense activity in
the physics that can be investigated at these new types of colliders and a
research and development programme is in full swing.

The aim of this paper is three-fold. First, we want to give the expected
cross sections for double Higgs production at TeV energies at the \epemt
colliders. In so doing we will present the details of the first calculation
of the process \gamgamwwhht and derive an approximate
anlytical expression for its high-energy behaviour through the use of the
structure function approach, in this case the $W$ content of the photon.
We will then compare
the effective luminosity for $W_L W_L$ in the \epemt and the \gamgamt modes.
 Another purpose is to put  forth the suggestion that for processes
with multi-W's especially in association with photons
it is by far advantageous
to calculate in a non-linear gauge\cite{Nonlineargauge}.
Till now these type of gauges have been used for
loop calculations, we will show how they can ease the
calculational task in the
case of tree-level amplitudes with many gauge bosons.
In passing, we will point to the
connection between the generalised non-linear gauge that we introduce
in this paper and the background field
inspired gauges\cite{Backgroundgaugeold}
applied to the \su\cite{BackgroundgaugeEW}
that  have been much discussed in the last year\cite{BckgDenner}.
As a third purpose we will discuss how to measure
the tri-linear Higgs self-coupling and the limit we could hope to extract
at the 2TeV \epemt collider in its different modes. We will also investigate
and compare with the limits that one may tentatively
set on this coupling from
indirect measurements. \\

The paper is organised as follows. In the next section we deal with the process
\ggwwhht and show how efficient a suitable gauge fixing condition
can be in tremendously
easing the computational task. We study various distributions and
comment on the effect
of polarisation, both initial and final. Section~3 compares
different processes for double
Higgs production at the \epemt and the \gamgamt modes including
the effect of photon spectra.
This comparison and the discussion in section ~2 lead us,
in section~4,  to inquire about the
validity of the effective W approximation both for
\epemt \cite{EWA}  and \gamgamt reactions
\cite{egnuwh,Parisgg}, in other words finding approximations
by considering the longitudinal
$W$ as a parton. Especially interesting is the longitudinal $W$ content
of the photon. We study both the
case of a heavy Higgs and a light Higgs.
When referring to a heavy Higgs we have in mind a Higgs
that decays into the weak
vector bosons and that is predominantly coupled to their
longitudinal parts. Our representative
example of a heavy Higgs throughout the paper will be $M_H=400GeV$.
As a by-product we give the full helicity amplitudes
for the hard sub-process $W^+W^- \ra HH$ including
the anomalous $H^3$ coupling.
In section~5 we show how we can test for the presence of a
tri-linear Higgs coupling
and give the limit one may hope to set on its strength
at a 2TeV \epemt machine and comment on the
improvement that a higher energy collider can bring.
We conclude in section~6. In an Appendix we discuss
in detail the ``generalised"
non-linear gauge fixing condition, show how the background
field gauge-fixing condition
can lead to a special case of the non-linear constraint.
Full Feynman rules, including the ghosts,
are presented for general values of the non-linear gauge parameters.

\section{\ggwwhht}
\subsection{Motivation}
\begin{figure*}[b]
\caption{\label{feynmanee}{\em Representative diagrams that contribute
to $\epem \ra ZHH$ and \eennhht. }}
\centerline{
\begin{picture}(15000,10000)
\drawline\fermion[3 0](0,6828)[4000]
\global\Xone=\fermionfrontx \global\Yone=\fermionfronty
\global\advance\Xone by -700 \put(\Xone,\Yone){$e^-$}
\drawline\fermion[5 0](\fermionbackx,\fermionbacky)[4000]
\global\Xtwo=\fermionbackx \global\Ytwo=\fermionbacky
\global\advance\Xtwo by -700 \put(\Xtwo,\Ytwo){$e^+$}
\drawline\photon[2 0](\fermionfrontx,\fermionfronty)[6]
\global\Xthree=\pmidx \global\Ythree=\pmidy
\global\advance\Ythree by -1200 \put(\Xthree,\Ythree){$Z$}
\drawline\scalar[1 0](\pmidx,\pmidy)[2]
\global\Xfour=\scalarbackx \global\Yfour=\scalarbacky
\global\advance\Xfour by -1000 \put(\Xfour,\Yfour){$H$}
\drawline\scalar[1 0](\photonbackx,\photonbacky)[2]
\global\Xfive=\scalarbackx \global\Yfive=\scalarbacky
\global\advance\Xfive by +1000 \put(\Xfive,\Yfive){$H$}
\drawline\photon[3 0](\photonbackx,\photonbacky)[4]
\global\Xsix=\photonbackx \global\Ysix=\photonbacky
\global\advance\Xsix by +1000 \put(\Xsix,\Ysix){$Z$}
\end{picture}
\hspace{2cm}
\begin{picture}(12000,10000)
\drawline\fermion[2 0](0,7000)[8000]
\global\Xone=\fermionfrontx \global\Yone=\fermionfronty
\global\advance\Xone by -1000 \put(\Xone,\Yone){$e^-$}
\global\Xtwo=\fermionbackx \global\Ytwo=\fermionbacky
\global\advance\Xtwo by 700 \put(\Xtwo,\Ytwo){$\nu$}
\drawline\photon[4 0](\pmidx,\pmidy)[2]
\global\Xthree=\pmidx \global\Ythree=\pmidy
\global\advance\Xthree by -1200 \put(\Xthree,\Ythree){$W$}
\drawline\scalar[2 0](\photonbackx,\photonbacky)[2]
\global\Xfour=\scalarbackx \global\Yfour=\scalarbacky
\global\advance\Xfour by 700 \put(\Xfour,\Yfour){$H$}
\drawline\photon[4 0](\scalarfrontx,\scalarfronty)[2]
\global\Xfive=\pmidx \global\Yfive=\pmidy
\global\advance\Xfive by -1200 \put(\Xfive,\Yfive){$W$}
\drawline\scalar[2 0](\photonbackx,\photonbacky)[2]
\global\Xsix=\scalarbackx \global\Ysix=\scalarbacky
\global\advance\Xsix by 700 \put(\Xsix,\Ysix){$H$}
\drawline\photon[4 0](\scalarfrontx,\scalarfronty)[2]
\global\Xseven=\pmidx \global\Yseven=\pmidy
\global\advance\Xseven by -1200 \put(\Xseven,\Yseven){$W$}
\drawline\fermion[2 0](\photonbackx,\photonbacky)[4000]
\global\Xeight=\fermionbackx \global\Yeight=\fermionbacky
\global\advance\Xeight by 700 \put(\Xeight,\Yeight){$\bar\nu$}
\drawline\fermion[6 0](\photonbackx,\photonbacky)[4000]
\global\Xone=\fermionbackx \global\Yone=\fermionbacky
\global\advance\Xone by -1000 \put(\Xone,\Yone){$e^+$}
\end{picture} }
\end{figure*}
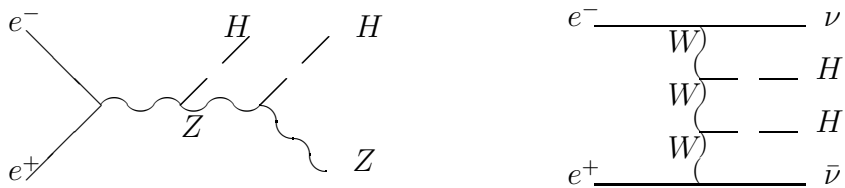
Double Higgs production in the classic mode of the \epemt has been
considered sometime ago\cite{Fernandhh,eeZhh,eennhh}. As with the
case of single $H$ production one has two different mechanisms obtained
by grafting another Higgs to the single Higgs production mechanism:
either as a double Higgs bremstrahlung off the $Z$: $e^+ e^- \ra Z HH$
 \cite{Fernandhh,eeZhh}  (see Fig.~\ref{feynmanee}) or
through WW fusion\cite{eennhh}  leading to
 $e^+ e^- \ra  \nu_e \bar \nu_e HH$ (Fig.~\ref{feynmanee}).
A similar mechanism\cite{eennhh} through $ZZ$ fusion
($e^+ e^- \ra  e^+ e^- HH$) gives much smaller cross sections due
to the much smaller coupling of the $Z$ to the electron than the $W$.

The one-loop induced
$e^+ e^- \ra   HH$\cite{eehh} has also been considered and
found to be dismal. The interest in \ggwwhht stems from the observation
that since the cross section for the basic process \ggwwt is
really enormous\cite{Nousggvv} with a value of about 80pb
at 400GeV and not decreasing with energy due to the spin-1 t-channel
$W$-exchange, one could use it as a backbone to hook yet more  particles,
especially the neutrals. This idea has been successfully applied to single
Higgs production through \ggwwht\cite{Nousgg3v}
where it was found that at high enough
energy, where one is not penalised by the reduced phase space, this
reaction occurs at a higher rate than its \epemt equivalent
$e^+ e^- \ra  \nu_e \bar \nu_e H$\cite{CahnnunuH} which the is dominant
mechanism for single Higgs production at \epemt. It is then hoped that
\ggwwhht can compete with double Higgs production in \epemt. \\

\subsection{Non-linear gauge fixing for tree-level $\gamma \gamma$ processes}
When switching from \epemt to \gamgamt production
the calculational task becomes much more arduous especially as
more gauge particles are involved. Not only does the number of
diagrams increase tremendously but each diagram has a much more
complicated structure brought about by the handling
of the non-Abelian vertices, as well as by the form of the
massive propagators in the unitary gauge. For
instance, in the case at hand, comparing
the $e^+ e^- \ra \nu_e \bar \nu_e HH$ with \ggwwhht, one notices that
for the former only four  diagrams  are present (neglecting
justifiably the electron mass) and that we can simply
leave out the ``longitudinal part" ($k_\mu k_\nu/M_W^2$ parts)
 of the $W$ propagators, since the W's couple to
conserved currents. One thus have a simple
calculation to perform with compact
formulae for the helicity amplitudes.
\begin{figure*}
\begin{center}
\caption{\label{feynmanggwwhh}{\em Different topologies of Feynman diagrams
contributing to \ggwwhht in the unitary gauge.
Figures (1a), (2) and (3) are the fusion type diagrams.
All others can be considered bremstrahlung. (1a-c) contain the triple
Higgs vertex.}}
\vspace*{-5cm}
\hspace*{-1.5cm}
\mbox{\epsfxsize=20cm\epsffile{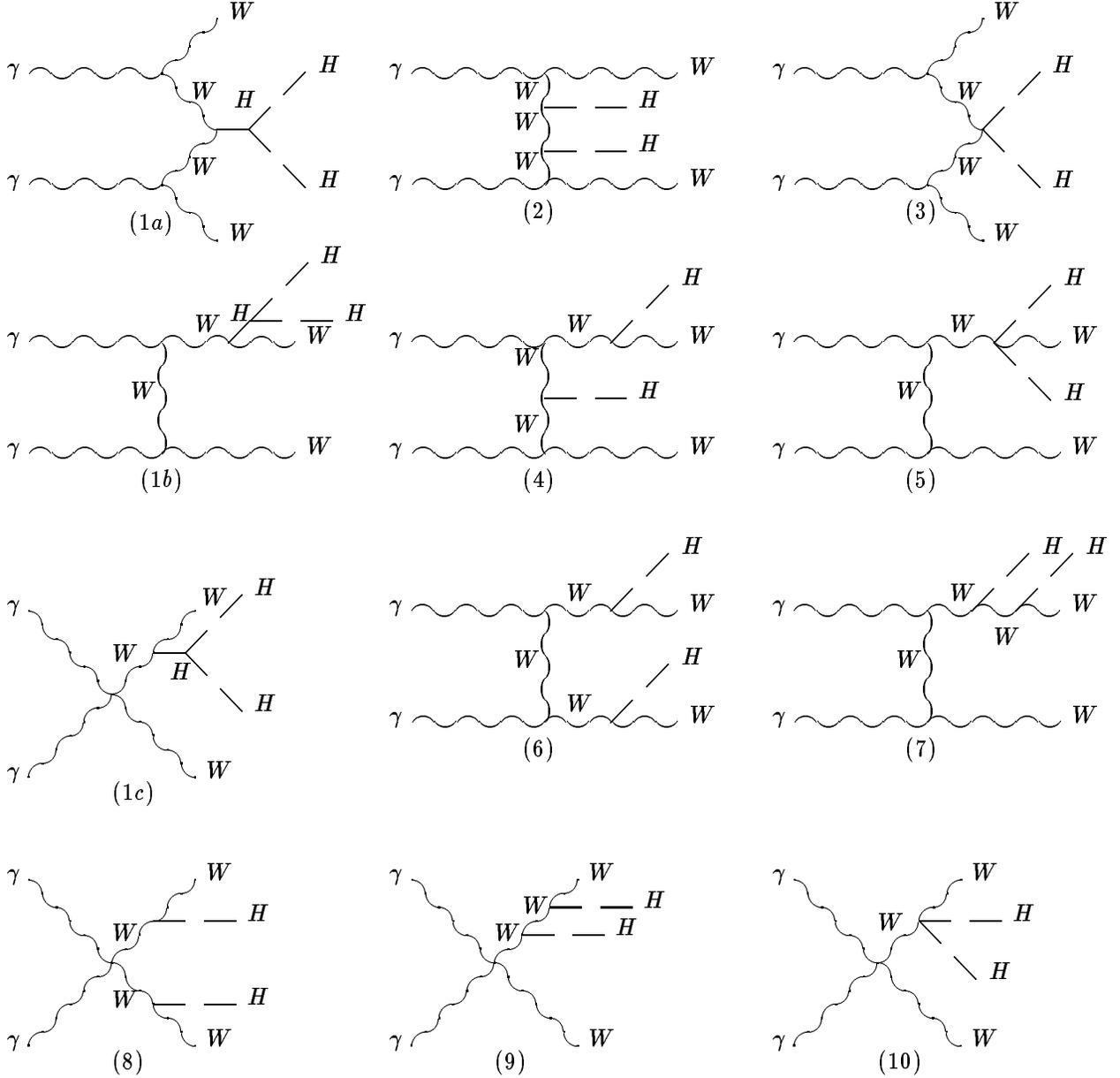}}
\end{center}
\end{figure*}
As can be seen from the list of diagrams
(Fig.~\ref{feynmanggwwhh})
contributing in the unitary gauge to \ggwwhht the situation is much more
complex: one has 12 possible graph topologies which upon symmetrisation bring
about a total of 46 diagrams. Moreover,
beside the non-Abelian vertices one has to
keep the ``longitudinal" terms which in the case of three $W$ exchanges
 ``eightuples" the number of terms!
What is worse for an already complex
situation
and more dangerous in a numerical implementation of the amplitudes,
is the  fact that these longitudinals which are  associated
with the very bad high energy behaviour
of the cross section, contribute large numerical factors:
$\sim (E_W/M_W)^{2n} \sim (s/M_W^2)^{n}$,
$n$ counting the number of propagators that can be large for
multiple vector bosons amplitudes. These factors  eventually  largely cancel
when the full set
of diagrams is taken into account. In a numerical
implementation of the amplitudes, the instability that
these terms might introduce is to be avoided while it is  essential to
make the calculation as tractable as possible and
with expressions that can be as compact as possible. \\
\noindent Almost invariably the way out is to revert to the standard
't~Hooft-Feynman gauge, thus eliminating the annoying
``longitudinal" terms. However, especially with the presence of many
$W$'s there is a price to pay, in that more
diagrams containing the compensating Goldstone
fields have to be added even though the
expressions derived from each diagram are more compact that in the unitary
gauge. The best of both worlds would be to keep the same
(or sensibly the same) number of
diagrams as in the physical gauge and at the
same time rendering the expressions much more compact. This is
possible if one judiciously chooses a non-linear
gauge fixing condition and selects the gauge parameter to correspond to the
't~Hooft-Feynamn gauge. The main reason, in  the non-linear gauges,
the Goldstone bosons do not appear in
most diagrams is
due to the fact that  the $W^\pm \varphi^\mp \gamma$ is not
present. The underlying principle behind the vanishing of this vertex is that
the gauge fixing is such that it still does not explicitely break the
$U(1)_{em}$ and as a consequence one should not expect an electromagnetic
``particle-changing" current such as  $W^\pm \varphi^\mp \gamma$.

\noi The widespread choice of the linear gauge fixing condition
\footnote{Our conventions are defined in the Appendix.}
\beq
\label{lineargauge}
{{\cal L}}^{Gauge-Fixing}_{linear}\;=\;-\frac{1}{\xi}
|\partial_\mu W^{\mu +}\,+\,i\xi M_W \varphi^{+}|^{2}
\eeq

\noi only eliminates the  $W^\pm \varphi^\mp$ mixing, while for a typical
 non-linear gauge

\beq
\label{nonlineargauge}
{{\cal L}}^{Gauge-Fixing}_{non-linear}\;=\;-\frac{1}{\xi}
|(\partial_\mu\;+\;i e A_\mu\;+\;ig \cos \theta_W Z_\mu) W^{\mu +}\,+\,
i\xi (M_W +\frac{g}{2} H)\varphi^{+}|^{2}
\eeq

\begin{figure*}
\begin{center}
\caption{\label{feynmanggwwhhnl}{\em Some extra topologies of Feynman diagrams
 that have to be added
in a renormalisable gauge even when the $W^\pm \varphi^\mp \gamma$ is absent.
The last 4 topologies are cancelled with the gauge-fixing that we take.}}
\vspace*{-5cm}
\hspace*{-1.5cm}
\mbox{\epsfxsize=20cm\epsffile{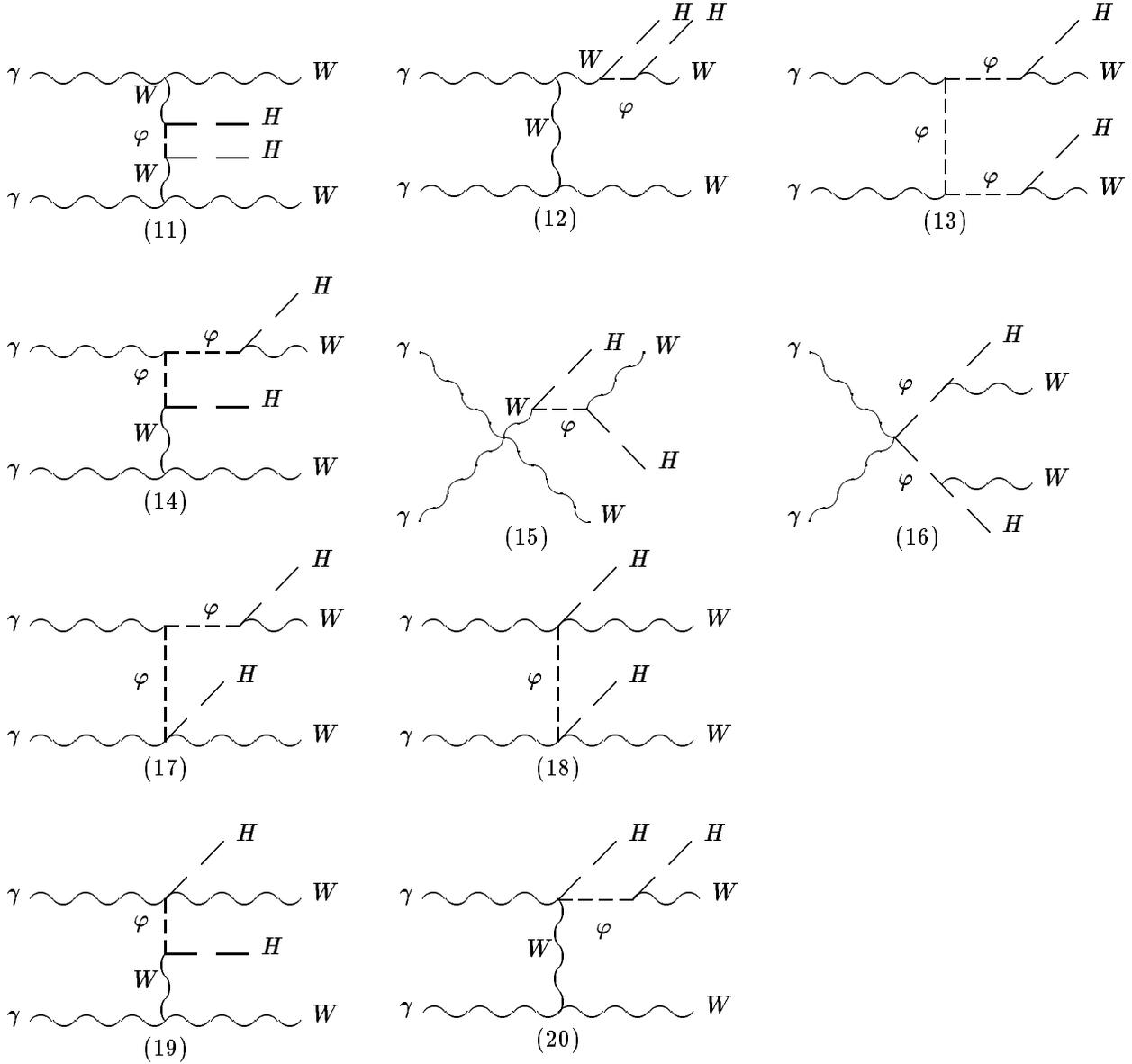}}
\end{center}
\end{figure*}

\noi the $W^\pm \varphi^\mp \gamma$ will disappear
due to the use of the covariant
derivative with respect to the $U(1)$ current. In fact, with  the above choice
the covariant derivative is along the full $T_3$ neutral direction of the SU(2)
group.
We have already used a slight variation of the above\cite{Nousgg3v,Parisgg}
gauge condition for
the calculations of $\gamma \gamma \ra W^+W^-Z, W^+W^-\gamma, W^+W^- H$ and
have found tremendous simplifications. For those processes the last term
involving the Higgs does not enter, however for \ggwwhht
it is essential because it eliminates the ``unnatural"
$W^\pm \varphi^\mp \gamma H$ that may enter at this
order (see Fig.~\ref{feynmanggwwhhnl}). The only other application of this
gauge-fixing for tree-level amplitudes in the electroweak theory that we are
aware of is the recent calculation of
$\gamma \gamma \ra W^+ W^- W^+ W^-,  W^+ W^-ZZ$ \cite{Jikia4w}. \\

\noi These are not the only simplifications that the non-linear gauge fixing
conditions (with the Feynman parameter $\xi=1$) bring. Both the tri-linear
$W^+W^-\gamma$ and  quartic $W^+W^-\gamma \gamma$ are simplified. The latter
now contains only one term out of three, moreover all diagrams containing this
coupling trivially vanish when the two
incident photons have opposite helicities ($J_Z=\pm 2$).
As for the tri-linear coupling (see Appendix) it is most usefully split into a
universal ``convection" current (present for the case of scalars and fermions)
and a spin current that is automatically transverse in the photon. To make full
use
of these gauges one has to combine them with the calculation of the helicity
amplitudes. In fact, recent developments in
the calculation of QCD processes\cite{Bernreview},
especially the so-called string inspired organisation, can be understood
as being (partly) based on the exploitation of similar
gauges. This type of gauges has an obvious connection and similarity with the
so-called background gauges\cite{Backgroundgaugeold} which,
within the $SU(2) \times U(1)$ model, have
been resurrected in the last year
\cite{BckgDenner}. The underlying idea is the same though: one
strives to maintain an explicit gauge covariance (after gauge-fixing) in the
classical fields (external fields or external currents).
For this one splits the
fields into a quantum and a classical part and take the covariant derivative
with respect to the classical gauge fields. \\
The above non-linear gauge can be generalised. For instance, for other
applications,  one can arrange for
the $W^\pm \varphi^\mp Z$ to vanish as well.
More generally we can tune the parameters of the
following gauge-fixing term for the $W$'s:
\beqn
\label{nonlineargaugegen}
{{\cal L}}^{Gauge-Fixing}_{non-linear}\;=\;-\xi^{-1}
|(\partial_\mu\;+\;i e \tilde{\alpha}
A_\mu\;+\;ig\cos \theta_W \tilde{\beta} Z_\mu) W^{\mu +}+
i\xi \frac{g}{2}(v +\tilde{\delta} H
-i\tilde{\kappa} \varphi_3)\varphi^{+}|^{2}
\nonumber\\
\;
\eeqn
In the Appendix we give the Feynman rules for this most general gauge fixing
term and will explicitely spell out the connection with the background gauge
applied to \su.

\subsection{Total cross section, energy dependence and polarisations}

With the above choice of gauge it is a relatively easy task to compute
the helicity amplitudes. As a check on our calculation we verified that
the amplitudes were transverse in the photon momenta. Moreover, since the
diagrams that involve the triple Higgs vertex constitue a gauge invariant
subset we have also checked that this subset is also transverse in the photons.
One way to argue that this subset is gauge invariant is to observe that this is
the same subset that constitutes the whole of the \ggwwht amplitude. Therefore,
 if one
``fuses" the two Higgses into a single state one has an effective
Higgs and thus one
is effectively calculating \ggwwht. Another way of seeing this is to observe
that the triple Higgs vertex
emerges from another part of the Lagrangian (scalar potential) than the $WWH$
(covariant derivative on the Goldstone field). Upon modifying the strength of
this vertex, transversality in the photon is still maintained. We will come
back to this point later when discussing the inclusion of a non-standard value
for the Higgs and how, again, the non-linear gauge fixing brings a welcome
simplication. \\

Our phenomenological analyses will be mainly  for centre-of-mass
energies up to 2TeV,
{\it i.e.}, in the range
considered for the next generation of linear colliders
 before including any spectra
for the $\gamma \gamma$ luminosity. However,
this process having quite interesting features
which are more easily revealed at high energies
we will also dicuss the case of much higher
energies and heavy Higgses so as to be able to
check whether a description in terms
of structure function, $W_L$ inside the photon, can be applicable. \\

We have taken $\alpha=1/137$ everywhere
with $M_W=80.2GeV$ and $sin^2\theta_W=s_W^2=0.2325$.
One may question why we decided
against the use of $\alpha=1/128$. For the photon vertex
(with an on-shell photon: $q^2=0$) this is perfectly justified.
Indeed $\gamma \gamma \ra W^+W^-$
has to be calculated with this value of $\alpha$, this has also been
confirmed by computing the one-loop corrections to
this reaction\cite{DennerRCggww}. As for
the $WWH$ couplings one may choose to use $\alpha=1/128$ by relating this
to $G_F$. Sticking with the latter choice our numbers have to be scaled by
a factor $(137/128)^2$.

\begin{figure*}
\begin{center}
\caption{\label{finalpolnoconv}{\em Contribution of the different polarisation
states of the $W$'s.}}
 \mbox{\epsfxsize=14cm\epsfysize=14cm\epsffile{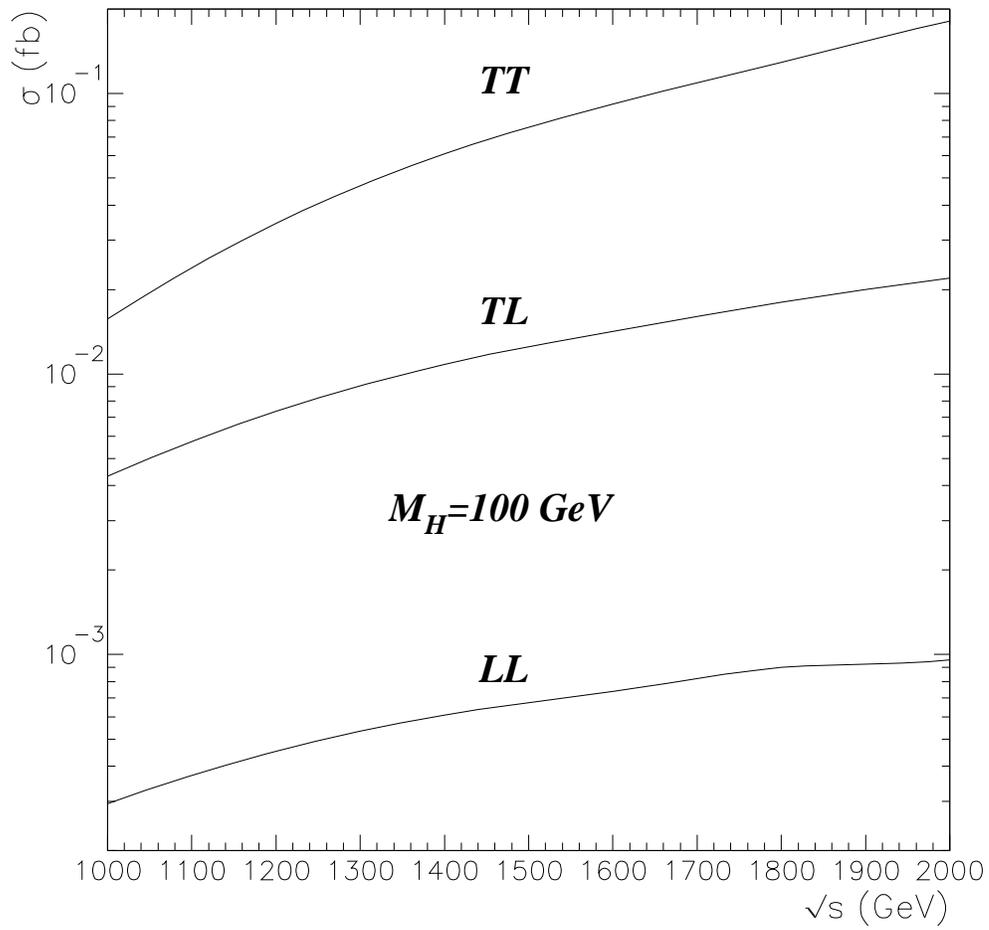}}
\end{center}
\end{figure*}

\begin{figure*}
\begin{center}
\caption{\label{initpolnoconv}{\em Contribution of the $J_Z=0$ and $J_Z=2$ to
the total cross section.}}
\mbox{\epsfxsize=14cm\epsfysize=14cm\epsffile{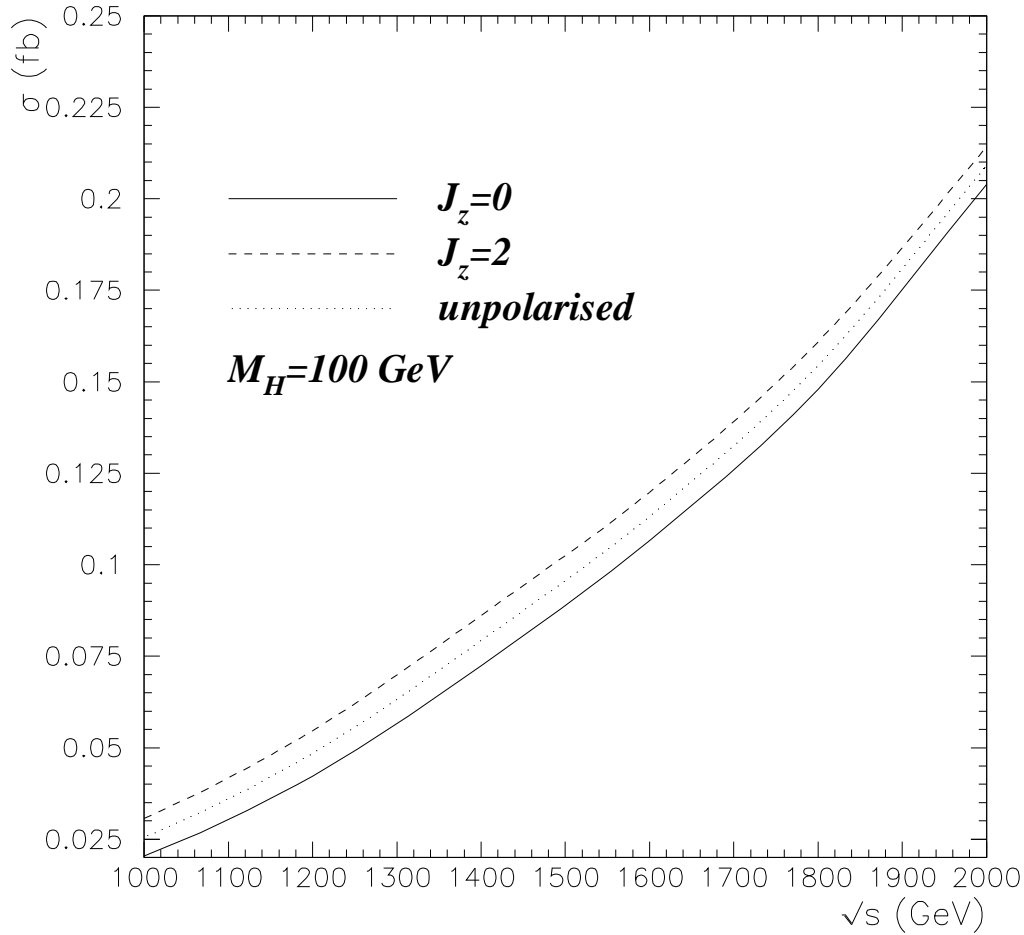}}
\end{center}
\end{figure*}

The first conclusion is that
the cross sections are rather small for the foreseable colliders.
For a Higgs mass of 100GeV and
with a total \gamgamt  centre-of-mass energy of 2TeV,
we find a total cross section of
about 0.21fb giving some 63 events for an integrated luminosity of
$\int \cl=300$fb$^{-1}$, before including the various W's branching ratios and
convoluting with the \gamgamt luminosity spectra. There is about a factor
of $\alpha$ as compared to \ggwwht\cite{Nousgg3v} for the same mass and energy.
These events will thus be rare events
that would need to be analysed after a long run. On the other hand, as
with all rare processes, this
cross section is therefore ideal for the investigation of New Physics. In this
respect, it is essential to study in detail all the characteristics of this
process and its salient signatures.  \\
The behaviour of the cross section reflects a few of the characteristics of the
\gamgamwwt process. The t-channel $W$ exchange enhancement, responsible for the
constant asymptotic \gamgamwwt cross section, brings for this
multiple particle production a
logarithmic growth as the energy increases (Fig.~\ref{finalpolnoconv}).
Moreover, one still has the dominance
of  the transverse $W$'s:  each time one
replaces a transverse $W$ by a longitudinal one looses an order of magnitude in
the total cross section. This recurring feature is well rendered in
Fig.~\ref{finalpolnoconv}.
As for the photon polarisations,  and again as is the case with the \gamgamwwt
at high energy,
there is very little dependence on the helicity
combinations of the initial photons already at 1TeV.
This is clearly seen in Fig.~\ref{initpolnoconv}.
where we note that there is a slight preference for a setting with opposite
photon helicities ($J_Z=2$) that tends to attenuate as the energy increases.
This feature persists for a higher Higgs mass.

\subsection{Higgs mass dependence}

At 2TeV \cms  the cross section drops precipitously with increasing Higgs mass.
For instance,
for a 200GeV Higgs mass it is twice as small as for
100GeV Higgs with  about a tenth of fb.
For a 300GeV Higgs there is again another
factor 1/2
reduction. In this process that involves the triple Higgs self-coupling
the Higgs dependence
does not only enter trivially in the phase space
(or to a lesser extent through the propagator) but it should be kept in
mind that the Higgs self
coupling itself is proportional to square of the Higgs mass.
We should thus expect this
feature to contribute in, somehow, balancing out the phase space reduction.
This however occurs only
slightly at much higher energies. The reason is the following:
in the minimal \sm the $W_L W_L H H$ is also of enhanced
strength ({\it i.e} proportional to $M_H^2/M_W^2$) as can be most easily
seen by reverting to the equivalence theorem\cite{Equivalencetheorem}
which, here, will tell us that this is $\varphi^\pm \varphi^\mp H H$.
The latter stems from
the Higgs potential also and has the same strength.
The other observation is that for a heavy enough Higgs, the fusion
diagrams are important and as we will show below are dominated
by the longitudinal quasi-real $W$'s.
Then a destructive interference takes place between the diagrams
involving the triple-Higgs vertex
in which we are most interested and the ones involving the $WWHH$ vertex.
This will
become clearer when we study $W^+W^- \ra HH$.
\begin{figure*}
\begin{center}
\caption{\label{mhggdep}{\em Higgs mass dependence of the
\ggwwhht cross section at 2TeV and 5TeV.
The contribution of the diagrams involving the triple Higgs
vertex (Signal) and the rest
(Background) is shown separetely. Also shown is the contribution
of the transverse $W$'s (TT). ``Total" stands for the contribution summed
over all helicities of the $W$'s.}}
\mbox{
\mbox{\epsfxsize=8.5cm\epsfysize=12cm\epsffile{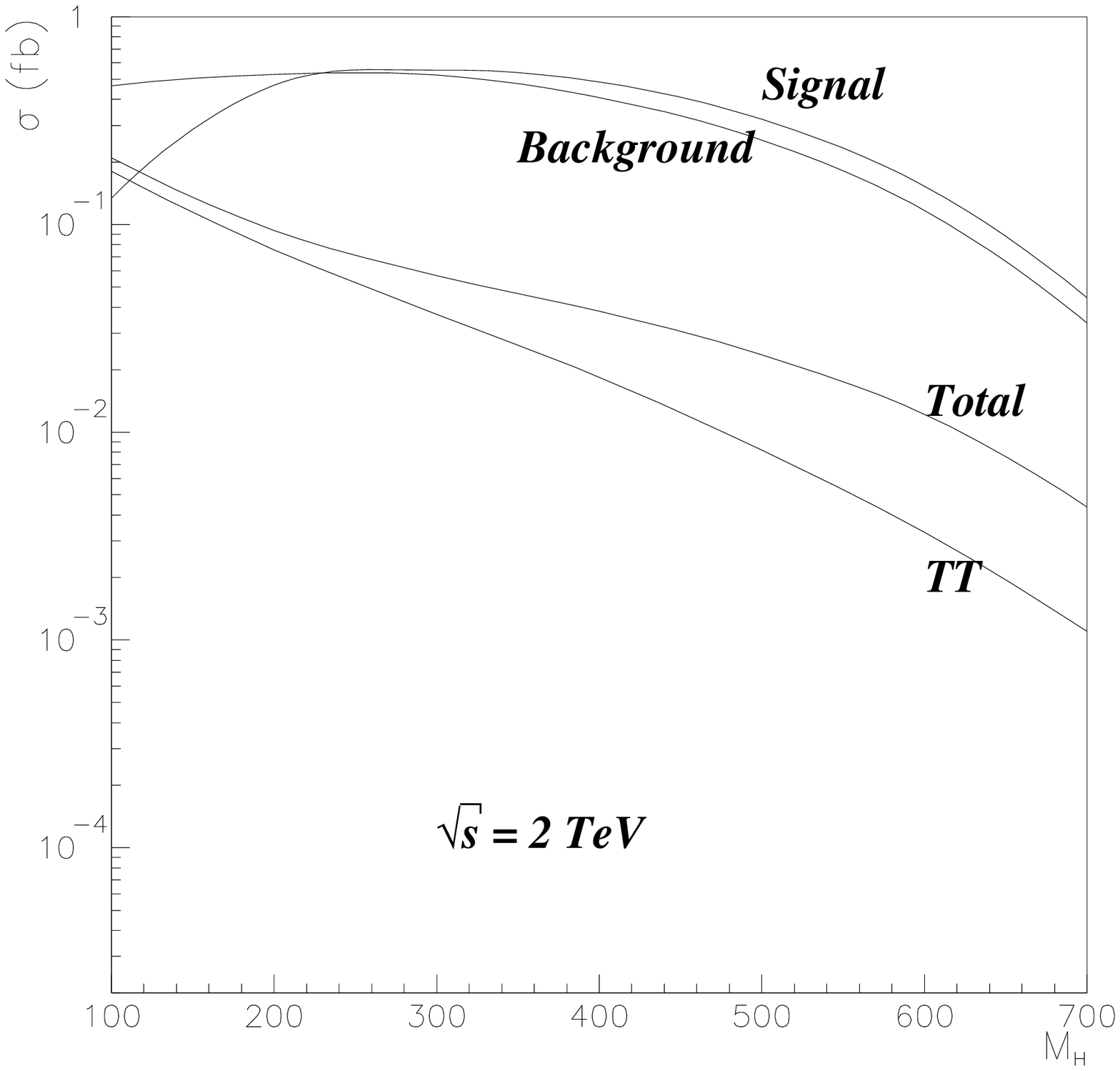}}
\mbox{\epsfxsize=8.5cm\epsfysize=12cm\epsffile{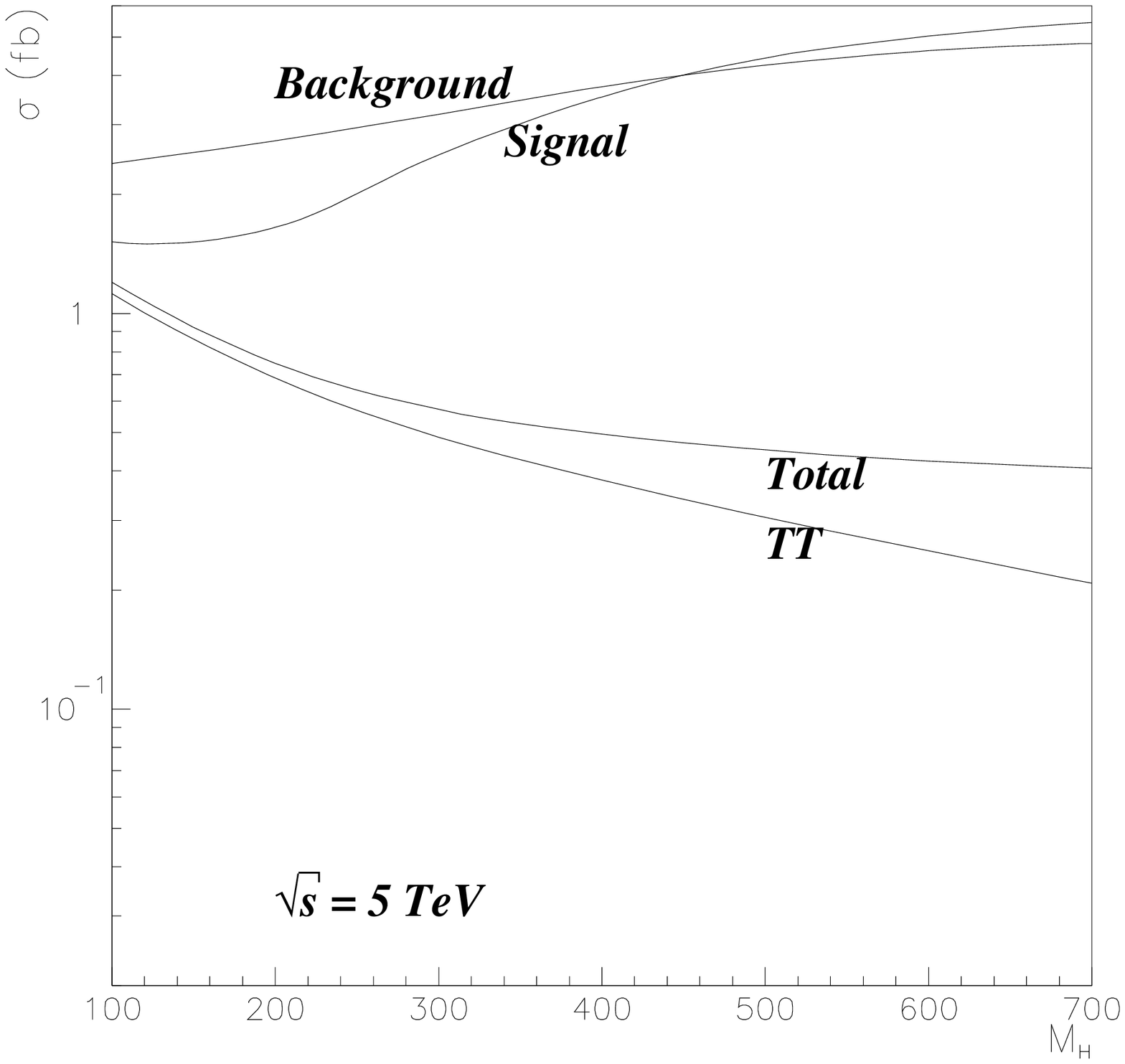}}}
\end{center}
\end{figure*}
\noindent To bring out this important feature in evidence,
and having in view the probing of the Higgs self-coupling,
it is instructive to divide the diagrams into
those containing the Higgs self-couplings (Signal)
and those where this coupling is absent
(Background). As figure~\ref{mhggdep} shows, at 2TeV  there is already
substantial destructive interference that is most
operative when the Higgs mass is in excess of 350GeV (heavy Higgs mass).
At $5TeV$ one sees clearly that both the ``Signal" and
``Background" increase with increasing Higgs
masses (taking $M_H$ up to $M_H=700GeV$),
however they conspire to give a much smaller cross
section. Note that the outgoing transverse $W$'s
dominate the cross section for all values of
$M_H$ when one is far from threshold.

\subsection{Distributions}
\begin{figure*}
\begin{center}
\caption{\label{cosw1tev}{\em Angular distribution of any $W$ with
respect to the beam
for $M_H=100GeV$ and \protect$\protect\sqrt{s}=1TeV$ \protect.}}
\mbox{\epsfxsize=10cm\epsfysize=11cm\epsffile{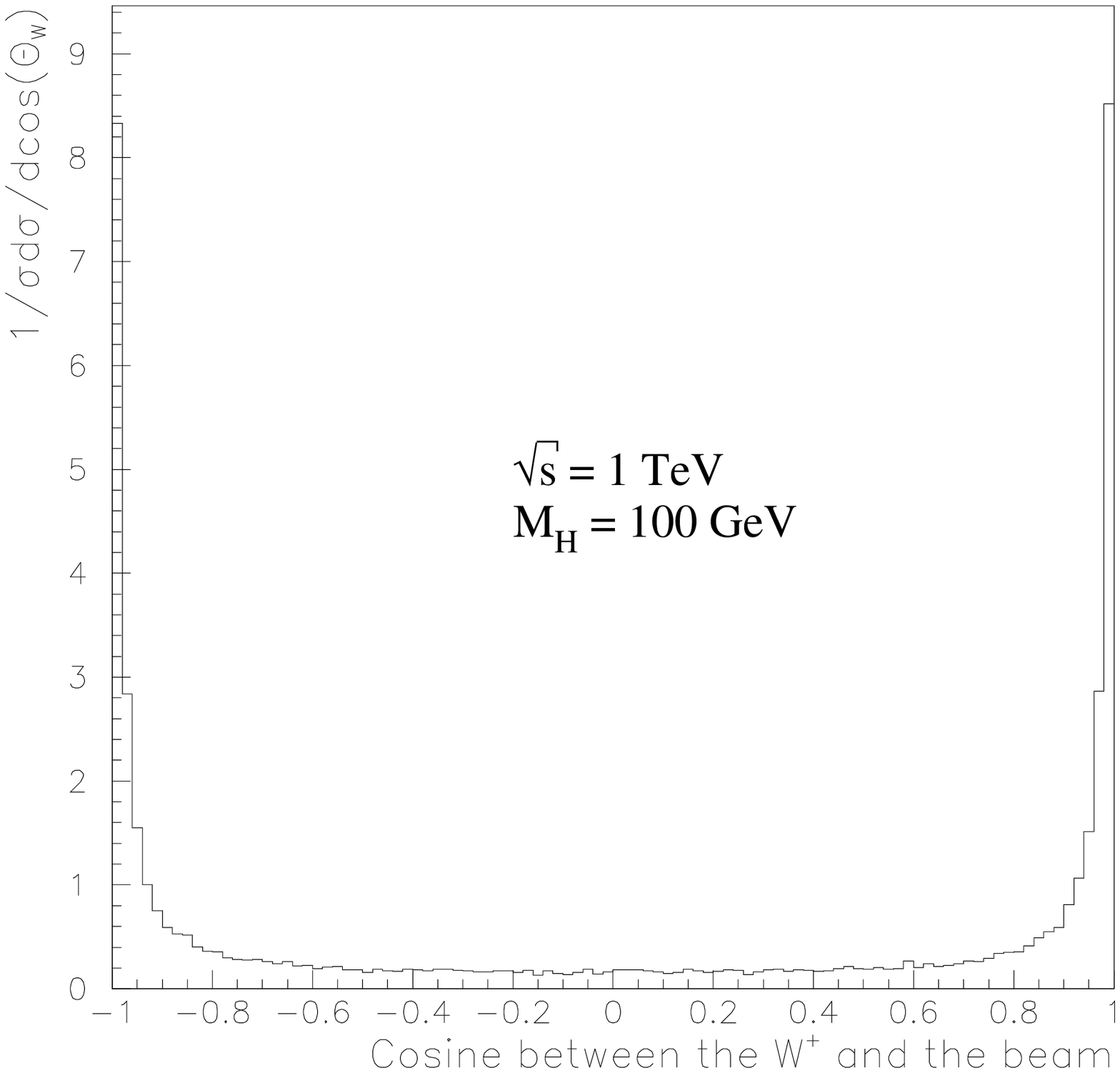}}
\end{center}
\end{figure*}

The logarithmic growth of the cross section, in particular for the transverse
modes of the $W$'s,  is no surprise. Again this stems from
the mutiple $W$ exchanges, this feature is also present in single Higgs
production\cite{Nousgg3v}. Indeed, the cross section is
dominated by the very forward (backward)
$W$'s. This can be seen in the angular distribution of the outgoing $W$'s.
Already at 1TeV
(where the cross section is very small) the forward peak is fantastic
(See Fig.~\ref{cosw1tev})

\begin{figure*}
\begin{center}
\caption{\label{angh}{\em Angular distribution of the least and most
energetic Higgses with respect to the beam .}}
 \mbox{\epsfxsize=15cm\epsfysize=15cm\epsffile{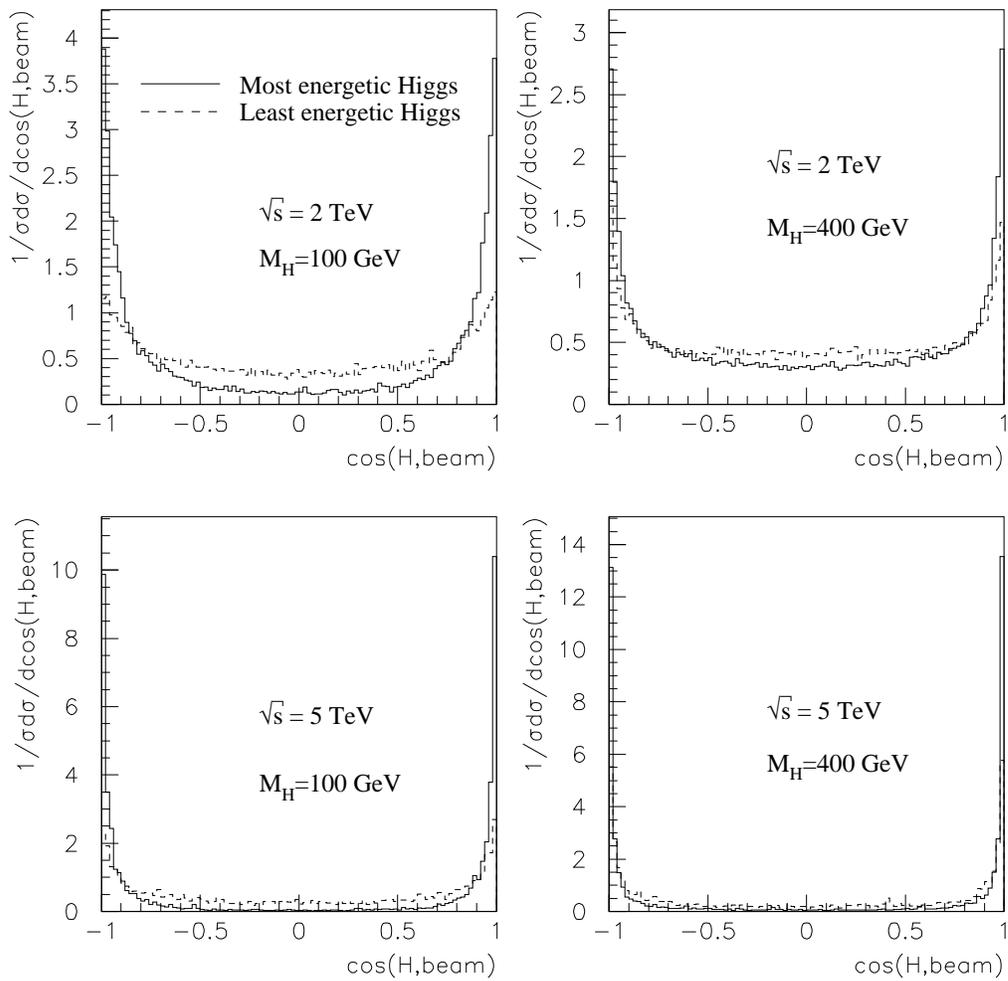}}
\end{center}
\end{figure*}

For a multiple particle production such as this one there are quite a number of
distributions that one may want to study. Knowing these distributions
will be most useful
for optimising the search for New Physics. We have another motivation
for analysing
the different distributions: we would like to find out which diagrams are most
dominant so that one could work out some approximation for the whole
process.
First, since we have two Higgses in the final state we have prefered
to single them out by labelling them as the least and most energetic Higgs.
These two have quite different angular distributions (measured with respect to
the beam) especially for the case of a light Higgs ($M_H~=100GeV$).
For the latter,
while the most energetic H is preferentially carried along the $W$ direction
(that is in the forward/backward), the least energetic has an almost
uniform distribution even for very high \cms energies, Fig.~\ref{angh}.
For a heavier Higgs, the two Higgses are both clearly preferentially forward.

\begin{figure*}
\begin{center}
\caption{\label{disew}{\em Energy distribution of the $W$ for different
Higgs masses at two typical energies. Both the transverse and the
longitudinal $W$ distributions are shown.}}
\mbox{\epsfxsize=15cm\epsfysize=15cm\epsffile{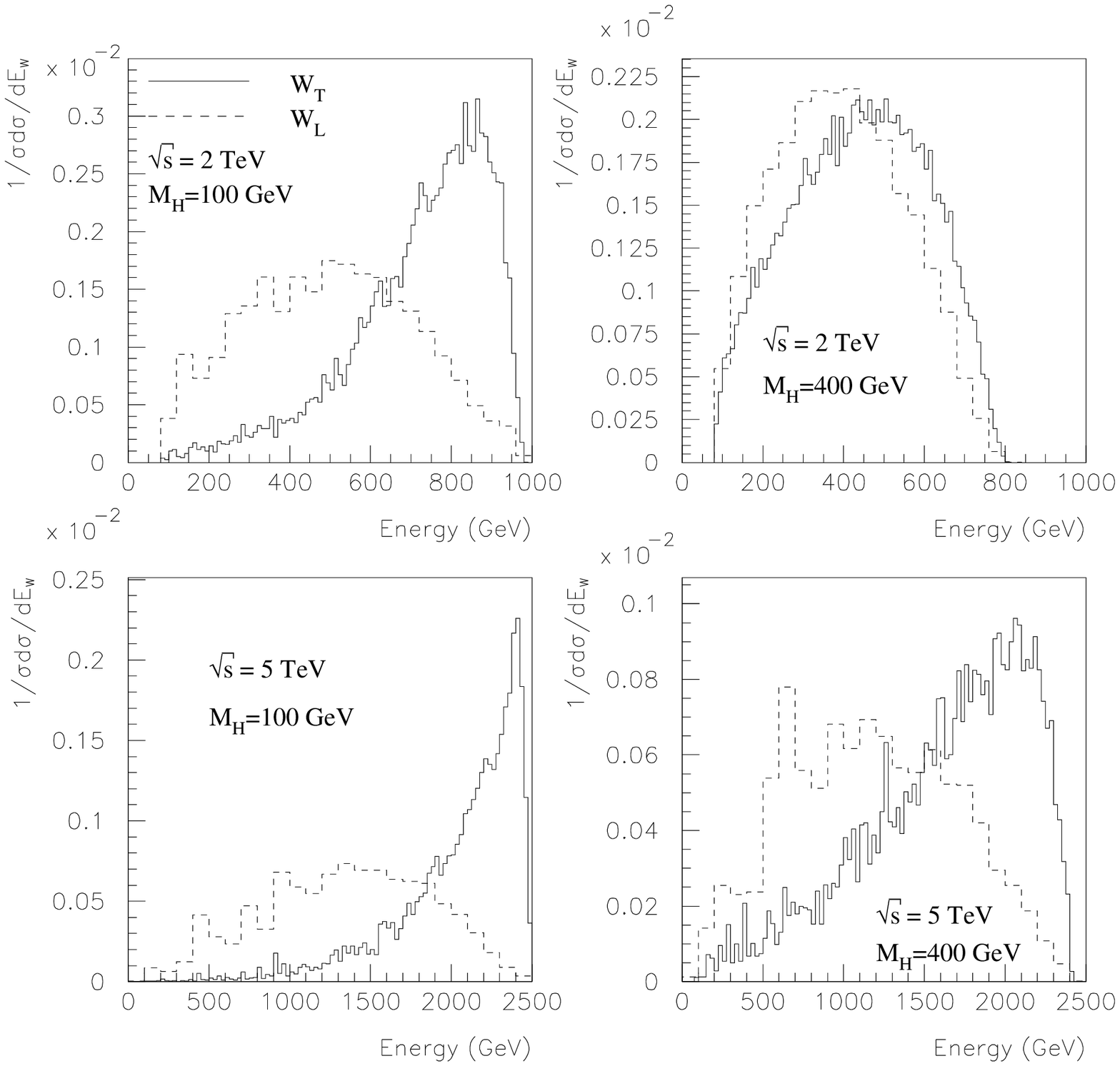}}
\end{center}
\end{figure*}

\begin{figure*}
\begin{center}
\caption{\label{diseh}{\em Distribution in the energy of the least energetic
($H_-$) and the most energetic ($H_+$) Higgs.}}
\mbox{
\mbox{\epsfxsize=8cm\epsfysize=8cm\epsffile{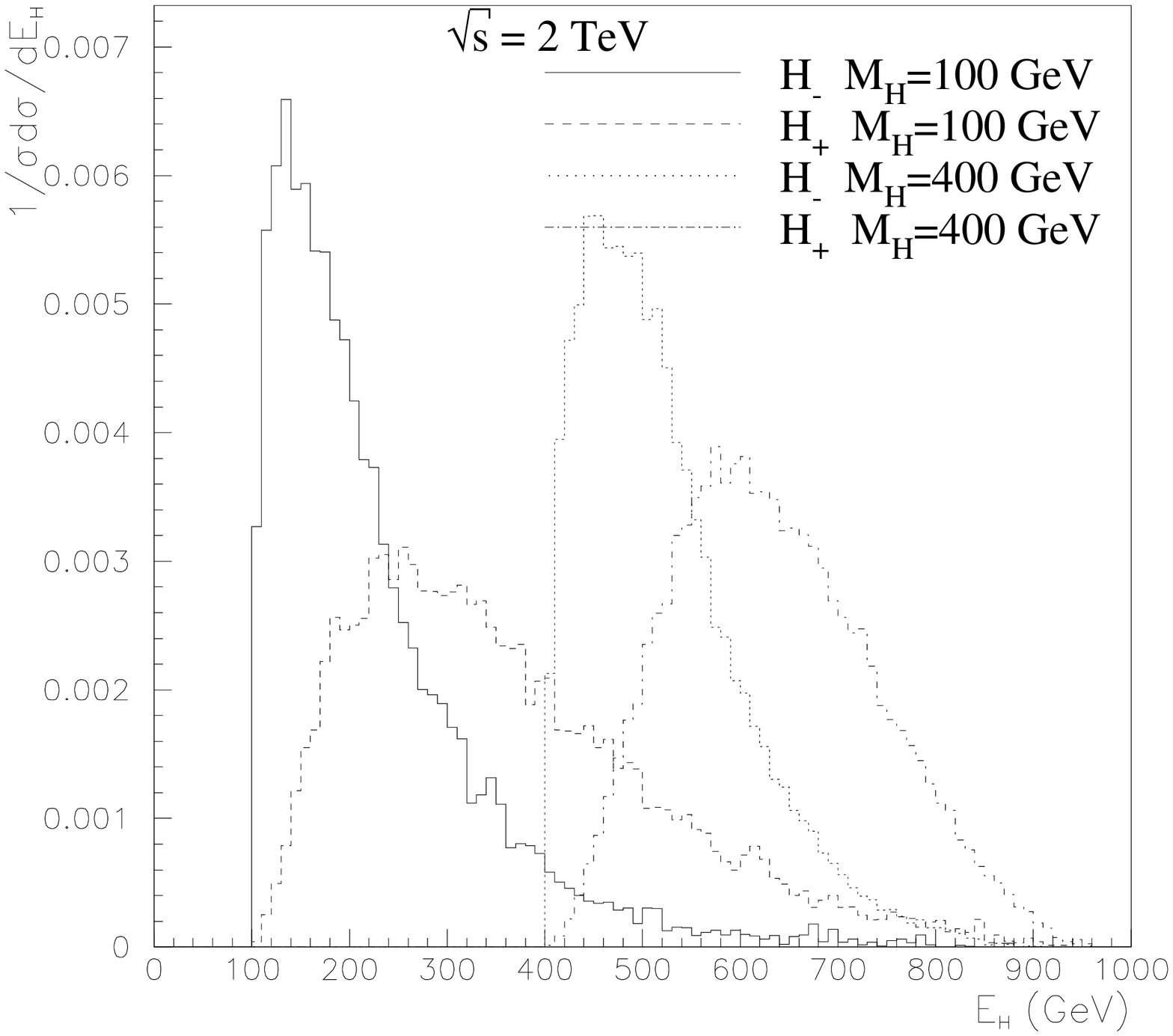}}
\mbox{\epsfxsize=8cm\epsfysize=8cm\epsffile{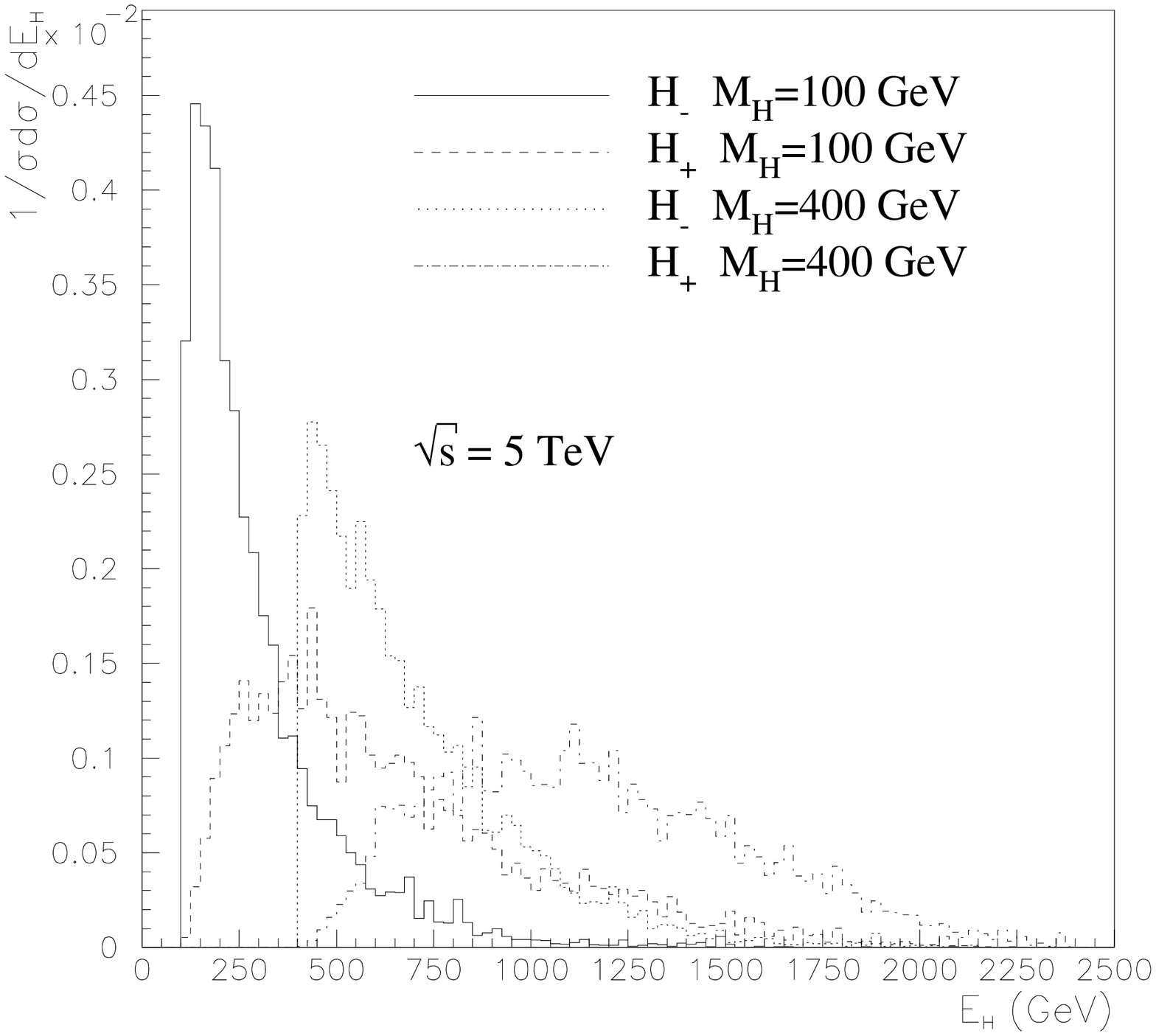}}}
\end{center}
\end{figure*}

As for the energy distributions, we see from Figs.~\ref{disew} that
it is the transverse $W$ that carries
on average most of the energy, the much less dominant longitudinal
$W$'s have a rather uniform
spectrum. Especially for the light Higgs, one of the two Higgses
is almost a bremstrahlung Higgs that
takes along very little momentum Figs~\ref{diseh}. This is the case
even at very high energies
(5TeV) and means that the decay products (the b's)
will be clearly separated.
In view of the characteristics of the energy and escape angle
of the Higgses, that show different
behaviour for the light and heavy Higgs, there is in the case
of the light Higgs some evidence
for the fact that the contribution of Higgses emitted as bremstrahlung
off the external $W$'s is not
negligible. This is because the least energertic Higgs carries little
energy while the most energetic
is along the $W$ direction, thus both Higgses end up
forming a small invariant $WH$ mass.

\begin{figure*}
\begin{center}
\caption{\label{Q2double}{\em Double distribution in the reduced variables
$X_\pm$ that measure the virtuality of the fusing $W$'s.}}
 \mbox{\epsfxsize=14cm\epsfysize=14cm\epsffile{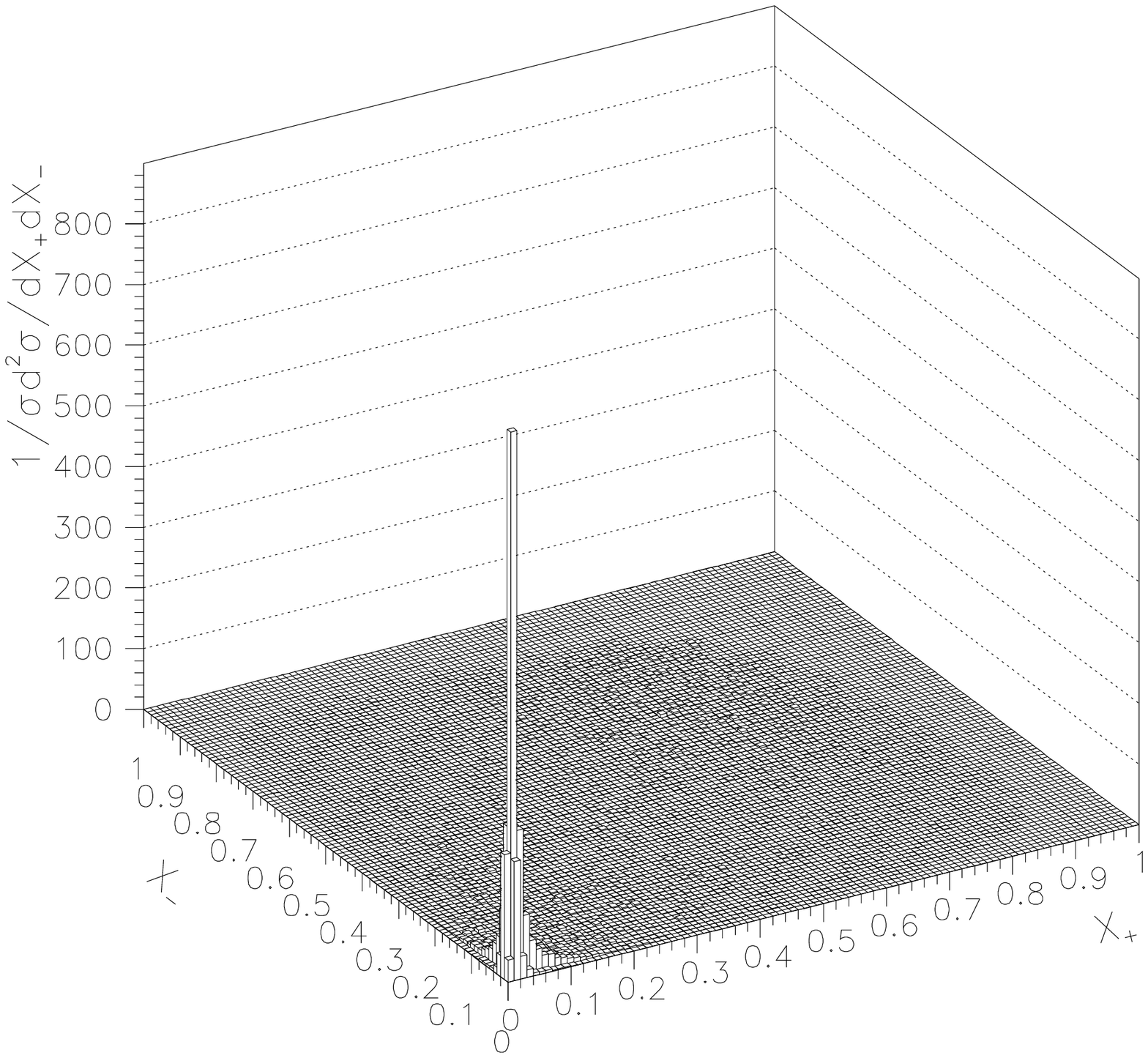}}
\end{center}
\end{figure*}

There are other types of diagrams that, especially for the heavy
Higgs mass, are dominating.
These are the fusion type diagrams where each photon splits
into two $W$, the external outgoing
$W$ (spectator $W$) going very forward and taking a large
amount of energy as we saw. The other,
internal
$W$, takes part in the hard process and triggers $W W \ra HH$.
If these internal $W$'s are
quasi-real (almost on shell) than the t-channel diagrams
are every much enhanced.
 To bring this important feature into prominance we introduce
the variables $X_\pm$ that measure the virtuality of the fusing $W$'s.
 $X_+=(m_W^2-q_+^2)/s$, where
$q_+ = p_{\gamma_1}-p_{W^+}$, ($q_+^2=-Q_+^2$)  and $X_-=(m_W^2-q_-^2)/s$
with $q_- = p_{\gamma_2}-p_{W^-}$. When the $W^+$ is emitted
with high energy in the
 ``parent"  photon   direction, the variable $X_+$ is very  small.
The domainance of the fusion diagrams  is well rendered by
Fig.~\ref{Q2double} that
clearly shows that the bulk of the events originate
from a very small region of
phase space tightly clustered around both $X_\pm \sim 0$.
It should also be noted
that for very light Higgs, when it is emitted either along the $W$ direction
or with litte energy (essentially a bremstrahlung Higgs)
the variables $X_\pm$
calculated for the bremsstrahlung diagrams can also be small.
This observation of the importance of the fusion diagrams, referred to earlier,
 will become crucial when we will
derive a high energy approximation for the process.

\begin{figure}
\begin{center}
\caption{\label{dismhh}{\em Distribution in the reduced
invariant mass of the Higgs system for $M_H=100,400GeV$
and $\protect\sqrt{s}=2TeV,5TeV$.}}
\mbox{
\mbox{\epsfxsize=8cm\epsfysize=8cm\epsffile{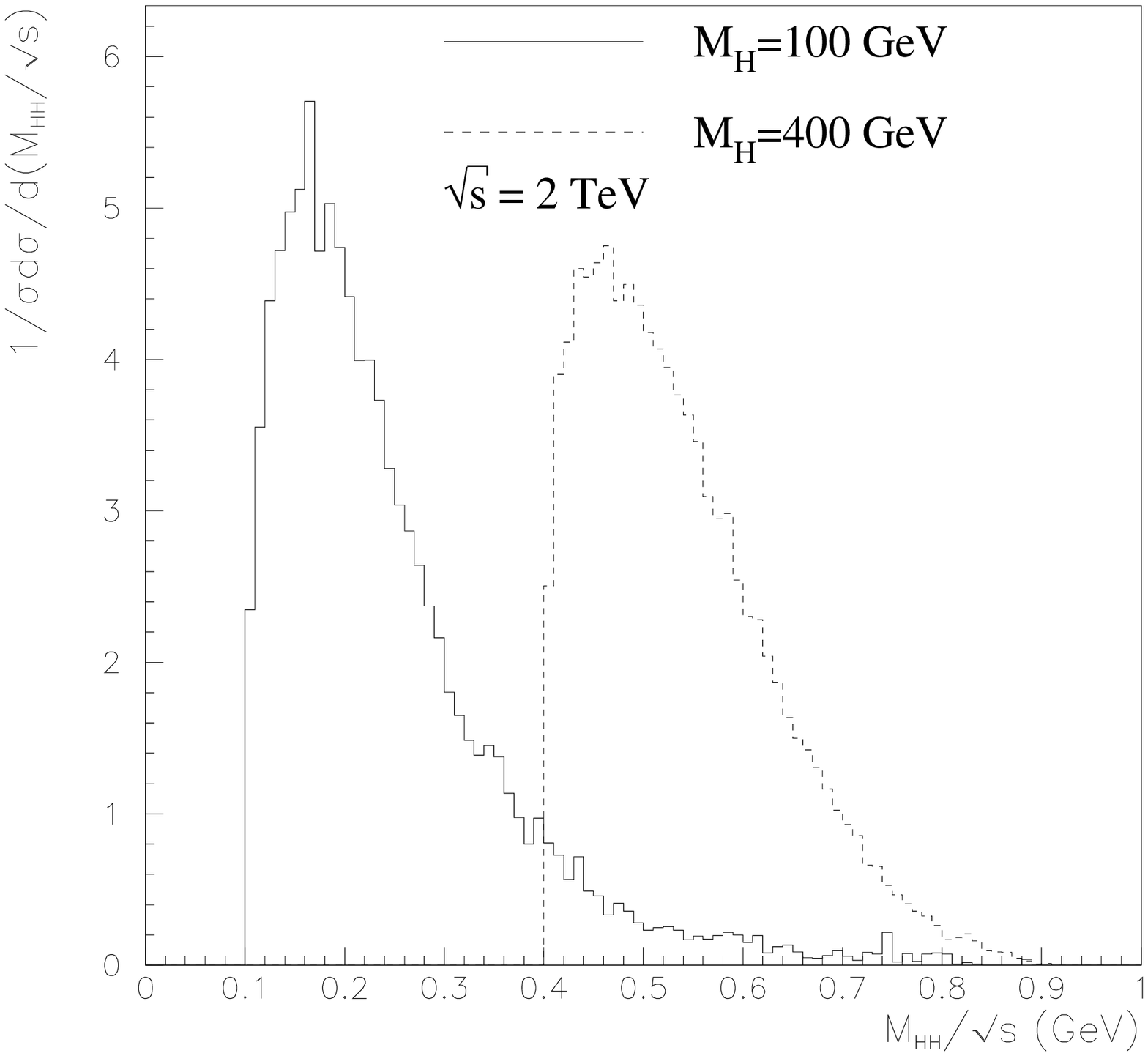}}
\mbox{\epsfxsize=8cm\epsfysize=8cm\epsffile{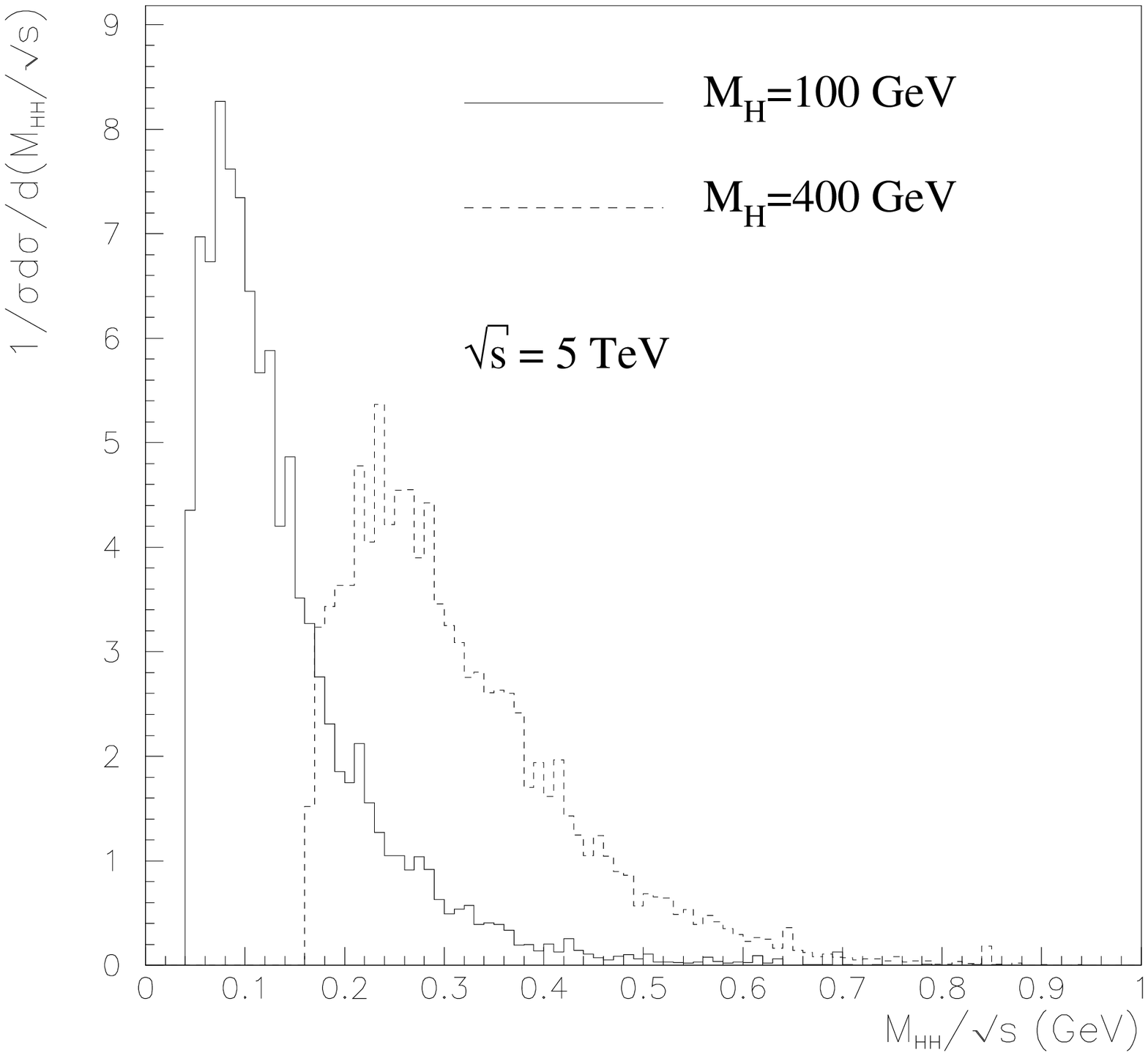}}}
\end{center}
\end{figure}

The above distributions together with the discussion we had
on the Higgs mass dependence
are an indication that the leading topology is one
with both outgoing $W$ being transverse. They are  emitted very forward along
the beam hence forcing the intermediate $W$'s to be quasi-real
especially at high energies.
All these characteristic are consistent with an interpretation,
especially for heavy Higgs,
 in terms of the dominance
of fusion diagrams with the photon splitting into a spectator
$W_T$ (transverse) and another
$W$ that takes part in the hard scattering process $W^+W^- \ra HH$.
We will see in
section~4 that this is dominated by $W_L W_L$. From a previous
study\cite{Parisgg,egnuwh}
we have given formulae for the structure function describing
a $W_L$ inside a photon (see below
also), these lead to an effective  $W_L W_L$ luminosity that
falls like $1/\hat{s}$, where $\hat{s}$ is the invariant mass
of the $W_L W_L$ subsytem. We thus
should expect that the largest contribution to the total cross
section to come from
values of the invariant $M_{HH}$ mass not too far from threshold,
if indeed the fusion diagrams
contribute substantially.
 As fig.~\ref{dismhh} confirms, the average
invariant mass is not sensibly above threshold especially for a heavy Higgs.
For a light Higgs
the invariant mass of the Higgs system is only a fraction of the total energy.

\setcounter{subsection}{0}
\setcounter{equation}{0}
\def\thesubsection {\thesection.\arabic{subsection}}
\def\theequation{\thesection.\arabic{equation}}
\def\thefigure{\arabic{figure}}

\section{Comparison with other processes at the linear colliders}
\subsection{Light Higgs at  \epemt colliders}
\begin{figure*}
\begin{center}
\caption{\label{eegghh100}{\em Comparison of cross sections for double Higgs
production at \epemt and $\gamma \gamma$ reactions
for a light Higgs $M_H=100GeV$.}}
\mbox{\epsfxsize=14cm\epsfysize=14cm\epsffile{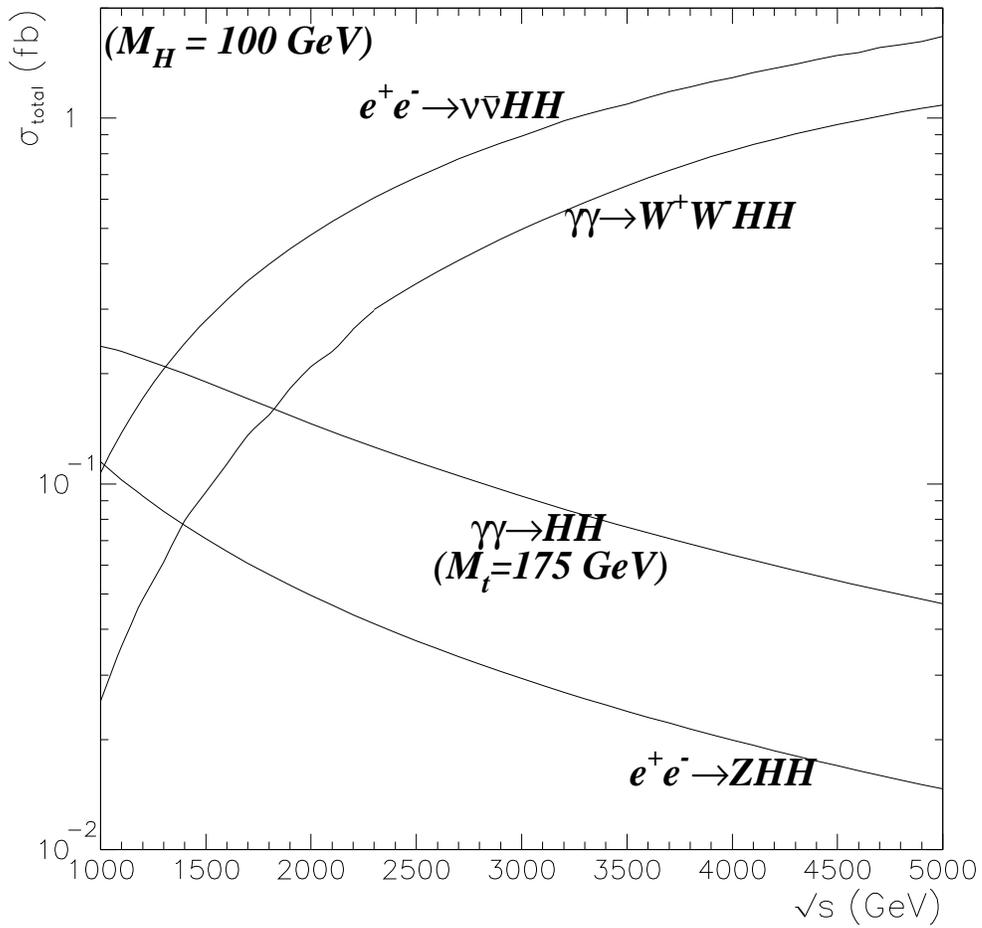}}
\end{center}
\end{figure*}

At the same order in the coupling constants as the process we are studying,
double Higgs production can be generated at the loop level via
$\gamma \gamma \ra HH$ and could allow the probing of the
triple Higgs vertex\cite{Jikiahh}.
However, this test will have to rely on a very precise value
of the $H \ra \gamma \gamma$
branching ratio.  At a 2TeV cms energy the cross
section is of the same order as the $WWHH$ production for a Higgs mass of
$100GeV$. When one is not penalised by phase space  $WWHH$ production
is larger, for instance at $5TeV$ there is an order of magnitude between
the two cross sections. The equivalent loop-induced double Higgs production
in \epemt has been found to be much too small\cite{eehh},
at best it has a cross section
of 0.1fb and decreases quickly with energy. However the most efficient
means for double Higgs production in the \epemt mode is via $WW$ fusion
leading to \eennhht. The double Higgs bremstrahlung
($\epem \ra ZHH$)
is only competing at relatively low energies where the event
sample is too low
to be useful. We have recalculated the \eennhht cross section.
As Fig.~\ref{eegghh100} shows,  \eennhht
exhibits the same
logarithmic growth and is slightly larger than \ggwwhht for the same
cms\footnote{We have not counted the part $e^+e^- \ra HHZ\ra HH\nu_e
\bar{\nu}_e$ with the fusion diagrams}. This
is partly due to the larger phase space allowed for the \epemt process.
 Another reason is that the  \eennhht cross section could
be interpreted solely in terms of the fusion of $WW$ that  rescatter into
HH, while the equivalent  \gamgamt involves bremstrahlung type diagrams
that for a light Higgs $\sim M_W$ interfere destructively with the fusion type
diagrams.  Note that these  fusion type diagrams $WW \ra HH$ are also
common to the loop induced $\gamma \gamma \ra HH$ and $\epem \ra HH$.
This last remark
also explains why $\gamma \gamma \ra HH$ is larger than $\epem \ra HH$,
 $\gamma \gamma \ra W^+ W^-$ is about an order of magnitude larger than
 $e^+ e^- \ra W^+ W^-$\footnote{Another reason, is that $\gamma \gamma \ra HH$
receives an important contribution from the top.}.
This said, all mechanisms for double Higgs production
give, for the foreseable future colliders, a cross section that is
below .5fb for the light Higgs scenario with the largest
cross section occuring in the classic \epemt.
The yield of double Higgs can be even larger
in the \epemt mode. If both the $e^+$ and $e^-$
are polarised one can gain as much as a factor of four!
No such beefing up can occur in the \gamgamt mode as all combinations
of the photon polarisations give sensibly
the same values as shown earlier.
If anything, \gamgamt cross sections will suffer
some reduction when one takes the photon spectra into consideration.

\subsection{Convoluting with the photon luminosity spectra}
\begin{figure*}
\begin{center}
\caption{\label{spectra}{\em (a) Photon-photon luminosity spectra
for different polarisation of the laser beams $P_c$ (circular)
and the electrons $\lambda_e$) of the linac. (b) shows the relative
contribution of the $J_Z=0$ and the $J_Z=2$ luminosities.}}
\mbox{\epsfxsize=15cm\epsfysize=10cm\epsffile{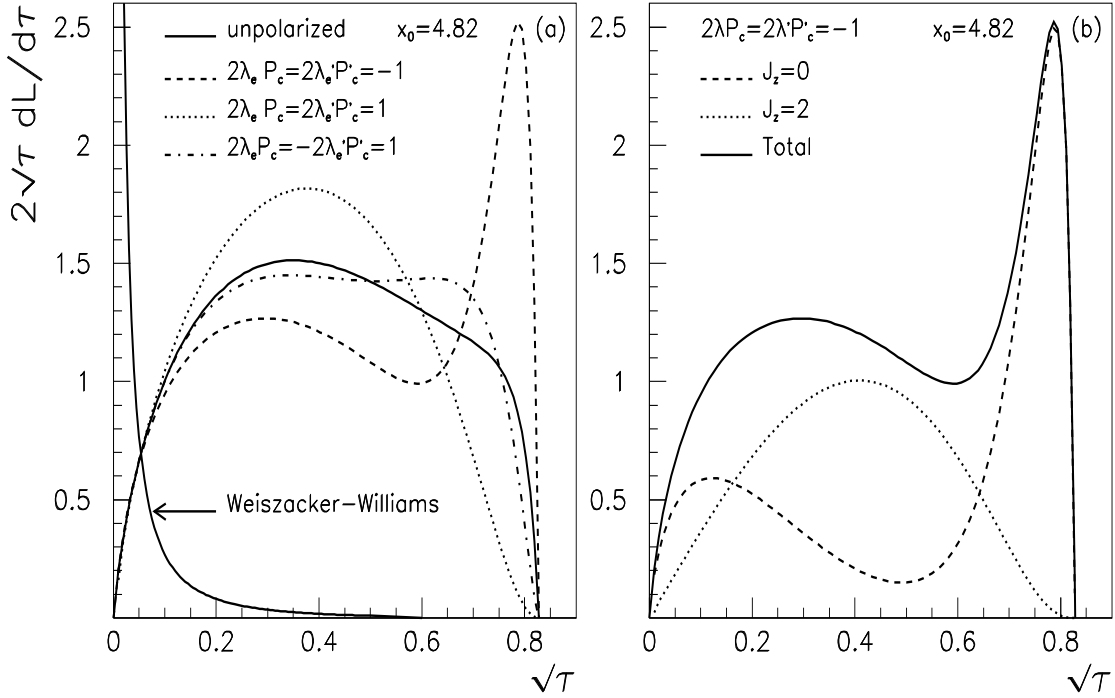}}
\end{center}
\end{figure*}

 At this point it is necessary to discuss how the cross section for double
Higgs production in $\gamma \gamma$ is affected by including the photon
luminosity spectra. The colliding high energy photons are obtained by Compton
backscattering of an intense laser light on the single-pass electrons.
The spectra used
by almost all physics analyses till now have been based on the much
adverstised luminosity
functions of\cite{Laser}. Nearly all studies have considered that
the conversion point coincided with
the interaction point. By increasing the conversion distance one
obtains more monochromatic
spectra peaked towards the highest possible energy but one looses
in luminosity as the
softer converted photons would not reach the interaction point.
To make the comparison with
other physics studies conducted for these types of machines we
will only illustrate the
case with a zero conversion distance and with a parameter
for the setting that allows
the highest possible \cms energy and yet  does not trigger
electron pair production
through the interaction of a laser photon and a converted photon.
This corresponds to the parameter $x_0=4.8$ \cite{Laser,Parisgg}
and means that \cms energies up to $83\%$ of the original
\epemt can be reached. The issue of polarisation is crucial.
Figure \ref{spectra} shows that by choosing the circular
polarisation of the laser, $P_c$, and that of the corresponding
electron, $\lambda_e$, such that for both ``arms" of the colliders one has
$2 \lambda_e P_c=2 \lambda_e' P_c'=-1$ one obtains a peaked spectrum toward
the highest
$\tau$: $\tau=s_{ee}/s_{\gamma \gamma}$. For this peaked spectrum one also
sees that if one takes
$P_c=P_c'=+1$ then the produced hardest photons are mainly in the $J_Z=0$.
 For our process where we
need to be at the highest possible $s_{\gamma \gamma }$ it is
clear that we should prefer a peaked spectrum.
Selecting between a $J_Z=0$ and a $J_Z=2$ dominated spectrum
is not terribly important as the process depends very litte
on the initial polarisation.

It has very recently been realised\cite{Sheffield95}
 that taking into account multiple scattering and non-linear
effects these ``ideal" spectra are disrupted.
The main characteristic is that the luminosity at the
higher end of the spectra is lowered while one
has an increase in the luminosity of
 the soft
photons. The detailed Monte Carlo simulations \cite{Sheffield95}
depend quite critically on the parameters of the beams
etc.., but because of the typical spectra arrived at, it would be
a good approximation to assume
that one has an almost monochromatic spectrum peaked at the highest possible
$\sqrt{s_{\gamma \gamma}}$
energy but with a luminosity of about a factor 2-5 lower than the original
\epemt\cite{Sheffield95}.

\begin{figure*}
\begin{center}
\caption{\label{convol}{\em The \ggwwhht cross section after convoluting with
the ``ideal" luminosity spectra for different combinations of the polarisations
of the electrons and the laser. The Higgs mass has been set at 100GeV.}}
\vspace*{-1.2cm}
\mbox{\epsfxsize=14cm\epsfysize=14cm\epsffile{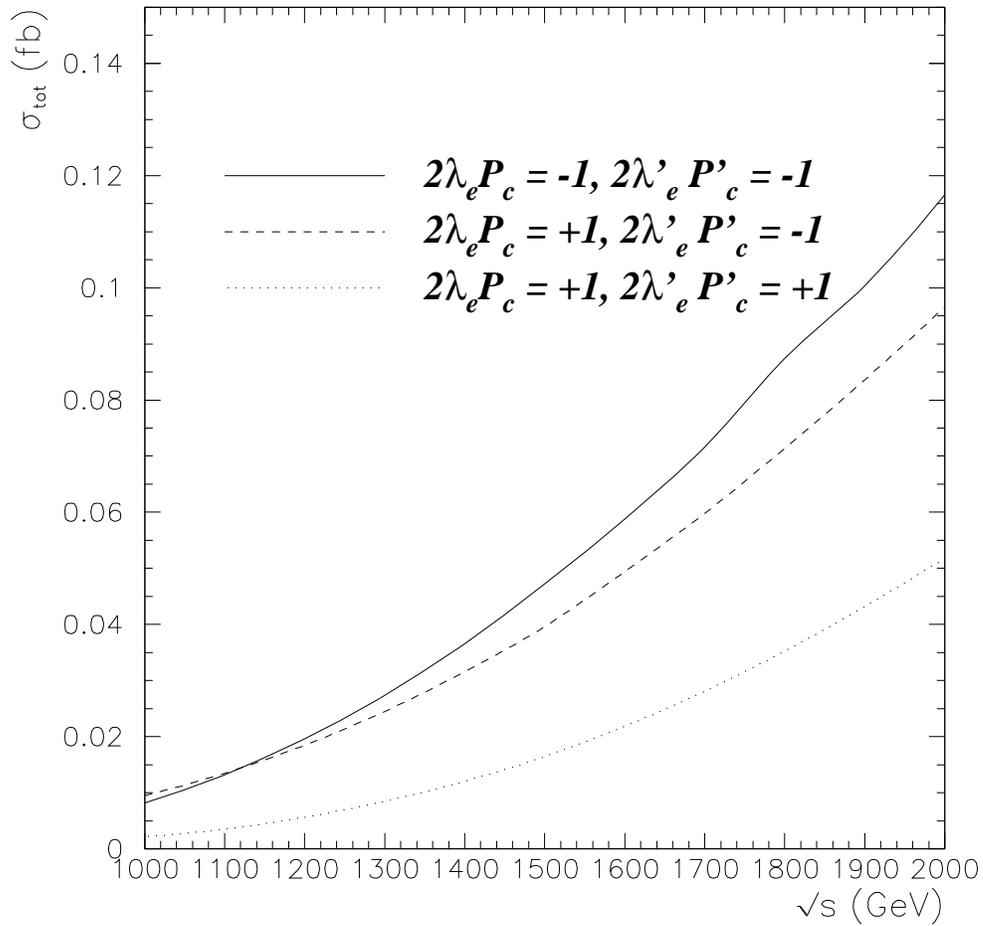}}
\end{center}
\end{figure*}
For our process, convoluting with the ``ideal" spectrum
\footnote{ This is  technically  more time
consuming as it introduces yet two more integration variables.}
we find in fact that at 2TeV
\epemt cms energy and for a Higgs mass of
$100GeV$ the cross section with the peaked
spectrum drops by about at least a factor of 2 compared with the result
 without convolution, see
Fig.~\ref{convol}. Of course
it is worse by a factor 4 if the broad spectrum
$2 \lambda_e P_c=2 \lambda_e' P_c'=1$ is taken.
Therefore, we see that had we taken the preliminary
results on the spectra that take into
account multiple scattering etc,  that suggests to include a factor 2
(at least) reduction in the
$\gamma \gamma$ luminosity we would have obtained
the correct order of magnitude of
the convoluted cross section. \\

Comparing now the \ggwwhht with double Higgs
production in \epemt for the foreseable
colliders it is clear that the yield of double Higgs
is larger at the \epemt option by
at least a factor 2. In  real life, there could easily
be an order of magnitude between the
two yields considering the availabilty of  polarisation.

\subsection{Higgs mass dependence of the \eennhht cross section}

\begin{figure*}
\begin{center}
\caption{\label{eemh}{\em Dependence of the \eennhht
cross section on the Higgs mass at
2TeV and 5TeV.The relative contributions of the signal
diagrams containing the
H triple vertex and the background are shown.}}
\mbox{
\mbox{\epsfxsize=8cm\epsfysize=12cm\epsffile{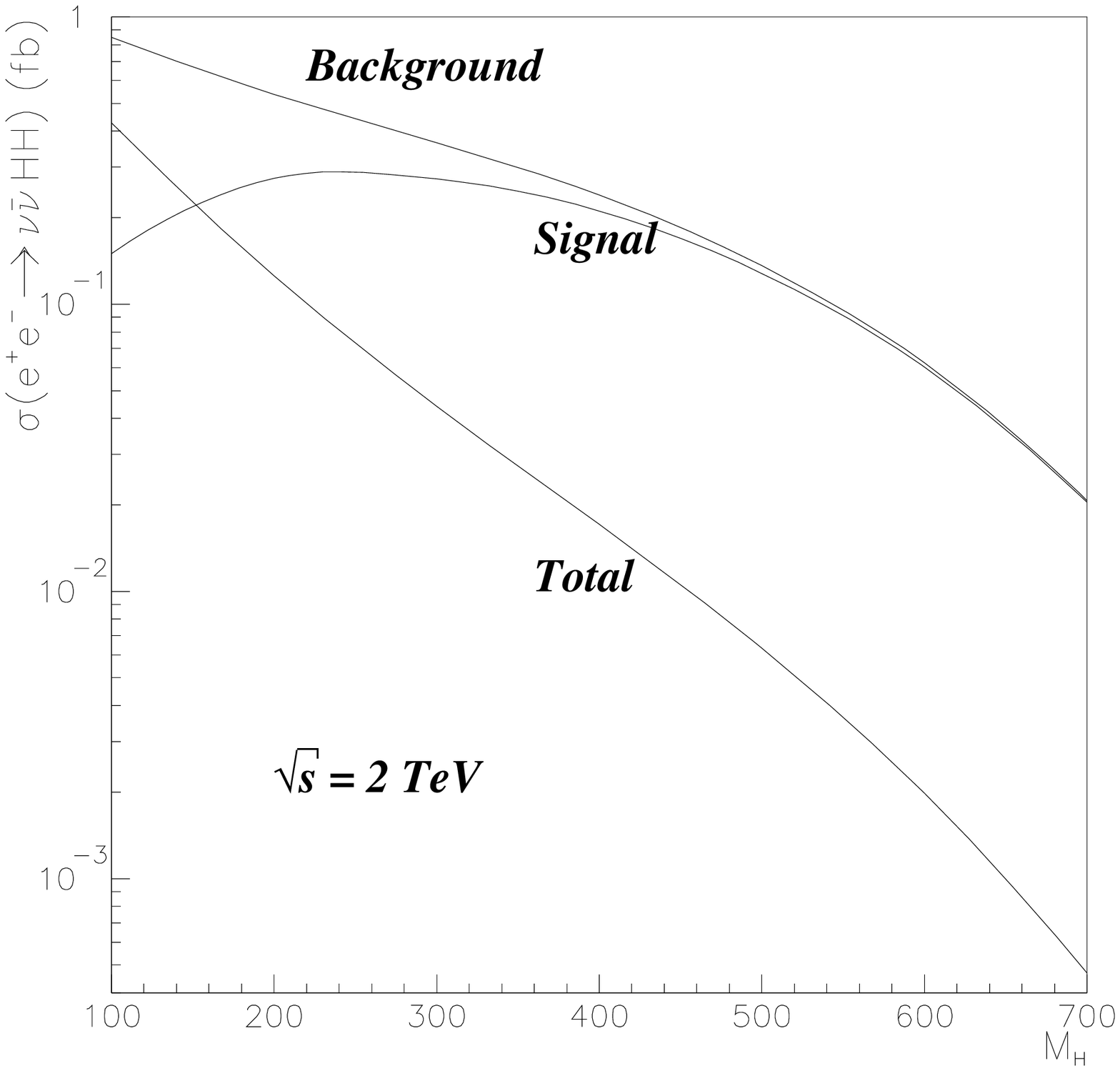}}
\mbox{\epsfxsize=8cm\epsfysize=12cm\epsffile{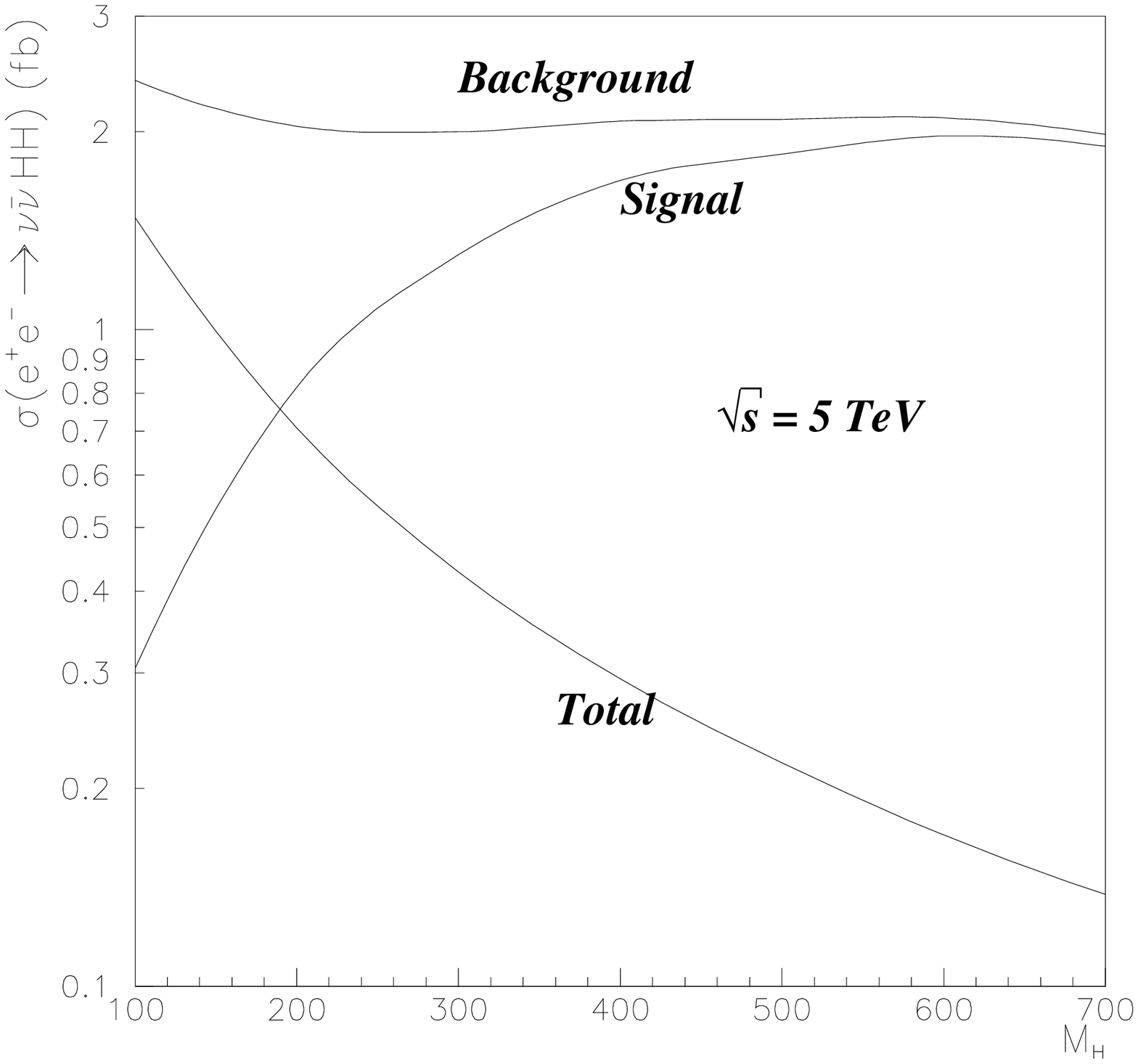}}}
\end{center}
\end{figure*}
\begin{figure*}
\begin{center}
\caption{\label{eemhh}
{\em Distribution in the reduced invariant mass of the
$HH$ system for} $M_H=100,400GeV$ {\em and} $\protect\sqrt{s}=2TeV,5TeV$.}
\mbox{\epsfxsize=12cm\epsfysize=12cm\epsffile{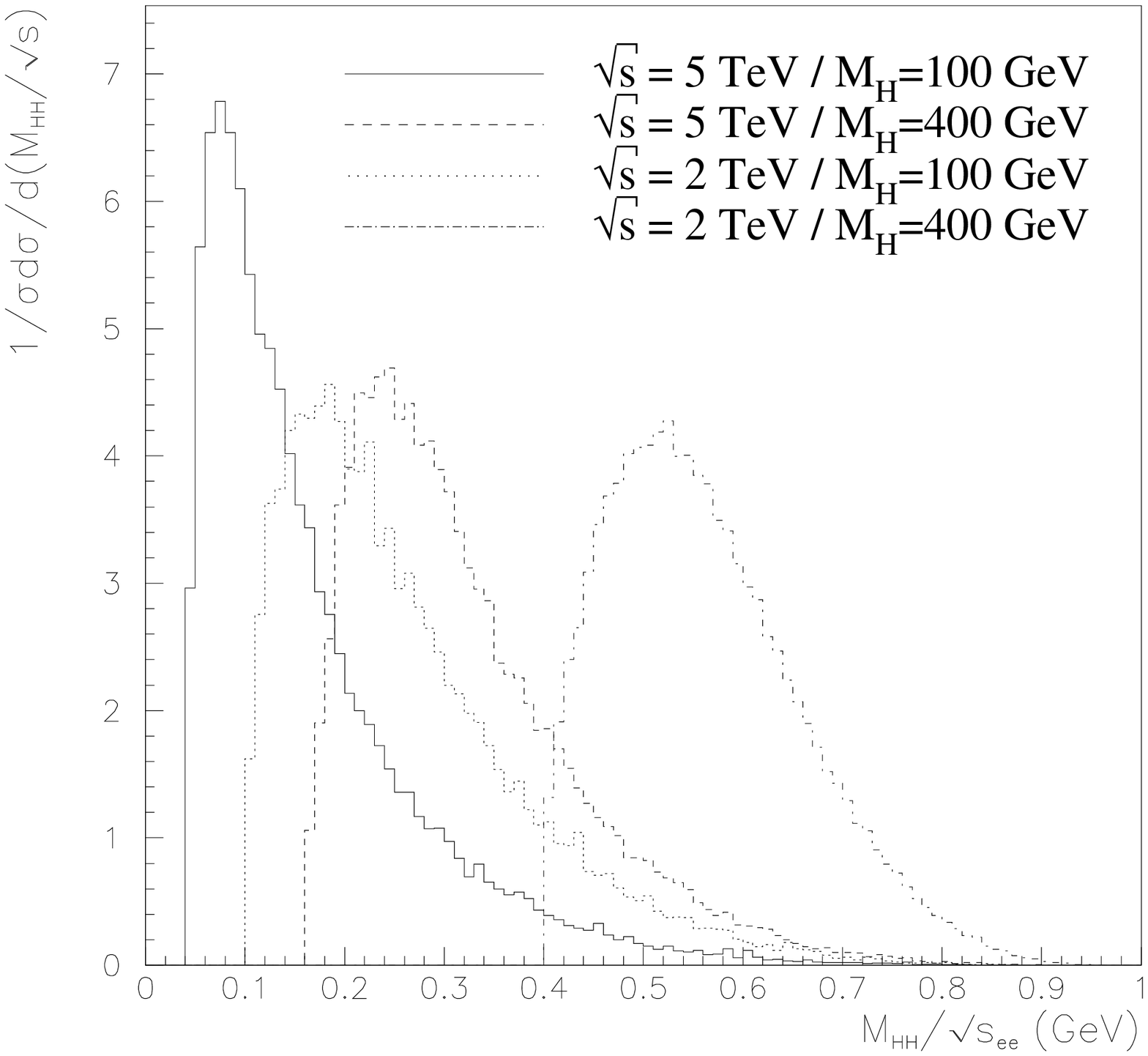}}
\end{center}
\end{figure*}

One revealing characteristics that we observed in the \ggwwhht was that
the accuracy in the probing of the triple Higgs vertex was
 acutely dependent on the mass of the
Higgs since there was destructive interference between the signal
diagrams that contain
the triple vertex and the rest. This destructive interference
becoming more severe as the
mass increases. As figures~\ref{eemh} show, this feature is still
present and it therefore
confirms that the background diagrams though not containing
the HHH vertex do
 get balanced out by
the $W_L W_L  H H$. This is a suggestion that the quasi-real
W's are preferentially
longitudinal. As we will see, $WW \ra HH$ is essentially
triggered by the longitudinal W.
Therefore we have strong indications to suspect that the
whole cross section can be well
recovered by the effective $W_L$ approximation.
Another supporting evidence is that this
approximation has been shown to work very (rather)
well for a heavy (light) Higgs for
$pp$ reactions at $40TeV$\cite{pphh}. Moreover, as Fig.~\ref{eemhh}
indicates the invariant mass of the
HH system is, for high enough energy,
clustered around small values of this invariant mass
so that the effective $W_L$ approximation (EWA) should work quite well.

\subsection{Heavy Higgs}
\begin{figure*}
\begin{center}
\caption{\label{compar400}{\em Energy dependence of
the main cross sections for double Higgs production for a heavy Higgs.}}
\mbox{\epsfxsize=12cm\epsfysize=12cm\epsffile{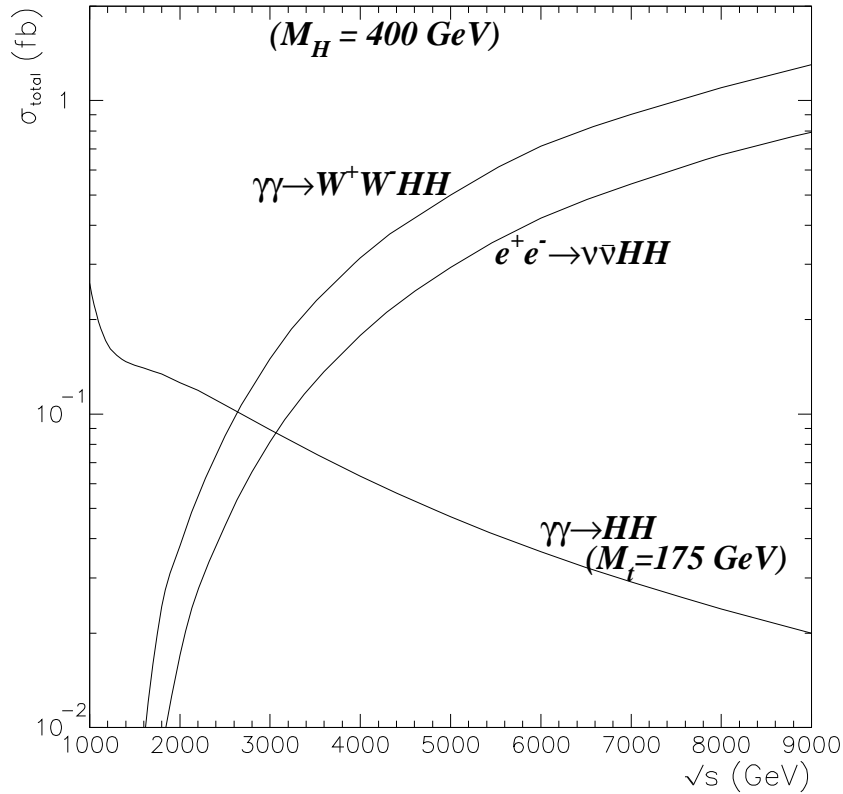}}
\end{center}
\end{figure*}

The comparison between all mechanisms for double
Higgs production when the Higgs is heavy
very much favours \ggwwhht as soon as phase space
factors (due to the accompanying $W$ pair)
are negligible.
For the 2TeV collider and a $400$GeV Higgs, there is enough phase space
to make \ggwwhht larger than \eennhht, though
both cross sections are far too small,
but not enough for  \ggwwhht  to be the dominant cross section.
The loop-induced
$\gamgam \ra HH$\cite{Jikiahh} is larger and could be
 measurable. However, very quickly \ggwwhht takes over
and one sees that it becomes the main mode for
double Higgs production already at
$3TeV$ (see Fig.~\ref{compar400}).

\section{The structure function approach and $W^+ W^- \ra HH$}
In order the check whether the equivalent $W$ approximation works,
one needs to know not only the
distributions of the $W$ inside the electron
and the photon but also the behaviour of the
the hard process, which for the case at hand is \wwhht. We will see
that it is dominated, by far,
by the longitudinal $W$'s in which case it is sufficient to only use
the longitudinal W distribution
function.

\subsection{ \wwhht }

The amplitude for the process $W_L W_L \ra HH$ has been calculated before
\cite{pphh}.
We have recalculated this amplitude by also
allowing for the triple Higgs vertex
to have a value different from that in the \sm and  have derived the other
helicity amplitudes.  We take the  $H^3$
vertex to have a strenght $h_3$ times
of what it is in the
\sm (see next section).
Denoting the velocities of the $W$ and $H$ as
$\beta_{W,H}=\sqrt{1-4M_{W,H}^2/s}$, we introduce the
``enhanced coupling" $r=M_H^2/M_W^2$ while we define
$x_0=(1+\beta_H^2)/2\beta_W \beta_H$.
$x$ is the cosine of the scattering angle $\theta$: $x=\cos\theta$. We find

\bea
\tilde{\cal M}_{LL} &=& \frac{g^2}{2} \left\{
\frac{1}{\beta_H \beta_W^3} \left( \frac{1}{x-x_0} - \frac{1}{x+x_0} \right)
(r \frac{M_H^2}{s} + \beta_H^2+\beta_W^4) \right. \nonumber \\
&+& \left. \frac{1}{\beta_W^2} (2-\beta_W^2-r) + \frac{3 h_3 r}{4}
\left(\frac{1+\beta_W^2}{1-M_H^2/s} \right)\right\}
\ena

It is important to note that there is  a cancellation between the terms
 of enhanced strength $\propto r=M_H^2/M_W^2$, as pointed out in section~2:
\bea
\tilde{\cal M}_{LL} \ra  \frac{g^2}{4} r (3h_3-2) +\dots
\ena
The term in $h_3$ comes solely from the s-channel and
for the \sm values there is indeed
a cancellation, though not complete. We clearly
see here that if there is a deviation
in the \sm value, it is made more conspicuous for
higher values of the Higgs mass,
since the terms
in the anomalous couplings are enhanced as the  Higgs mass increases.

Obviously the $h_3$ dependence only occurs for like-sign
$W$ helicities.Thus the remaining
 transverse modes that do not have an enhanced coupling
factor $r$ are not so conducive
to testing these couplings.  The other helicity amplitudes are

\beqn
\tilde{\cal M}_{+L} &= &\frac{g^2}{4 \beta_W^2}
\sqrt{ \frac{2 M_W^2}{s}} \sin\theta
\left(\frac{1}{x-x_0}+\frac{1}{x+x_0}\right) \ (r-2) \nonumber \\
\tilde{\cal M}_{+-} &  =&g^2 \frac{\beta_H}{4\beta_W} \sin^2\theta
\left(\frac{1}{x-x_0}-\frac{1}{x+x_0}\right) \nonumber \\
\tilde{\cal M}_{++}& = &g^2 \left\{
\frac{1}{\beta_H \beta_W^3} \left( (2\beta_W^2-\beta_H^2)\frac{M_W^2}{s}-
\left(\frac{M_H^2}{s} \right)^2 \right)
\left( \frac{1}{x-x_0}-\frac{1}{x+x_0} \right) \right. \nonumber \\
&&\left. \;\;\;\;\;\; +\frac{2 \beta_W^2 -\beta_H^2-1}{4 \beta_W^2} +
\frac{3h_3}{2}\frac{M_H^2}{s}\left(\frac{1}{1-M_H^2/s}\right) \right\}
\eeqn

\begin{figure*}
\begin{center}
\caption{\label{wwhh}{\em Energy dependence of the
$W^+ W^- \ra HH$ cross section . The different helicity contributions are
shown for comparison against the unpolarised cross section.
Thick lines are for the total
cross sections while the thin lines correspond to a cut on
the scattering angle
with $|\cos \theta|<0.8$.}}
\mbox{\epsfxsize=14cm\epsfysize=17cm\epsffile{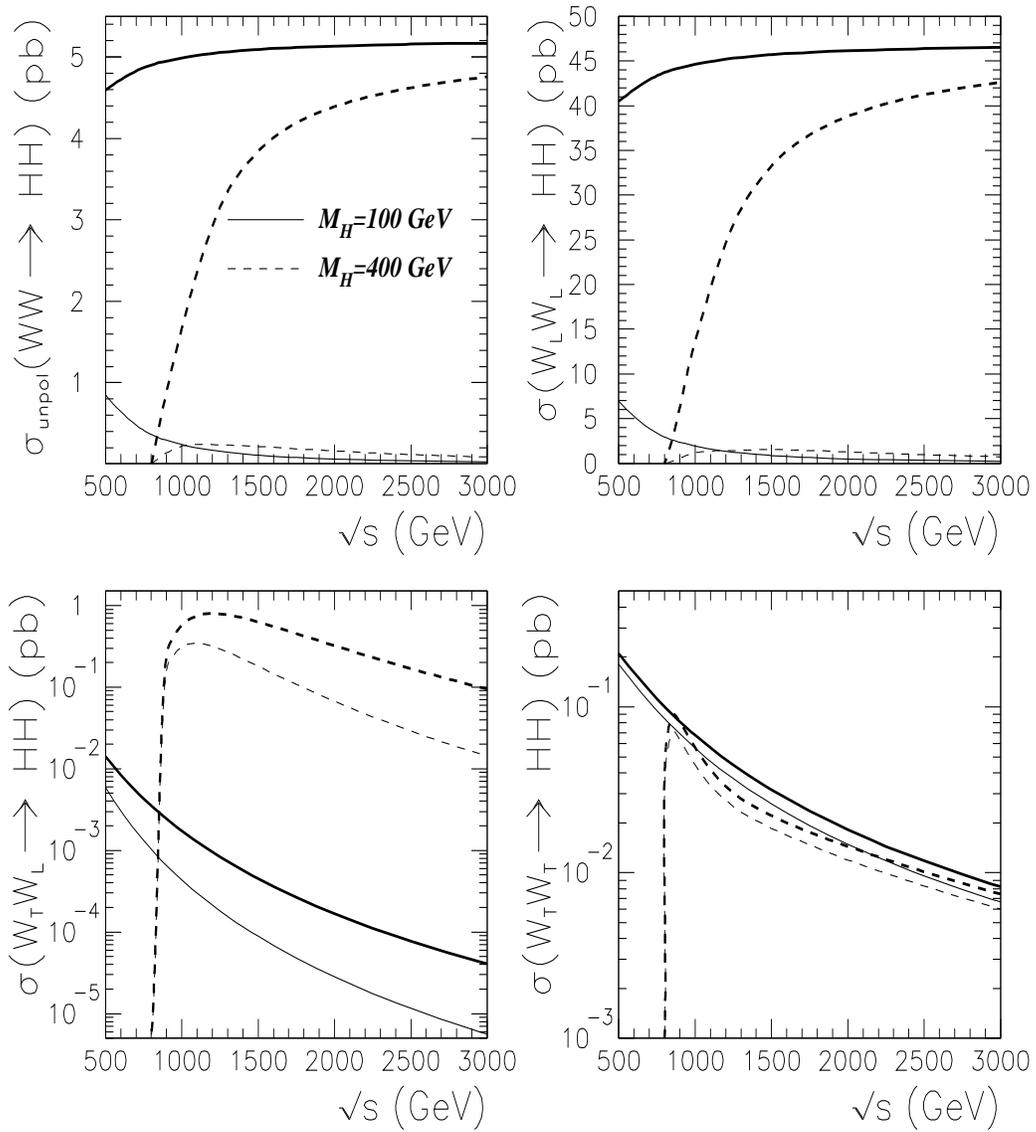}}
\end{center}
\end{figure*}

As Figures~\ref{wwhh} show the cross sections are indeed essentially
produced through both $W$ being longitudinal, for both a light and a heavy
Higgs. Nonetheless the leading contribution at high energies is from the
t-channel $W$ diagrams and comes from the $W$ being extremely forward.
The asymptotic form of the total cross section is not sensitive to the
$H^3$ coupling and is given by
\beqn
\label{wwhhinf}
\sigma_{LL}& &\stackrel{s \gg M_H^2, M_W^2}{\Longrightarrow}
\sigma_{\infty} \equiv \frac{\pi\alpha^2}{s_W^4}
\frac{1}{4m_W^2}  \nonumber \\
\sigma_{TL}& &\stackrel{s \gg M_H^2, M_W^2}{\Longrightarrow}
\frac{M_W^2}{2s}
\frac{M_H^2-2M_W^2}{s} \left(\ln(s/M_W^2)-3 \right)
\sigma_{\infty}\nonumber \\
\sigma_{TT}& &\stackrel{s \gg M_H^2, M_W^2}{\Longrightarrow}
\frac{M_W^2}{4s} \sigma_{\infty}
\eeqn
 To bring the effect of the $H^3$ coupling into prominence
a cut on the forward/backward
directions is essential. Introducing an angular cut $\theta_0$,
such that all Higgses within
this angle (measured  with respect to the $W$ direction)
are rejected, and with
$ 2M_W/\sqrt{s} \ll \theta_0 \ll 1$, the leading behaviour
becomes very sensitive to
the value of the triple Higgs vertex. To wit, with $r\gg 1$
\beq
\sigma_{LL} \sim \frac{\pi\alpha^2}{s_W^4}
\frac{1}{16 s} \left( \frac{M_H^2}{M_W^2} \right)^2 ( 3h_3-2 )^2
\eeq

Note also that the $W_L W_L$ cross section is dominant even after angular cuts
are applied, see Fig.~\ref{wwhh}.

Having confirmed the overwhelming dominance of the $W_L W_L$
cross section, we now
turn to a discussion on the structure functions of the $W_L$
inside the electron and
photon.

\subsection{Comparisons betwen the distribution functions for the $W_L$
inside the electron and the photon}
There have been numerous derivations of the distribution
(or structure function) of the $W$ inside
the light fermions (quarks and electrons)
\cite{EWA}. For the effective $W$ approximation,
the most interesting aspect concerns the $W_L$ content,
which in combination with the
equivalence theorem\cite{Equivalencetheorem}
 has been used to investigate manifestations of models of symmetry breaking
and Higgs production. All the available distributions
reproduce the same leading function
that exhibits scaling behaviour. For the {\em unpolarised} electron one has

\beq
D_{W_L/e}(y)=\frac{\alpha}{4\pi s_W^2} \frac{1-y}{y}
\eeq

\noindent where $y$ is the momentum fraction
of the electron transfered to the $W$.

The $W_L$ distribution inside the photon has only very recently been studied
\cite{egnuwh,Parisgg,Jikia4w}. This distribution consists of two different
parts. The first represents the situation where the spectator
$W$\footnote{The one that does not take part in the hard scattering
and plays the role of the neutrino in the electron case.} is transverse.
The second component
takes into account the contribution of the longitudinal spectator $W$.
The latter does not have an equivalent in the
fermion splitting case. As for the former we have shown that\cite{Parisgg}
, modulo an overall factor, it has the
same universal structure as the one for the electron
\beqn
D_{W_L/\gamma_\lambda}^{(W_\lambda)}(y)=\frac{\alpha}{\pi} \frac{1-y}{y}
\eeqn
Note that in this case the photon transfers its helicity, $\lambda$, to
the spectator $W$. The contribution from the $W$ with the opposite helicity
, $-\lambda$, is non leading. One important
difference with the electron case is that
whatever the helicity of the photon one gets the same probability for having
a $W_L$. Thus we may just as well write the {\em unpolarised} distribution:
\beqn
\label{sfgwtwl}
D_{W_L/\gamma}^{(W_T)}(y)=\frac{\alpha}{\pi} \frac{1-y}{y}
\eeqn
As for the spectator-$W_L$ component we have shown\cite{Parisgg}
 that an approximation that works
very well for \ggwwht is
\beq
D_{W_L/\gamma_\lambda}^{(W_L)}(y,Q^2_p)=
\frac{\alpha}{\pi} \frac{y (1-y)}{2}
\left(-2 + \ln \frac{Q^2_p}{\mwmw} \right)
\label{strucfq2}
\eeq
Where $Q^2_p$ is a typical $Q^2$ value for  the hard process.
Despite its $Q_p^2$
{\em logarithmic} enhancement this contribution is only a
fraction of the $W_T$ component
as evidenced by the fact that the yield of external $W_L$'s
is always an order of magnitude
below that of the transverse. We will not consider the
contribution from this component
in the present paper, therefore when looking at the validity
of the approximation we will only
compare it with the doubly transverse exact \ggwwhht
cross section that may also be
compared to the equivalent \eennhht.

The issue of the $W_L$ content of the photon is of importance. Since
one of the aims of the
\gamgamt colliders is to study symmetry breaking, it is essential to know
whether \gamgamt collisions can yield a higher luminosity in $W_L W_L$.
Looking at the
$W_L$ distributions in both the photon and the
electron one is tempted to conclude that
there are equally as many $W_L$ in the photon as in the electron. However,
the photon being
``democratic"  produces, regardless of its polarisation, both $W^+$ and $W^-$
 with an equal probability.   We can then study $W^\pm_L W^\pm_L$, while only
$W^+_L W^-_L$ is possible in \epemt. Moreover, for the latter channel (
$W^+W^-$) one gains a factor of
two in the \gamgamt mode. However, we should not
forget that one has to take into account
the convolution with the real spectra, moreover
for the structure function approximation
to be trusted one needs to go to a regime where
the $W$ mass can be neglected, since in
\gamgamt one has less phase space.
Furthermore, we have seen that, in order to have a
\gamgamt luminosity peaked at high values of the \gamgamt \cms,
polarisation of the
electron was essential. Now, if one polarises the
electrons in the \epemt mode
to be left-handed (and the positrons right-handed),
one gains a factor of 4 in the
\epemt convolution. Therefore, it is fair to argue
that there could be a complementarity
as different channels for effective $WW$ scattering are open
($W^\pm W^\pm, W^+ W^-$) in the \gamgamt mode,
but it is far from certain
that the effective lumininosity after convolution
with the \gamgamt spectra will be enough
for the foreseable future colliders to fare better
in the \gamgamt than the \epemt mode.

\subsection{Approximating the \eennhht cross section}
We are now in a position to evalute the \eennhht
cross section through the effective $W_L$
inside the electron:

\beqn
\sigma_{\eennhh} &\simeq &\int_{4M_H^2/s}^{1}
{\rm d}\tau \;\;\int_{\tau/y_m}^{y_m}\;\;
\frac{{\rm d}y}{y}\;\;D(y)D(\tau/y){\rm d}
\sigma(W_LW_L\rightarrow HH) \nonumber \\
&\simeq& \int_{4M_H^2/s}^{1} {\rm d}\tau \;
\cl^{\epem}_{W^+_L W^-_L}(\tau)\sigma_{W_LW_L\rightarrow HH}(\tau s)
\eeqn

\noi with $y_m$ the largest $y$ allowed kinematically and
where the luminosity for $W^+_L W^-_L$ is

\beq
\cl^{\epem}_{W^+_L W^-_L} (\tau)\simeq \left(
\frac{\alpha}{4\pi s_W^2} \right)^2
\frac{1}{\tau} \left( (1+\tau) \ln
\frac{1}{\tau} -2 (1-\tau)\right)
\eeq
where we have neglected the $W$ mass compared
to the electron energy to define the
luminosity of the $W^+_L W^-_L$ in \epemt.

As known and as we argued above, this effective luminosity is peaked for
lower values of $\tau$.
One may try to find an analytical approximation for the \eennhht total
cross section since we just have one convulation to
perform. We have not attempted to do this exactly. However, we have
looked at how good
an approximation one could obtain if one  takes the constant
 asymptotic value for the \wwhht cross section
as given by \ref{wwhhinf}. We obtain

\beqn
\sigma_{\infty}^{EWA}(\eennhh) \simeq \left(
\frac{\alpha}{4\pi s_w^2} \right)^2
\left( \frac{1}{2} \ln^2 (\frac{4M_H^2}{s}) + 2
\ln \frac{4M_H^2}{s} +3 \right)
\sigma_\infty
\eeqn

We should not expect this additional  approximation to work well as it is clear
from the energy
dependence of the \wwhht cross section (Figs~\ref{wwhh}), that
this limiting value is quite higher than the
values not far from  threshold where the effective luminosity will pick up much
of the cross section.
Nonetheless, we should expect the energy behaviour to be fairly well reproduced
as well as the order of magnitude.

\begin{figure*}
\begin{center}
\caption{\label{eeapprox}{\em Comparing
the result of the $W_L$ effective approximation
($\sigma^{EWA}$) to the exact result $\sigma^{exact}$ for \eennhht for a
light Higgs and a heavy Higgs.
Also shown is the asymptotic analytical
cross section $\sigma^{EWA}_{\infty}$.}}
\mbox{
\mbox{\epsfxsize=8cm\epsfysize=15cm\epsffile{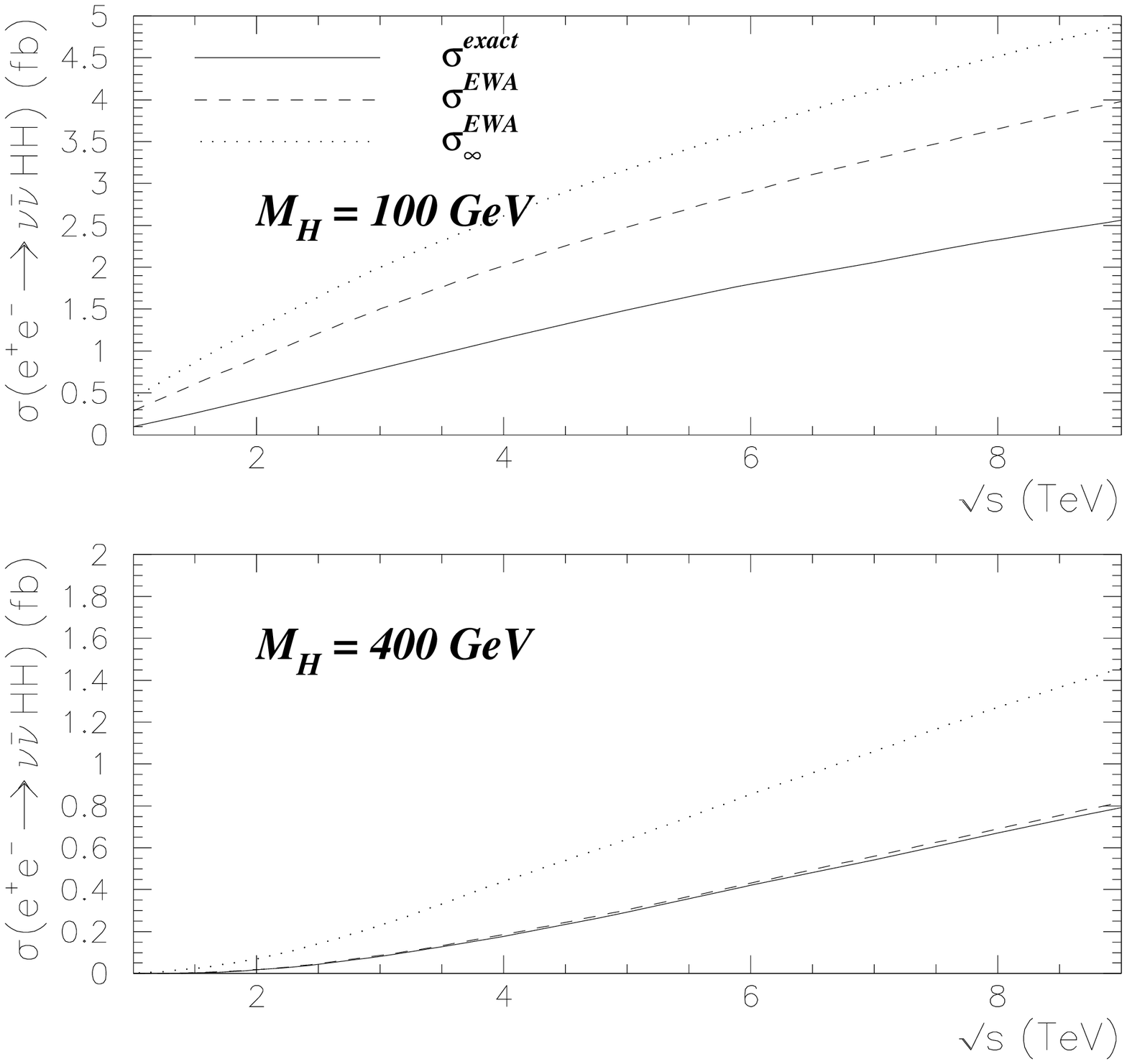}}
\mbox{\epsfxsize=8cm\epsfysize=15cm\epsffile{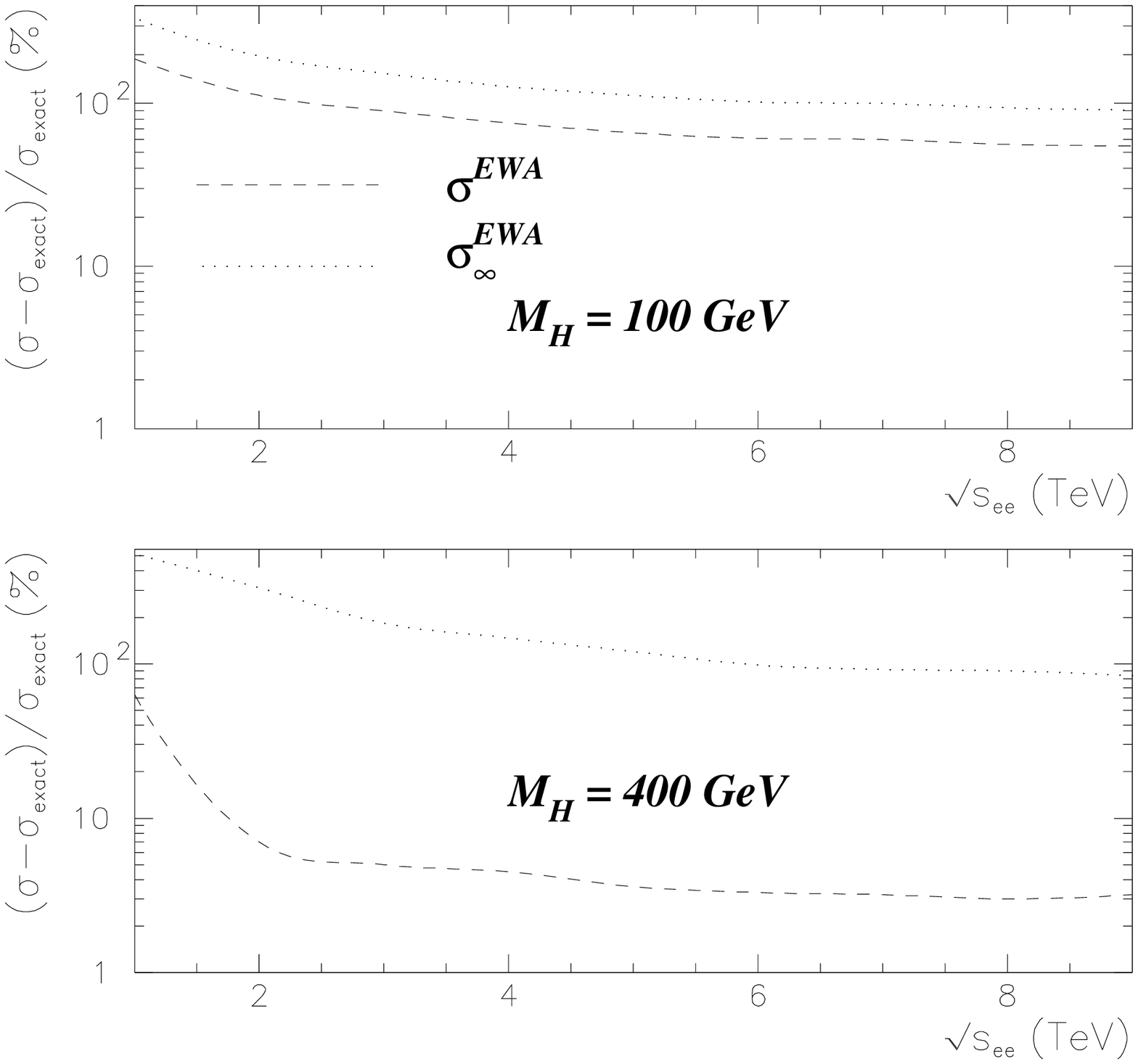}}}
\end{center}
\end{figure*}

We see from Figs.\ref{eeapprox} that for a Higgs
mass of 100GeV, the effective $W_L$
approximation is never within $50\%$, the latter
attained only around 10TeV. However the
energy dependence is well reproduced. For the case
of the heavy Higgs the approximation is
excellent, already at 2TeV it is within $10\%$ while
at 10TeV the agreement is almost
perfect, reaching $3\%$. Very similar conclusions
were obtained for the case of
$pp$ collisions at $40TeV$\cite{pphh}.
Note that the approximation of taking the limiting
high energy constant cross section
(for $M_H=400$GeV)  gives  a $100\%$ overestimate.

\subsection{Approximation in \ggwwhht}
\begin{figure*}
\begin{center}
\caption{\label{ggeeapprox}{\em The effect of
the bremstrahlung diagrams in
\ggwwhht.}}
\mbox{
\mbox{\epsfxsize=8cm\epsfysize=11cm\epsffile{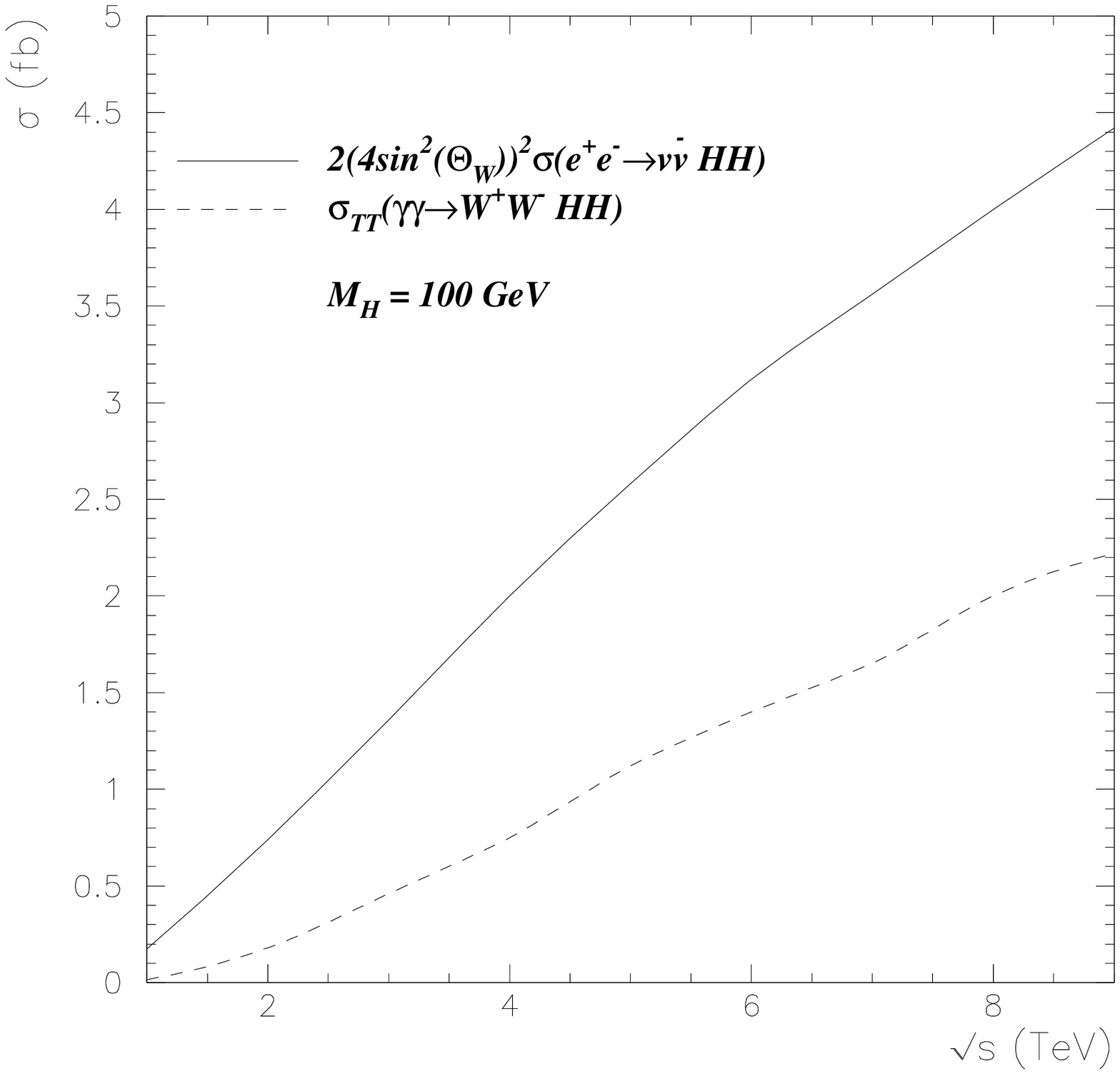}}
\mbox{\epsfxsize=8cm\epsfysize=11cm\epsffile{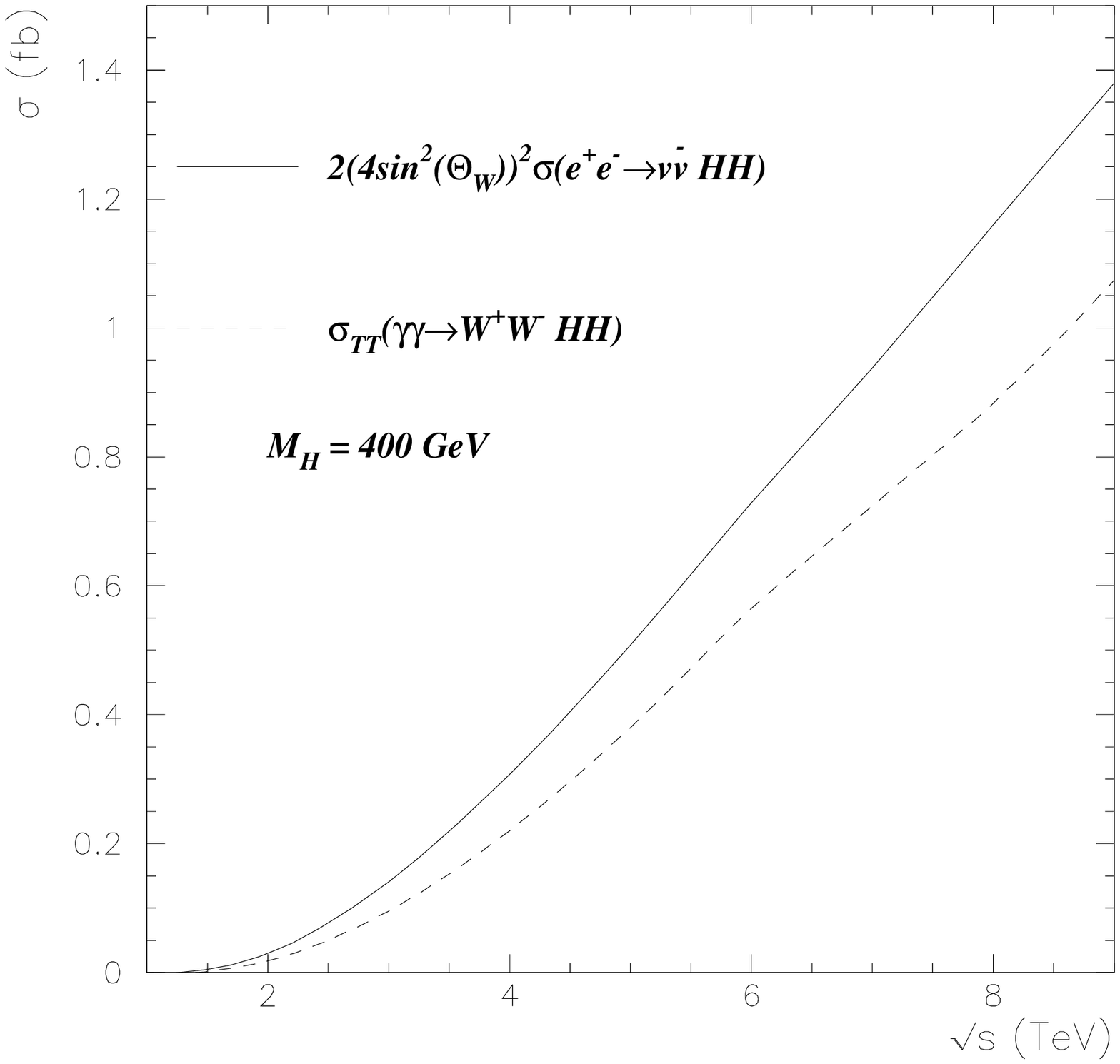}}}
\end{center}
\end{figure*}

\begin{figure*}
\begin{center}
\caption{\label{ggapprox}{\em \ggwwhht:  comparing the result
of the $W_L$ effective approximation,
$\sigma^{EWA}$, to the exact result for both $W$'s being transverse (
$\sigma^{TT}$)  for a light Higgs and a heavy Higgs.
$\sigma^{total}$ is the exact result
including all helicity modes of the $W$'s.
Also shown is the asymptotic analytical cross section
$\sigma^{EWA}_{\infty}$. For the
heavy Higgs the percentage deviation is given.}}
\mbox{
\mbox{\epsfxsize=8cm\epsfysize=15cm\epsffile{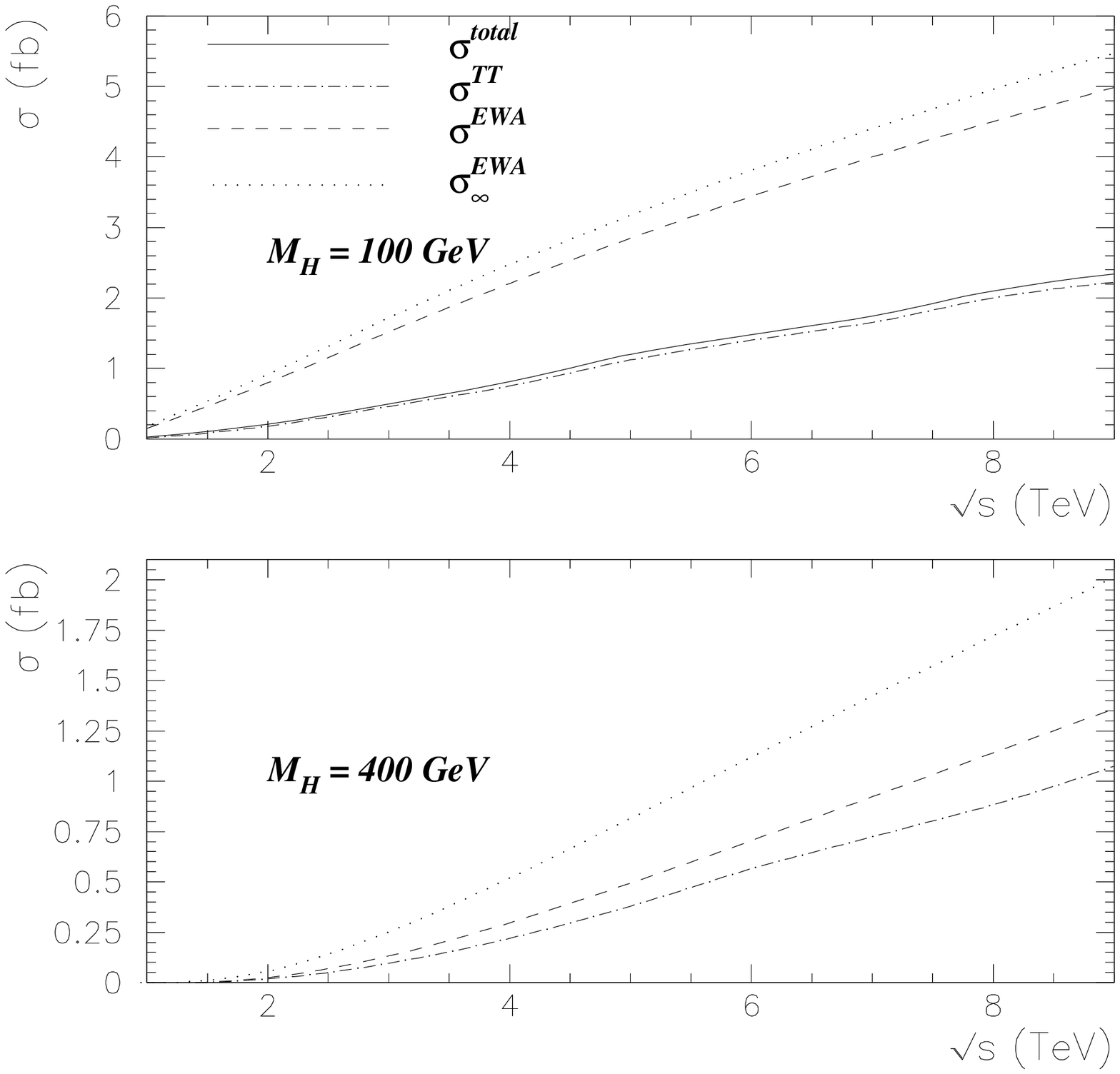}}
\mbox{\epsfxsize=8cm\epsfysize=15cm\epsffile{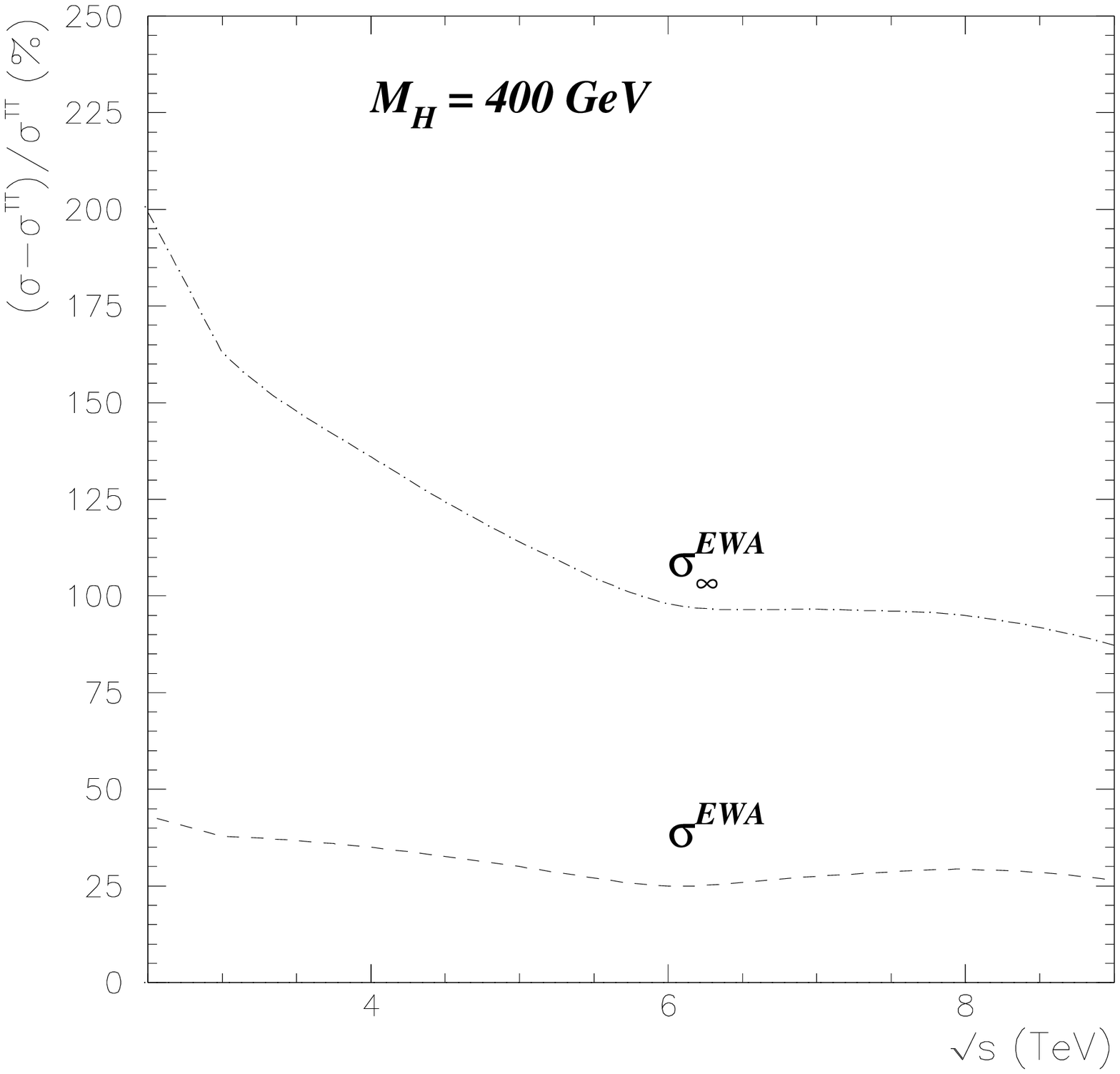}}}
\end{center}
\end{figure*}

Taking the distribution function for the $W_L$ inside
the photon given in \ref{sfgwtwl}
and taking into account the factor of two in the convolution we
would predict $\gamgam \ra W_T^- W_T^+ H H$ to be about
$2 (4 s_W^2)^2$ times the \eennhht cross section at high
enough energy where the
difference in phase space does not play a role. We have
compared these two cross
sections allowing for this factor. This comparison is
therefore a measure of the
bremstrahlung contributions in \ggwwhht especially that
the structure function approach
in \epemt has been verified to be a very good approximation.
The bremstrahlung diagrams in \gamgamt
can not be
deleted at a stroke since they do not form a
gauge invariant sub-set, thus the comparison
we propose should be a more trustworthy way
of extracting the effect of bremstrahlung.
It is gratifying to see that the arguments
we gave in section~2 when analysing  a combination
of distributions are born out by Figs.~\ref{ggeeapprox}.
We see that the effect of the
genuine part of the bremstrahlung is not negligible at
all for $M_H=100GeV$, while for
$M_H=400GeV$, their contribution gets smaller,
for example it is about $30\%$ at
$8TeV$.  Keeping this feature in mind it is no
wonder that the effective
$W_L$ does not reproduce the result as well as
in the case of the \epemt. Nonetheless
we see (Fig.~\ref{ggapprox}) that for 400GeV
the approximation is no worse than
what it is for single Higgs production\cite{Nousgg3v}.
At $8TeV$ the cross section is reproduced
within $20\%$ of its exact value.

\section{Identifying and measuring the Higgs triple vertex}

\subsection{The triple Higgs vertex}
We have, in the introduction, given ample motivation
for the importance of checking
the triple Higgs vertex, especially if no sign of New
Physics at TeV energies has
been revealed such as supersymmetry. The origin of
symmetry breaking and the naturality argument
would be a real
puzzle, especially if only a light scalar particle has been found.
One possibility, if this scalar is the Higgs, is to scrutinise the
 Higgs potential. Its most general form beside respecting the $SU(2)$
custodial symmetry must
lead to the correct value of the vacuum expectation
value\cite{Bijhh,Jikiahh}, then

 \beqn
\label{generalpotential2}
V_{\ssbe}= \lambda \left\{
\sum_{n=2} \frac{g^{2(n-2)}}{\Lambda^{2(n-2)}} \frac{a_n}{(n-1)^2}
\left[ \Phi^\dagger \Phi - \frac{v^2}{2} \right]^{n} \right\}
\eeqn

Compared to the form we have given in~\ref{generalpotential},
the first term in this series ($n=2$)
normalised with $\kappa_2=a_2=1$, is the usual minimal \sm potential;
whereas for $n>2$,
$\kappa_n=g^{2(n-2)}\frac{a_n}{(n-1)^2}$. Because of the new scale,
this expansion suggests a
hierarchy such that the dominant term (besides the \sm contribution)
 would correspond to $n=3$. Truncating the expansion at this
order, the leading parts of the potential that contribute to the
processes we have been studying write:
\beqn
\label{generalpotentiald}
V_{\ssbe}&=&\frac{1}{2} M_H^2
\left\{ H^2 + \frac{g}{M_W} H (\varphi^+ \varphi^- +\frac{\varphi_3^2}{2})+
\frac{g}{2 M_W}  h_3 H^3 \right. \nonumber \\
&& \;\;\;\;\;\left. h_4 \left( \frac{g}{4 M_W} \right)^2 H^4 +
h_3' \left( \frac{g}{2 M_W} \right)^2  H^2 (\varphi^+ \varphi^-
+\frac{\varphi_3^2}{2}).....\right\}
\eeqn

Note that the $H \varphi^+\varphi^-$ is unaffected, while not only the strength
parameterising the $H^3$ coupling , $h_3$, but also
 the quartic $h_4$ ($H^4$) as well as $H^2 \varphi^+\varphi^-$ get a
 contribution at this order, such that
their  resulting  strengths are not in the same ratio as in the \sm.

\beqn
h_3&=&1+ a_3 \frac{M_W^2}{\Lambda^2}=1+\delta h_3 \nonumber \\
h_4&=&1+ 6\delta h_3 \nonumber \\
h_3'&=&1+ 3\delta h_3
\eeqn

It is worth observing that had we done the calculation for \ggwwhht
with the usual linear gauge fixing condition in a renormalisable gauge
 we would have had to add an anomalous $H^2 \varphi^+\varphi^-$
contribution. No such addition is needed
with the gauge fixing we have exploited.

There is yet another gauge invariant implementation
of the triple Higgs vertex, that also
maintains the custodial SU(2) symmetry. Symmetry
breaking can be  realised non-linearly
as one would be led to assume in the limit of a very
heavy (or no) Higgs scenario.
In this picture the Higgs would have to be interpreted
as a scalar that has to be
coupled in a chirally invariant way to the Goldstones.
The latter may be collected in the matrix $\Sigma$ with
\beqn
\Sigma=exp(\frac{i \omega^i \tau^i}{v}) \;\;\;\;\;
{{\cal D}}_{\mu} \Sigma=\partial_\mu \Sigma + \frac{i}{2}
\left( g \Ww_{\mu} \Sigma
- g'B_\mu \Sigma \tau_3 \right)
\eeqn

The most general lowest order Lagrangian that
represents the symmetry breaking sector
is\cite{Chivukulascalar}

\beqn
\cl_{SSB}^{(2)}&=&\frac{1}{4}
\left(v^2 + 2 h_1 v H + h_2 H^2 + ..... \right)
\tr(\cd^\mu \Sigma^\dagger \cd_\mu \Sigma) +
\frac{1}{2} (\partial_\mu H)^2 \nonumber \\
& -&
\frac{1}{2} M_H^2
\left( H^2 + \frac{g}{2 M_W}  h_3 H^3 + h_4
\left(\frac{g}{4 M_W} \right)^2 H^4 + ..... \right)
\eeqn
In the standard model all $h_i=1$. $h_{1,2}$
represent $VVH,VVHH$ interactions which in our study we
took to be standard.  $h_1$ can be probed in other reactions
(such as Higgs decays or single Higgs production) where $h_3$ does not take
part.
taking only $h_3 \ne 1$ would be unnatural.

\subsection{Identifying the triple Higgs vertex in double Higgs production}
\begin{figure*}
\begin{center}
\caption{\label{thetastar}{\em The distribution
in the reconstructed angle $\theta^{\star}$
for the signal, background and the interference
in the case of \ggwwhht without convolution
with photon spectra.}}
\mbox{\epsfxsize=15cm\epsfysize=15cm\epsffile{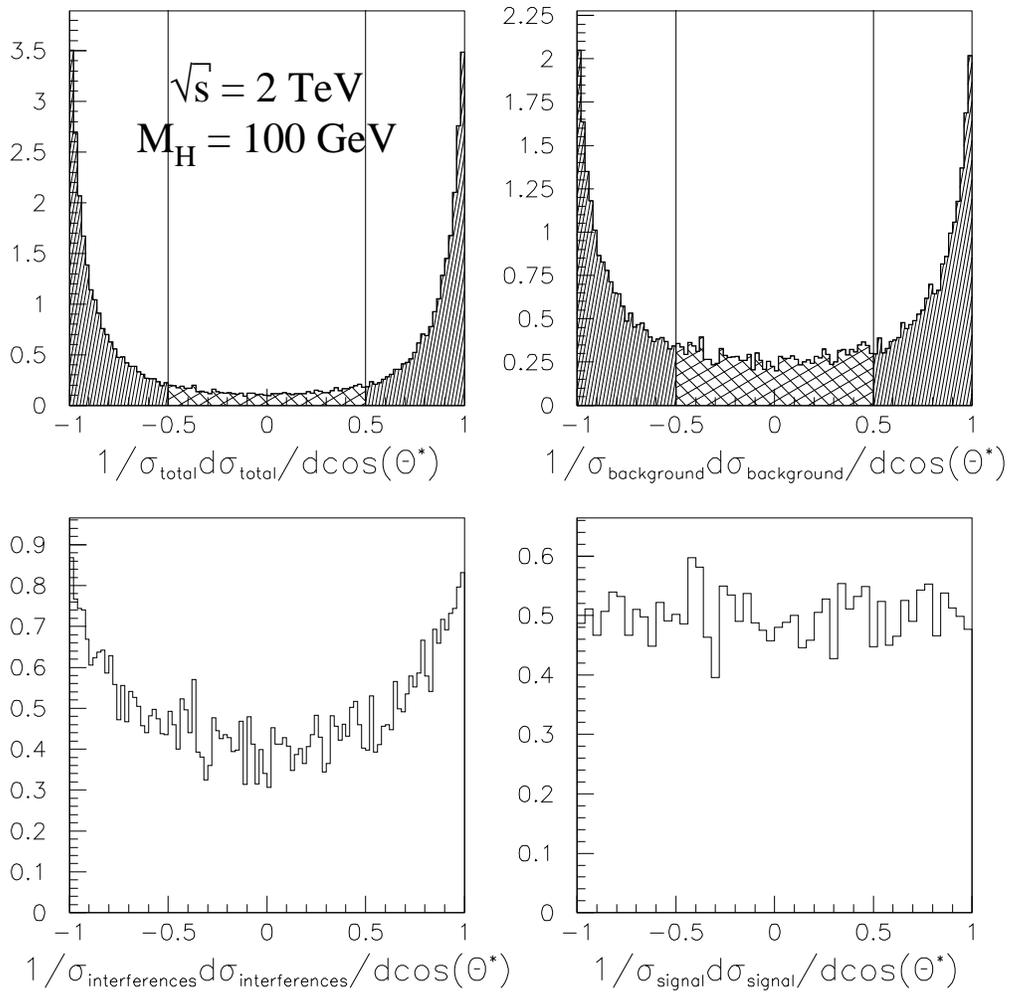}}
\end{center}
\end{figure*}

Beside the fact that the cross sections for double Higgs production are quite
small, the extraction of the $h_3$ part is not so easy. We have seen
that already at the level of $WW \ra HH$ the cross section is dominated by
forward/backward
events and the effect of $H^3$ is blurred.
There is though a specific signature of the
$H^3$ coupling in all processes that we have studied.
Once we note that the two Higgses that
originate from this vertex can be regarded
as produced by a scalar $H^\star$ then
in the centre of mass system of the pair,
the angular distribution of the
 Higgses is flat.
Therefore, we suggest to reconstruct the angle,
$\theta^\star$, measured
in the \cms of the pair, between the
Higgs direction and
 the boost axis (or for that matter the reference direction Oz of the beam).
For the signal events the distribution is flat,
while the background events are peaked in
the forward/backward direction. In Figs.~\ref{thetastar} we show,
for \ggwwhht,  the distributions
in this variable for the signal as well as the background and the
interference terms, that clearly
make and confirms the point: The signal
does not show any dependence on the angle
 $\theta^\star$, while the ``background" is clearly peaked in the
 forward direction and
the interference shows some angular dependence.

For \eennhht boosting back one obviously obtains
the same angular distribution as for
\wwhht.
It is clear that to improve the detection of the
$H^3$ or get a better limit on its
self-coupling
one should keep to the central region in $\theta^{\star}$.
One possibility is to consider
the ratio of events within a central region, characterised by an optimised
angle $\theta^{\star}_0$,
over the number of events outside this region:
\beq
R=\frac{\sigma_{HH}(|\cos \theta^\star| < |\cos \theta^\star_0| )}
{\sigma_{HH}(|\cos \theta^\star| >|\cos \theta^\star_0| )}
\eeq

 As with almost all ratios, this has the
 advantage of being free of many of the uncertainties in measuring
the cross section as well
as some of the theoretical uncertainties.
Here we have in mind the choice of the input
parameters which for a 6-particle amplitude in the electroweak theory can
introduce large uncertainties
as we pointed out in section~2.

\subsection{Measuring $h_3$ in \eennhht}
\begin{figure*}
\begin{center}
\caption{\label{h3limitee}{\em Dependence of the
\eennhht cross section on the
self-coupling $h_3$ at 2TeV for $M_H=100,400GeV$. Both the exact
calculation and the effective $W_L$ are shown.}}
\mbox{\epsfxsize=14cm\epsfysize=14cm\epsffile{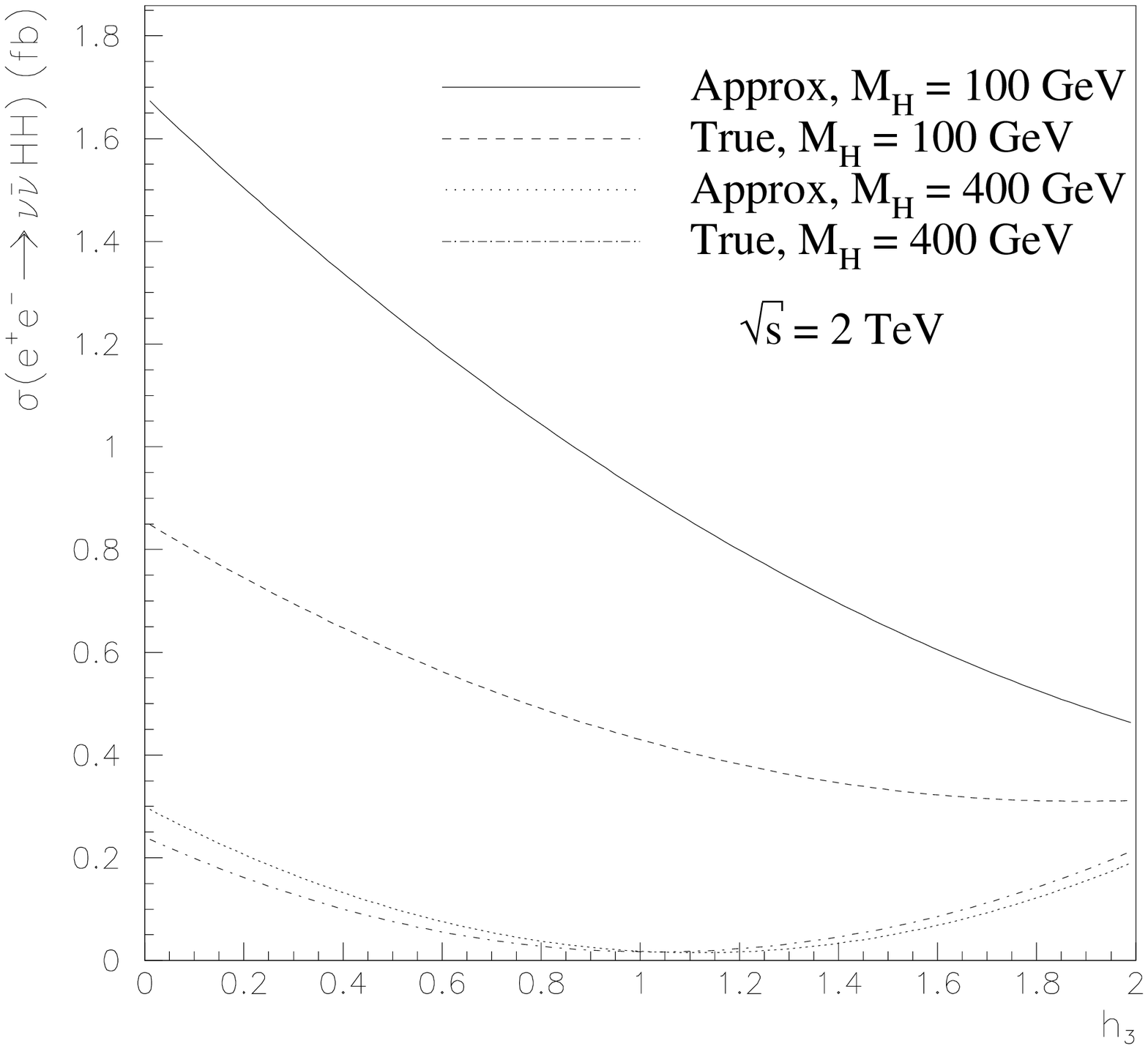}}
\end{center}
\end{figure*}
For the \eennhht, the $h_3$ coupling appears in only
one of the diagrams. We have also
looked at whether the $W_L$ approximation works when
we include a non-standard coupling.
The answer is that the approximation is as good as with what we found
for the \sm. That is, at
100GeV the approximation is not to be trusted since it overestimates
the cross section by a factor
of 2, while for $400$GeV the approximation is excellent, see
Fig.~\ref{h3limitee}. For the measurements of the coupling
we concentrated essentially on the 2TeV collider. We will content
ourselves   with a brief
comment about  what would happen with a higher energy ``futuristic" machine.

We assume that, at $\sqrt{s}=2TeV$ one will be able to collect a total
integrated luminosity of
$300 fb^{-1}$. Moreover, we assume  an efficiency for the reconstruction
 of the double Higgs
events to be $50\%$. We will be conservative in the sense that we base
our results as if no
initial polarisation were available, remembering that if beam polarisation
 were available
one could gain a factor of four in statistics.

At 2TeV this leads to an event sample of $\sim 68$ HH events for
$M_H=100GeV$ whose signature is $4b$ events plus large missing energy.
One obvious background is
$\nu \nu ZZ$, but we assume that invariant mass
constraints should get rid of these as they  should
for $WW \nu \bar \nu$ if no $b$ tag were available.
Of course, for a Higgs with a mass closer to  $M_Z$
the situation would be problematic if not hopeless.
Considering this not so high statistics we further {\em conservatively}
take as a criterion for detection of an anomaly
in $h_3$ that  on has a $50\%$ deviation
in the expected number of events, provided one has at least $30$ events.
 With this
setting, we conclude from Fig.~\ref{h3limitee} that  with
 the total cross section one would  only be able to claim New Physics if
$\delta h_3<-0.75$ (for positive $\delta h_3$ one needs values in exces
of $\simeq 2$). For $M_H=400GeV$, SM values will not
lead to a measurement (in this case the
signal is $4W$ plus large missing energy), however if
 $|\delta h_3| > 1$ a signal will be
recorded (there will ``unexpectedly be more than 30 events)
and would be a clear indication for an anomalous $h_3$ coupling.

\begin{figure*}
\begin{center}
\caption{\label{Reegg}{\em Dependence of the ratio $R$ on $h_3$ both for
\ggwwhht and \eennhht.}}
\mbox{\epsfxsize=14cm\epsffile{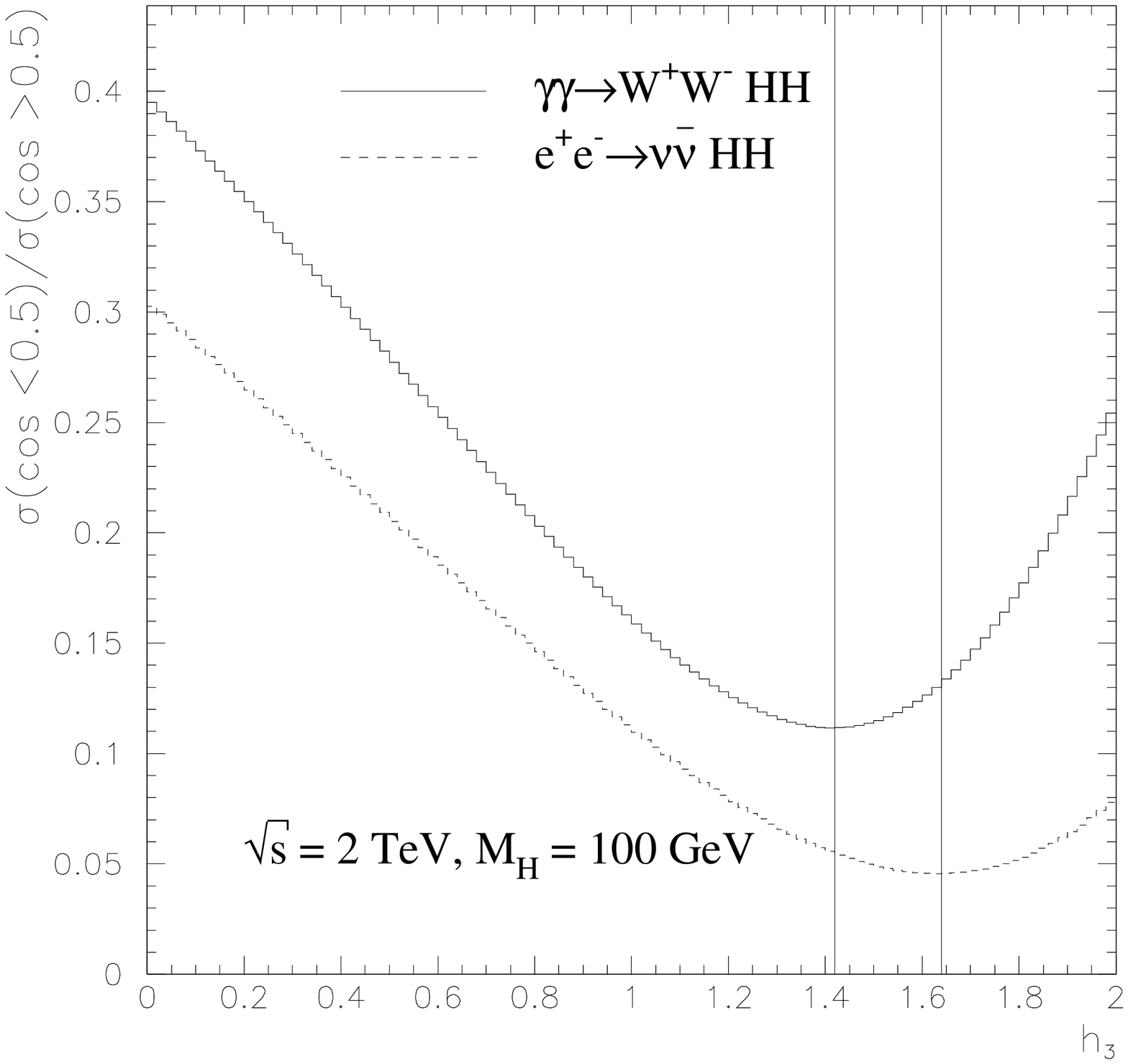}}
\end{center}
\end{figure*}
\noindent
For $M_H=100GeV$ where one has enough events for a \sm value, the ratio $R$
that we introduced
earlier is much more powerful in constraining the coupling.
For $\theta^\star$ we took $|\cos \theta^\star|<0.5$.
We have not made any effort to optimise this value.
First the event sample (with the luminosity and efficiency taken above) within
$|\cos \theta^\star|<0.5$  is about 7 out of 60 outside this region.
Assuming that the ratio can be measured
at $20\%$, we find $-.10 < |\delta h_3| < .15$
(see Fig.~\ref{Reegg}) which means a precision of about $10\%$ on $h_3$.

Let us comment briefly about how much better a higher
energy \epemt machine can do. To see
 this we
refer to Fig.~\ref{eemh} that shows, at $5TeV$, the Higgs
mass dependence of the cross section
for both the signal and the background.
We can conclude that for $M_H=700GeV$ switching off
the triple Higgs vertex ($h_3=0$, that corresponds to taking into account only
the ``Backgrounds diagrams) leads to a 10 fold increase of the cross section!

\subsection{Measuring $h_3$ in \ggwwhht}
\begin{figure*}
\begin{center}
\caption{\label{h3limitgg}{\em Dependence of the \ggwwhht cross section on the
self-coupling $h_3$ at 2TeV for $M_H=100,400GeV$. }}
\mbox{\epsfxsize=14cm\epsfysize=14cm\epsffile{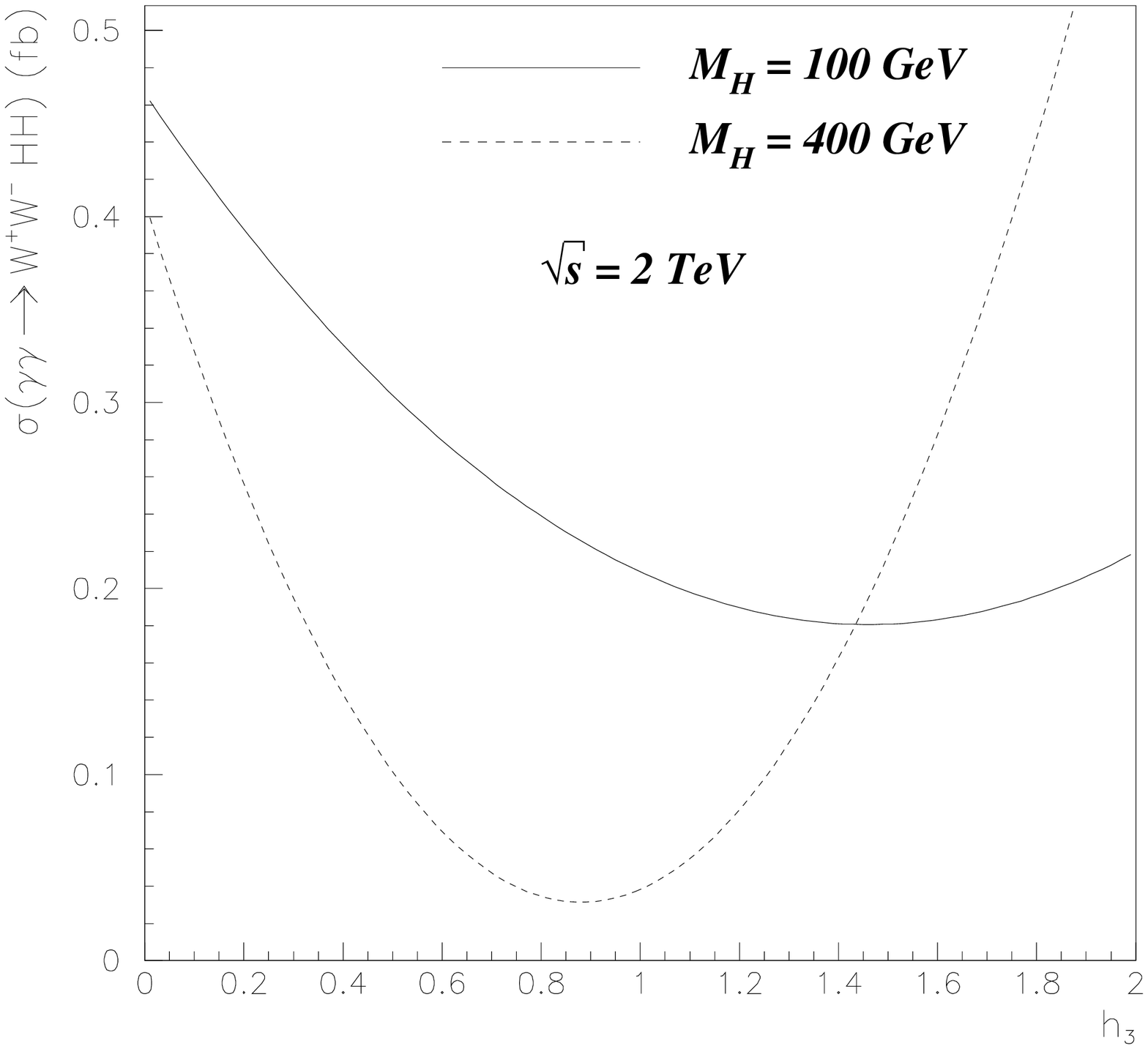}}
\end{center}
\end{figure*}
{}From the detailed analysis on the \ggwwhht process, we learnt that at 2TeV
the
 cross section
is lower than in \eennhht and this even before taking into account the
 reduction introduced by the
convolution and the branching ratios of the $W$'s
(if not all $W$ decays are used for the
analysis). Of course, at much higher energies and with higher masses the
 \gamgamt mode fares
better. In any case, as concerns the triple Higgs
vertex one advantage of the \gamgamt
mode is that this coupling takes part not only
in the fusion diagrams but also in
the bremstrahlung type diagrams which are not negligible at 2TeV,
 especially for a light Higgs.
Taking the effective
\gamgamt luminosity to be only half that of
the \epemt and with the same efficiency for
reconstruction of double Higgs events, for
$M_H=100GeV$, one can hope to collect 15 events.
In view of this number our criterion for
detection of non-standard values is $100\%$
deviation in the number of events which,
here means that we require a doubling of events.
In this case there is  a very slight improvement
on the limit extracted from the total
\eennhht cross: $\delta h_3<-0.5$ which in any
case does not compete with the limit one
extracts from the ratio $R$ in \epemt. However,
we find that for a Higgs mass of $400$GeV
the effect of an anomalous $H^3$ coupling are
dramatic and, by far, much more interesting than in \epemt.
Requiring observation of at least 15 events for $M_H=400$GeV
(where within the \sm one expects
only about 3)  useful constraint on the coupling can be set:
$-.7<\delta h_3 <0.5$. There is thus a complementarity between the \epemt
and the \gamgamt depending on the Higgs mass
in probing the Higgs triple vertex even at
2TeV. As for the ratio R, taking $M_H=100GeV$
it is unlikely that with the number of total
$WWHH$ events at \gamgamt one would be able
to make such a measurement, nonetheless even
if this ratio were measured with the same
precision as in \epemt one would not constrain the
couplings further than what is achieved in the classic \epemt mode.

\noindent The limits on $h_3$ we have extracted for a light Higgs
from a measurement of $R$ in \epemt
and the ones for heavy Higgs up to a mass around 400GeV are much
better than what could be achieved
with the reaction $\gamgam \ra HH$ which is dominated by a
$J_Z=2$ contribution that is insensitive
to the the $J_Z=0$ s-channel Higgs exchange. However,
with $\sqrt{s}=2TeV$, for $M_H >600GeV$
$\gamgam \ra HH$  is the only reaction
where useful limits, of the order of what we
extracted for lower Higgs in \eennhht and
\ggwwhht, can be set. Thus, there is at a 2TeV collider a
 very nice coverage of the $h_3$ sensitivity
by all three reactions. Of course, as the energy increases
 and especially if the Higgs mass is
large,
\ggwwhht is the ideal laboratory for testing the $h_3$ coupling,
provided there is enough
luminosity in the \gamgamt mode and other ``technical"
 problems for the would-be novel colliders
are under control. \\

\noindent One may wonder
whether it is at all possible to extract an indirect
limit on $h_3$ from low-energy experiments.
This is difficult since present data are not  very
 sensitive to even the Higgs mass,  let alone to its
tri-linear coupling. The latter would only enter
at two-loop for current observables
\footnote{At one-loop, it only enters
for the renormalisation of the Higgs mass.}. Considering
that we would be dealing with a non
renormalisable theory if the Higgs coupling is anomalous,
 an unambiguous limit is not possible.
Only an expected order of magnitude can be given based on
``naturality argument"
\cite{Georginatural}. This is like trying to constrain the
anomalous weak vector bosons couplings
from radiative corrections compared to a direct limit through $W$
pair production. The indirect
limit on $h_3$ has been investigated in \cite{Bijhh} by considering the effect
 of $h_3$ on the
$\rho$ parameter. Even with the proviso that the estimate from the
two-loop calculation
 is only an indication on the order of
magnitude, the indirect limit is an order of magnitude worse than
the direct limits covered
by the three main double Higgs cross section at 2TeV.

\setcounter{subsection}{0}
\setcounter{equation}{0}
\def\thesubsection {\thesection.\arabic{subsection}}
\def\theequation{\thesection.\arabic{equation}}
\def\thefigure{\arabic{figure}}

\section{Conclusions}
The investigation of the scalar potential
or the nature of the Higgs, as revealed
through its self-couplings, can be done
most unambiguously through double
Higgs production. For the planned colliders
in the forseeable future the
expected cross sections are rather small
thus the need to study these signatures
in an environment with the least and best understood backgrounds.
We have seen that \epemt colliders
operating at 2TeV \cms offer great possibilities
especially if use is  made to run them
in both the usual \epemt mode as well as the
\gamgamt mode. We have found that with
the realistic luminosities expected for these
machines one may hope to achieve
a measurement of the tri-linear couplings at
the level of $10\%$ (for a light Higgs).
The results are also
encouraging in the sense that the \epemt and
the \gamgamt modes can cover different
ranges of the Higgs mass if a 2TeV machine is built.
We find that for a light Higgs
(up to 250GeV) the best limits on the $H^3$ couplings come from $\eennhh$.
 However, for heavier Higgses up to mass of $500GeV$,
the best channel is the
associated double Higgs production in \gamgamt.
For still heavier masses, the one-loop induced
$\gamgam \ra HH$ is by far better, mainly because
of its larger phase space. We have
proposed the variable R that clearly helps in
discriminating the triple Higgs
vertex. As a by-product we have verified the
validity of the distribution function
describing the longitudinal $W$ content of
the photon. We find, as with the case of single $H$
production\cite{Nousgg3v}, that for a heavy Higgs ($M_H>400GeV$) the
approximation works quite well, although not
as well as for the electrons. This is due to
the bremstrahlung contributions that accompany
\gamgamt production modes and which are often absent in the \epemt mode.
Another critical issue
that our study has addressed and which comes as a by-product of our results,
is how much better
 a \gamgamt collider fares in probing models of strongly
interacting $W$'s that would have
to manifest themselves in the absence of a light Higgs.
The answer is directly related
to the luminosity of $W_L W_L$ that one effectively gets at the two modes.

Translating our
comparisons for the heavy Higgs production and combining
them with our previous findings
about single heavy Higgs production\cite{Parisgg}, the
conclusion is that we can indeed
arrive at a higher luminosity in the \gamgamt mode.
However, when we  include realistic photon spectra the
factor two enhancement provided in the \gamgamt mode is
lost in the convolution, especially
if we take into account the new results on the spectra
that simulate multiple scattering
and include non-linear QED effects\cite{Sheffield95}.
Moreover, this factor $2$ can easily
go in favour of the \epemt mode if both the electron
and the positron are polarised. Nonetheless,
the \epemt mode will only be open to the $W_L^+ W_L^-$
channel, while \gamgamt provides all
combinations of charges. \\
Another result of our study concerns the computational part.
We have advocated the use of
non-linear gauge fixing conditions\cite{Nonlineargauge} for {\em tree-level}
 mutiple bosons processes in
\gamgamt reactions and have made the connection with the background field
 gauge-fixing constraints.
These types of gauges tremendously ease the calculational task. We foresee
their use for all
calculations in the framework of the laser
induced high-energy \gamgamt collider.

\subsection*{Note added}
During the writing up of this paper, there
appeared\cite{Gracehh} a new ``automatic"
calculation of \ggwwhht in the unitary gauge.
Our reults agree with the ones found
in\cite{Gracehh} when we take the same input
parameters. Note, however, that most of the
results of this paper have been presented at the
{\em 2nd European Workshop on Physics with
\epemt Colliders}\cite{Chopinitalie} and
in\cite{Chopinthese} prior to \cite{Gracehh}.

\subsection*{Aknowledgements}

It is  with great pleasure that we wish to thank
Genevi\`eve B\'elanger and Marc Baillargeon
for their ever unfailing help and comments.
We are also very grateful to George Jikia
for the many discussions and for most kindly
providing us with his program for
$\gamma\gamma\rightarrow HH$.
One of us (E.C) thanks Xavier Artru for interesting
discussions about multiperipheral processes.

\setcounter{section}{0}
\setcounter{subsection}{0}
\setcounter{equation}{0}
\def\thesection{\Alph{section}}
\def\thesubsection {\thesection.\arabic{subsection}}
\def\theequation{\thesection.\arabic{equation}}

\section{Appendix:
{\small Feynman Rules for the Generalised Non-Linear Gauge Fixing
Condition}}

We begin by presenting our conventions and notations for the bosonic sector
of the $SU(2)\times U(1)$ model.

The $SU(2)$ gauge fields  are $\Ww_\mu=W^{i}_{\mu}\tau^i$,
while the hypercharge
field is denoted by $\B_\mu=\tau_3 B_\mu$. The normalisation for the Pauli
matrices is $\tr(\tau^i \tau^j)=2\delta^{i j}$.
The radiation Lagrangian is expressed {\em via}
field strength,
$\Ww_{\mu \nu}$
\beqn
\Ww_{\mu \nu}&=&\frac{1}{2} \left(
\partial_\mu \Ww_{\nu}- \partial_\nu \Ww_{\mu} +\frac{i}{2}
g[\Ww_\mu, \Ww_\nu]
\right) \nonumber \\
&=&\frac{\tau^i}{2} \left(\partial_\mu W^{i}_\nu-\partial_\nu W^{i}_\mu
-g \epsilon^{ijk}W_{\mu}^{j}W_{\nu}^{k} \right)
\eeqn

and
\beqn
\B_{\mu \nu}=\frac{1}{2} \left(
\partial_\mu B_{\nu}- \partial_\nu B_{\mu} \right)\tau_3
\eeqn

such that the pure gauge kinetic term writes
\beqn
\cl_{{\rm Gauge}}=- \frac{1}{2} \left[
\tr(\Ww_{\mu \nu} \Ww^{\mu \nu}) + \tr(\B_{\mu \nu} \B^{\mu \nu}) \right]
\eeqn

The Higgs doublet, $\Phi$, with hypercharge $Y=1$, is written as
$$ \left(\begin{array}{c} \varphi^+ \\ \frac{1}{\sqrt{2}}(v+H+i\varphi_3)
\end{array}\right) $$
and the covariant derivative acting on this doublet is such that
\beqn
\cd_\mu \Phi=\left(\partial_\mu +\frac{i}{2} (g \Ww_\mu +g'Y B_\mu)
\right)\Phi
\eeqn

The Higgs potential is introduced
\beqn
{\cal V}_{SSB}=\lambda \left[ \Phi^\dagger \Phi - \frac{\mu^2}{2 \lambda}
\right]^2
\eeqn

with $\mu^2, \lambda >0$ such that spontaneous symmetry breaking ensues.
The $W^\pm$ and Higgs masses are
\beqn
M_W=\frac{g v}{2}\;\;\;\;\; M_H^2=2 \mu^2=2\lambda v^2 \;\;\;\; v=246GeV
\eeqn

Our convention for the fields and couplings is
\beqn
Z_\mu &=& c_WW^{(3)}_\mu-s_W B_\mu \\
A_\mu &=& s_WW^{(3)}_\mu+c_W B_\mu \\
g &=& e/s_W \\
g'&=& e/c_W \\
g_Z &=& e/s_Wc_W
\eeqn

We propose, when one is dealing with multiparticle final states that involve
photons and weak bosons, to use the generalised non-linear gauge fixing
condition, both for the $W$
\beqn
\label{nonlineargauge}
{{\cal L}}_{\xi_W}\;&=&\;-\frac{1}{\xi_W}
|(\partial_\mu\;+\;i e \tilde{\alpha} A_\mu\;+\;ig \cos_W
\tilde{\beta} Z_\mu) W^{\mu +}
+ i\xi_W \frac{g}{2}(v +\tilde{\delta} H -i \tilde{\kappa}
\varphi_3)\varphi^{+}|^{2} \nonumber \\
&&
\eeqn
and $Z$
\beqn
\label{nlz}
{{\cal L}}_{\xi_Z}=-\frac{1}{2 \xi_Z}
(\partial.Z +\; \xi_Z \frac{g}{ 2 \cos_W} (v+\tilde\varepsilon H)
\varphi_3)^2
\eeqn

The most practical choice for the $\xi_i$ is $\xi_W=\xi_Z=\xi=1$.
We do not touch the gauge fixing for the photon:

\beq
{{\cal L}}_{\xi}=-\frac{1}{2 \xi}
(\partial.A )^2
\eeq

As we  pointed earlier these gauge fixing
conditions have to be paralleled with
the gauge fixing constraints one imposes in
the background-field method.
In the latter, upon splitting the fields $\psi$
into into their
classical,  $\psi_{cl}$,  and quantum,  $\psi_Q$, parts:
$\psi=\psi_{cl}+\psi_Q$
and specialising to the case where the gauge parameters
$\xi$ are all equal,
one has for the \su theory (see for instance\cite{BckgDenner})

\beqn
\label{bckgrdgauge}
{{\cal L}}^{bckgrd}\;&=&\;-\frac{1}{\xi}
|(\partial.W^+_Q+\;i g (W^{(3)}_{cl}.W^+_{Q}
-W^{(3)}_Q.W^+_{cl})+\frac{i}{2}\xi
S^+|^{2} \nonumber \\
&-&\frac{1}{2 \xi} |(\partial.A_Q+ie (W^{+}_{cl}.W^-_{Q}-W^{+}_{Q}.W^-_{cl})
+ie\xi(\varphi^{+}_Q\varphi^-_{cl} - \varphi^{+}_{cl}
\varphi^-_{Q})|^2  \nonumber \\
&-&\frac{1}{2 \xi} |(\partial.Z_Q+ig c_W ( W^{+}_{cl}.W^-_{Q}
-W^{+}_{Q}.W^-_{cl} )\nonumber \\
&+&i\xi\frac{1}{2 s_W c_W}\left( (c_W^2 -s_W^2)(\varphi^{+}_Q\varphi^-_{cl}
 - \varphi^{+}_{cl}\varphi^-_{Q})
+i (H_Q\varphi_{3cl}-(v+H_{cl})\varphi_{3Q})\right)|^2  \nonumber \\
S^+&=&\varphi^{+}_{Q} (v+H_{cl}-i\varphi_{3cl}) - \varphi^{+}_{cl}
(H_Q-i\varphi_{3Q})
\eeqn

The identification with the non-linear gauge-fixing
constraint is the following.
For the
$\gamma \gamma$ processes we have been studying,
one does not need  a gauge-fixing
for the photon (and the $Z$). These are then
considered purely classical as is the
corresponding neutral Goldstone and the Higgs.
On the other hand, since there is no
separation, in  the non-linear
gauge,  between classical and quantum fields one interprets the
$W^\pm$ and their Goldstones
as ``quantum". Then making
$W^\pm_{cl}, \varphi^\pm_{cl} \ra 0$ (but $W^{(3)}_Q, H_Q \ra  0$)
leads to the charged part of the
 non-linear gauge constraint with

\beq
\tilde{\alpha}=\tilde{\beta}=\tilde{\delta}=\tilde{\kappa}=1
\eeq
Which are the values that bring the most
simplifications in practical calculations.
Note, however that if we {\em also} fix the gauge in the neutral
sector then the identification is  less
transparent, as mixtures necessarily occur. In \ref{nlz},
for instance, and with
$\tilde\varepsilon=1$, $H$ is to be interpreted classical
whereas $\varphi_3$ is necessarily quantum like the $Z$.

This is the gauge we have taken in this paper
(although we did not need to
specify $\tilde{\varepsilon}$).

At this point it is worth comparing with specific examples of
non-linear gauges that have been used for loop calculations.
The condition used in\cite{Nonlineargauge} can be recovered by setting
\beqn
\tilde{\alpha}=\tilde{\beta}=1 \;\;\;\;\
\;\;\;\tilde{\delta}=\tilde{\kappa}=\tilde{\epsilon}=0
\eeqn
This condition gets rid of $W^\pm \varphi^\mp \gamma$
and has the advantage of
keeping the same Lorentz structure for the tri-linear
$WW\gamma$ and $WWZ$ vertices. However, the
vertices $W^\pm \varphi^\mp Z$ and $W^\pm \varphi^\mp H (A,Z)$ are present.
The condition taken in\cite{Jikia4w} corresponds to
\beq
\tilde{\alpha}=\tilde{\delta}=\tilde{\epsilon}=1 \;\;\;
\tilde{\beta}=\tilde{\kappa}=0
\eeq
Here both  $W^\pm \varphi^\mp \gamma$ and $W^\pm \varphi^\mp H A$ vanish.

In\cite{Dicusnlgauge} both $W^\pm \varphi^\mp \gamma$
and $W^\pm \varphi^\mp Z$ are made to
vanish. One can see that this is arrived at by taking $\tilde{\alpha}=1$,
$\tilde{\delta}=\tilde{\epsilon}=\tilde{\kappa}=0$, while
\beq
\tilde{\beta}=-\frac{s_W^2}{c_W^2}
\eeq
This corresponds to a $U(1)_Y$ covariant derivative.
However, contrary to what is
claimed in \cite{Dicusnlgauge}, with this choice,
$W^\pm \varphi^\mp H (Z,\gamma)$ still remain (but luckily
these
vertices have no incidence on the calculation
$\gamma \gamma \ra ZZ$ in\cite{Dicusnlgauge}).

We now give the Feynman rules for the generalised
non-linear gauge fixing
condition:

\subsection*{Propagators}
\bea
\Pi^W_{\mu\nu} &=& \frac{-i}{k^2-M_W^2}\left[g_{\mu\nu}+
\frac{(\xi_W-1)k_\mu k_\nu}{k^2-\xi_WM_W^2}\right] \\
\Pi^Z_{\mu\nu} &=& \frac{-i}{k^2-M_Z^2}\left[g_{\mu\nu}+
\frac{(\xi_Z-1)k_\mu k_\nu}{k^2-\xi_ZM_Z^2}\right] \\
\Pi^\gamma_{\mu\nu} &=& \frac{-i}{k^2}\left[g_{\mu\nu}+
\frac{(\xi_\gamma-1)k_\mu k_\nu}{k^2}\right] \\
\Pi^H &=& \frac{i}{k^2-m_H^2} \\
\Pi^{\varphi_3} &=& \frac{i}{k^2-\xi_ZM_Z^2} \\
\Pi^{\varphi^\pm} &=& \frac{i}{k^2-\xi_WM_W^2} \\
\Pi^{\eta_\gamma} &=& \frac{i\xi_\gamma}{k^2} \\
\Pi^{\eta_Z} &=& \frac{i\xi_Z}{k^2-\xi_ZM_Z^2} \\
\Pi^{\eta_\pm} &=& \frac{i\xi_W}{k^2-\xi_WM_W^2}
\ena

As is obvious, for all calculations $\xi_i=1$ is to be prefered.
\subsection*{Trilinear vertices}
In the following all momenta are taken to be incoming.

\noindent
\centerline{
\begin{picture}(30000,10000)
\drawvertex\photon[2 3](0,5000)[4]
\global\Xone=\vertexonex \global\Yone=\vertexoney
\global\advance\Yone by 1000 \global\advance\Xone by -3000
\put(\Xone,\Yone){$[\gamma^\mu;Z^\mu](k)$}
\global\Xtwo=\vertextwox \global\Ytwo=\vertextwoy
\global\advance\Xtwo by -2500 \global\advance\Ytwo by 500
\put(\Xtwo,\Ytwo){$W^{+\alpha}(p_+)$}
\global\Xthree=\vertexthreex \global\Ythree=\vertexthreey
\global\advance\Xthree by -2500 \global\advance\Ythree by -800
\put(\Xthree,\Ythree){$W^{-\beta}(p_-)$}
\global\Xfour=\vertexthreex \global\Yfour=\vertexmidy
\global\advance\Xfour by 3000 \global\advance\Yfour by 1500
\put(\Xfour,\Yfour){$ -ie\left[ 1;\displaystyle\frac{c_W}{s_W}\right]
\left[g_{\alpha \beta}(p_- - p_+)_\mu
\;+\;\left(1+\frac{\tilde\alpha}{\xi_W}\right)
(k_\alpha g_{\mu \beta}-k_\beta g_{\mu \alpha})\right.$}
\global\Yfive=\Yfour \global\advance\Yfive by -3000
\global\advance\Xfour by 7000
\put(\Xfour,\Yfive){$\displaystyle\left. \;+\;\left(1-
\frac{\tilde\alpha}{\xi_W}\right)
(g_{\mu \alpha}p_{+\beta}-g_{\mu \beta}p_{-\beta});
\tilde\alpha \ra \tilde\beta
\right]$}
\end{picture}}

The form of this vertex calls for some comments. First,
when  $\tilde\alpha$
and $\tilde\beta$ are equal the vertices have, apart from
an overall constant, the same
Lorentz structure. The first term, that does not depend
on any of the gauge-fixing  parameters, corresponds to
the convection current. This is the
same current that one obtains for scalars and indeed,
apart from the $g_{\alpha \beta}$
term that counts the vector degrees of freedom, this
is exactly as in scalar
electrodynamics. When we further take
the most ``practical values" $\tilde\alpha=\tilde\beta=1$
(that correspond to taking a covariant derivative along
the $T_3$ direction)
and with $\xi_W=1$ the third term vanishes and the
second is nothing else but
the spin current with the correct value for the
magnetic moment of a spin-1 gauge
particle.

\centerline{\begin{picture}(20000,10000)
\drawline\photon[2 0](0,5000)[4]
\global\Xone=\pfrontx \global\Yone=\pfronty \global\Yeight=\Yone
\global\advance\Yone by 700 \global\advance\Xone by -2000
\put(\Xone,\Yone){$[\gamma^\mu;Z^\mu]$}
\drawline\photon[1 0](\pbackx,\pbacky)[4]
\global\Xtwo=\pbackx \global\Ytwo=\pbacky
\global\advance\Xtwo by -2000 \global\advance\Ytwo by 500
\put(\Xtwo,\Ytwo){$W^{\pm\nu}(p_W)$}
\drawline\scalar[3 0](\pfrontx,\pfronty)[2]
\global\Xthree=\pbackx \global\Ythree=\pbacky
\global\advance\Xthree by -2000 \global\advance\Ythree by -800
\put(\Xthree,\Ythree){$\varphi^{\mp}(p_\varphi)$}
\global\Xfour=\pbackx \global\advance\Xfour by 1000
\put(\Xfour,\Yeight){$i g^{\mu\nu}[eM_W(1-\tilde\alpha);
-gM_Z(1-c_W^2 (1-\tilde\beta))] $}
\end{picture}}
One can make both the $W^\pm \varphi^\mp Z$ and
$W^\pm \varphi^\mp\gamma$ vanish. While the vanishing of the
photon part is for the value $\alpha=1$ that makes
the $WW\gamma$ (and as we will see the
the $WW\gamma\gamma$) simple, the vanishing of the
$W^\pm \varphi^\mp Z$ requires
$\tilde\beta=-s_W^2/c_W^2$ that does not make the
other vertices simpler. Note that these
vertices do not depend on $\xi_W$. The remaining
tri-linear vertices that we list
below can not be made to vanish.

\centerline{\begin{picture}(15000,10000)
\drawline\scalar[2 0](0,5000)[2]
\global\Xone=\scalarfrontx \global\Yone=\scalarfronty \global\Yeight=\Yone
\global\advance\Yone by 700
\put(\Xone,\Yone){$H$}
\drawline\photon[1 0](\pbackx,\pbacky)[4]
\global\Xtwo=\pbackx \global\Ytwo=\pbacky
\global\advance\Xtwo by -3500 \global\advance\Ytwo by 500
\put(\Xtwo,\Ytwo){$[W^{+\rho}(p_+);Z^\rho]$}
\drawline\photon[3 0](\pfrontx,\pfronty)[4]
\global\Xthree=\pbackx \global\Ythree=\pbacky
\global\advance\Xthree by -3500 \global\advance\Ythree by -800
\put(\Xthree,\Ythree){$[W^{-\sigma}(p_-);Z^\sigma]$}
\global\Xfour=\pbackx \global\Yfour=\Yeight
\global\advance\Xfour by 1000
\put(\Xfour,\Yfour){$ig^{\rho\sigma}[gM_W;g_ZM_Z]$}
\end{picture}}
\centerline{\begin{picture}(25000,10000)
\drawline\scalar[2 0](0,5000)[2]
\global\Xone=\pfrontx \global\Yone=\pfronty \global\Yeight=\Yone
\global\advance\Yone by 700 \global\advance\Xone by -2000
\put(\Xone,\Yone){$H$}
\drawline\photon[1 0](\pbackx,\pbacky)[4]
\global\Xtwo=\pbackx \global\Ytwo=\pbacky
\global\advance\Xtwo by -2000 \global\advance\Ytwo by 500
\put(\Xtwo,\Ytwo){$W^{\pm\mu}(p_W)$}
\drawline\scalar[3 0](\pfrontx,\pfronty)[2]
\global\Xthree=\pbackx \global\Ythree=\pbacky
\global\advance\Xthree by -2000 \global\advance\Ythree by -800
\put(\Xthree,\Ythree){$\varphi^{\mp}(p_\varphi)$}
\global\Xfour=\pbackx \global\advance\Xfour by 1000
\put(\Xfour,\Yeight){$\pm i\displaystyle\frac{g}{2}(
(1-\tilde\delta) p_\varphi
-(1+\tilde\delta) p_H )^\mu$}
\end{picture}}
\centerline{\begin{picture}(25000,10000)
\drawline\scalar[2 0](0,5000)[2]
\global\Xone=\pfrontx \global\Yone=\pfronty \global\Yeight=\Yone
\global\advance\Yone by 700 \global\advance\Xone by -2000
\put(\Xone,\Yone){$\varphi_3$}
\drawline\photon[1 0](\pbackx,\pbacky)[4]
\global\Xtwo=\pbackx \global\Ytwo=\pbacky
\global\advance\Xtwo by -2000 \global\advance\Ytwo by 500
\put(\Xtwo,\Ytwo){$W^{\pm\mu}(p_W)$}
\drawline\scalar[3 0](\pfrontx,\pfronty)[2]
\global\Xthree=\pbackx \global\Ythree=\pbacky
\global\advance\Xthree by -2000 \global\advance\Ythree by -800
\put(\Xthree,\Ythree){$\varphi^{\mp}(p_\varphi)$}
\global\Xfour=\pbackx \global\advance\Xfour by 1000
\put(\Xfour,\Yeight)
{$-\displaystyle\frac{g}{2}( (1-\tilde{\kappa})p_\varphi
-(1+\tilde{\kappa})p_{\varphi_3} )^\mu$}
\end{picture}}
\centerline{\begin{picture}(25000,10000)
\drawline\photon[2 0](0,5000)[4]
\global\Xone=\pfrontx \global\Yone=\pfronty \global\Yeight=\Yone
\global\advance\Yone by 700 \global\advance\Xone by -2000
\put(\Xone,\Yone){$[\gamma^\mu;Z^\mu]$}
\drawline\scalar[1 0](\pbackx,\pbacky)[2]
\global\Xtwo=\pbackx \global\Ytwo=\pbacky
\global\advance\Xtwo by -2000 \global\advance\Ytwo by 500
\put(\Xtwo,\Ytwo){$\varphi^{+}(p_+)$}
\drawline\scalar[3 0](\pfrontx,\pfronty)[2]
\global\Xthree=\pbackx \global\Ythree=\pbacky
\global\advance\Xthree by -2000 \global\advance\Ythree by -800
\put(\Xthree,\Ythree){$\varphi^{-}(p_-)$}
\global\Xfour=\pbackx \global\advance\Xfour by 1000
\put(\Xfour,\Yeight){$-i\left[e;g_Z\displaystyle\frac{c_W^2-s_W^2}{2}\right]
(p_+-p_-)^\mu$}
\end{picture}}
\centerline{\begin{picture}(25000,10000)
\drawline\scalar[2 0](0,5000)[2]
\global\Xone=\pfrontx \global\Yone=\pfronty \global\Yeight=\Yone
\global\advance\Yone by 700
\put(\Xone,\Yone){$H$}
\drawline\photon[1 0](\pbackx,\pbacky)[4]
\global\Xtwo=\pbackx \global\Ytwo=\pbacky
\global\advance\Xtwo by -1500 \global\advance\Ytwo by 500
\put(\Xtwo,\Ytwo){$Z^\mu$}
\drawline\scalar[3 0](\pfrontx,\pfronty)[2]
\global\Xthree=\pbackx \global\Ythree=\pbacky
\global\advance\Xthree by -1500 \global\advance\Ythree by -800
\put(\Xthree,\Ythree){$\varphi_3$}
\global\Xfour=\pbackx \global\advance\Xfour by 1000
\put(\Xfour,\Yeight){$\displaystyle\frac{g_Z}{2}((1+\tilde\varepsilon)p_H
-(1-\tilde\varepsilon)p_{\varphi_3})^\mu$}
\end{picture}}
\centerline{\begin{picture}(12000,10000)
\drawline\scalar[2 0](0,5000)[2]
\global\Xone=\pfrontx \global\Yone=\pfronty\global\Yeight=\Yone
\global\advance\Yone by 700
\put(\Xone,\Yone){$H$}
\drawline\scalar[1 0](\pbackx,\pbacky)[2]
\global\Xtwo=\pbackx \global\Ytwo=\pbacky
\global\advance\Xtwo by -1000 \global\advance\Ytwo by 500
\put(\Xtwo,\Ytwo){$H$}
\drawline\scalar[3 0](\pfrontx,\pfronty)[2]
\global\Xthree=\pbackx \global\Ythree=\pbacky
\global\advance\Xthree by -1000 \global\advance\Ythree by -800
\put(\Xthree,\Ythree){$H$}
\global\Xfour=\pbackx \global\advance\Xfour by 1000
\put(\Xfour,\Yeight){$-\displaystyle\frac{3igM_H^2}{2M_W}$}
\end{picture}}
\begin{picture}(5000,10000)
\drawline\scalar[2 0](0,5000)[2]
\global\Xone=\pfrontx \global\Yone=\pfronty \global\Yeight=\Yone
\global\advance\Yone by 700
\put(\Xone,\Yone){$H$}
\drawline\scalar[1 0](\pbackx,\pbacky)[2]
\global\Xtwo=\pbackx \global\Ytwo=\pbacky
\global\advance\Xtwo by -2500 \global\advance\Ytwo by 500
\put(\Xtwo,\Ytwo){$[\varphi_3;\varphi^+]$}
\drawline\scalar[3 0](\pfrontx,\pfronty)[2]
\global\Xthree=\pbackx \global\Ythree=\pbacky
\global\advance\Xthree by -2500 \global\advance\Ythree by -800
\put(\Xthree,\Ythree){$[\varphi_3;\varphi^-]$}
\global\Xfour=\pbackx \global\advance\Xfour by 1000
\put(\Xfour,\Yeight){$\left[-\displaystyle\frac{ig}{2M_W}
(M_H^2+ 2 \xi_W \tilde\delta M_W^2);
(g\ra g_Z,M_W\ra M_Z,\xi_W \ra \xi_Z,
\tilde\delta \ra \tilde\varepsilon\right]$}
\end{picture}\\
\subsection*{Quartic vertices}
\noindent
\begin{picture}(5000,10000)
\drawvertex\photon[3 4](0,5000)[4]
\global\Xone=\vertexonex \global\Yone=\vertexoney
\global\advance\Yone by 500 \global\advance\Xone by -1000
\put(\Xone,\Yone){$W^{+\rho}$}
\global\Xtwo=\vertextwox \global\Ytwo=\vertextwoy
\global\advance\Xtwo by -3000 \global\advance\Ytwo by 500
\put(\Xtwo,\Ytwo){$[\gamma^\mu;\gamma^\mu;Z^\mu]$}
\global\Xthree=\vertexthreex \global\Ythree=\vertexthreey
\global\advance\Xthree by -3000 \global\advance\Ythree by -800
\put(\Xthree,\Ythree){$[\gamma^\nu;Z^\nu;Z^\nu]$}
\global\Xfour=\vertexfourx \global\Yfour=\vertexfoury
\global\advance\Xfour by -1000 \global\advance\Yfour by -800
\put(\Xfour,\Yfour){$W^{-\sigma}$}
\global\Xfive=\vertexthreex \global\Yfive=\vertexmidy
\global\advance\Xfive by 3000 \global\advance\Yfive by 500
\put(\Xfive,\Yfive){$-i e^2 [1;c_W/s_W;c_W^2/s_W^2]
(2g^{\mu\nu}g^{\rho\sigma}$}
\global\Ysix=\Yfive \global\advance\Ysix by -1500
\put(\Xfive,\Ysix){$ -(g^{\mu\sigma}g^{\nu\rho}+g^{\mu\rho}g^{\nu\sigma})
[(1-\tilde\alpha^2/\xi_W);
(1-\tilde\alpha\tilde\beta/\xi_W);(1-\tilde\beta^2/\xi_W)])$}
\end{picture}\\

Again for the values that correspond to the $T_3$ covariant derivative
and $\xi_W=1$ there only remains the same part that one finds for the
scalars (apart from the factor counting the vector degrees of freedom).

\begin{picture}(5000,10000)
\drawvertex\photon[3 4](0,5000)[4]
\global\Xone=\vertexonex \global\Yone=\vertexoney
\global\advance\Yone by 500 \global\advance\Xone by -1000
\put(\Xone,\Yone){$W^{+\rho}$}
\global\Xtwo=\vertextwox \global\Ytwo=\vertextwoy
\global\advance\Xtwo by -1000 \global\advance\Ytwo by 500
\put(\Xtwo,\Ytwo){$W^{+\mu}$}
\global\Xthree=\vertexthreex \global\Ythree=\vertexthreey
\global\advance\Xthree by -1000 \global\advance\Ythree by -800
\put(\Xthree,\Ythree){$W^{-\nu}$}
\global\Xfour=\vertexfourx \global\Yfour=\vertexfoury
\global\advance\Xfour by -1000 \global\advance\Yfour by -800
\put(\Xfour,\Yfour){$W^{-\sigma}$}
\global\Xfive=\vertexthreex \global\Yfive=\vertexmidy
\global\advance\Xfive by 3000 \global\advance\Yfive by 500
\put(\Xfive,\Yfive){$ig^2(2g^{\mu\rho}g^{\nu\sigma}-
(g^{\mu\sigma}g^{\nu\rho}+g^{\mu\nu}g^{\rho\sigma}))$}
\end{picture}\\
\begin{picture}(5000,10000)
\drawline\photon[3 1](0,5000)[4]
\global\Xone=\pfrontx \global\Yone=\pfronty \global\Yeight=\pbacky
\global\advance\Yone by 500 \global\advance\Xone by -1000
\put(\Xone,\Yone){$\gamma_\mu$}
\drawline\scalar[1 0](\pbackx,\pbacky)[2]
\global\Xtwo=\pbackx \global\Ytwo=\pbacky \global\Xeight=\Xtwo
\global\advance\Xtwo by -1000 \global\advance\Ytwo by 500
\put(\Xtwo,\Ytwo){$\varphi^\mp$}
\drawline\scalar[3 0](\pfrontx,\pfronty)[2]
\global\Xthree=\pbackx \global\Ythree=\pbacky
\global\advance\Xthree by -1500 \global\advance\Ythree by -800
\put(\Xthree,\Ythree){$[H;\varphi_3]$}
\drawline\photon[5 0](\pfrontx,\pfronty)[4]
\global\Xfour=\pbackx \global\Yfour=\pbacky
\global\advance\Xfour by -1000 \global\advance\Yfour by -800
\put(\Xfour,\Yfour){$W^\pm_\nu$}
\global\advance\Xeight by 2000
\put(\Xeight,\Yeight){$[i(1-\tilde\alpha\tilde\delta);
\mp(1-\tilde\alpha\tilde\kappa)]\displaystyle\frac{ge}{2}g_{\mu\nu}$}
\end{picture}\\

Note that it is not sufficient to take $\tilde\alpha=1$ to get rid of this
vertex. \\
\begin{picture}(30000,10000)
\drawline\photon[3 1](0,5000)[4]
\global\Xone=\pfrontx \global\Yone=\pfronty \global\Yeight=\pbacky
\global\advance\Yone by 500 \global\advance\Xone by -1000
\put(\Xone,\Yone){$Z_\mu$}
\drawline\scalar[1 0](\pbackx,\pbacky)[2]
\global\Xtwo=\pbackx \global\Ytwo=\pbacky \global\Xeight=\Xtwo
\global\advance\Xtwo by -1000 \global\advance\Ytwo by 500
\put(\Xtwo,\Ytwo){$\varphi^\mp$}
\drawline\scalar[3 0](\pfrontx,\pfronty)[2]
\global\Xthree=\pbackx \global\Ythree=\pbacky
\global\advance\Xthree by -1500 \global\advance\Ythree by -800
\put(\Xthree,\Ythree){$[H;\varphi_3]$}
\drawline\photon[5 0](\pfrontx,\pfronty)[4]
\global\Xfour=\pbackx \global\Yfour=\pbacky
\global\advance\Xfour by -1000 \global\advance\Yfour by -800
\put(\Xfour,\Yfour){$W^\pm_\nu$}
\global\advance\Xeight by 2000
\put(\Xeight,\Yeight){$[-i\left( 1-c_W^2(1-\tilde\beta\tilde\delta)\right);
\pm \left( 1-c_W^2(1-\tilde\beta\tilde\kappa)\right)]
\displaystyle\frac{gg_Z}{2}g_{\mu\nu}$}
\end{picture}\\

Note that if  $\tilde\delta=\tilde\kappa=1$ then the same condition that makes
the $W^\pm \varphi^\mp Z$ vanish, eliminates this vertex too.\\
\noindent\begin{picture}(14000,10000)
\drawline\photon[3 1](0,5000)[4]
\global\Xone=\pfrontx \global\Yone=\pfronty \global\Yeight=\pbacky
\global\advance\Yone by 500 \global\advance\Xone by -1000
\put(\Xone,\Yone){$W^{+\rho}$}
\drawline\scalar[1 0](\pbackx,\pbacky)[2]
\global\Xtwo=\pbackx \global\Ytwo=\pbacky \global\Xeight=\Xtwo
\global\advance\Xtwo by -2500 \global\advance\Ytwo by 500
\put(\Xtwo,\Ytwo){$[H;\varphi_3;\varphi^+]$}
\drawline\scalar[3 0](\pfrontx,\pfronty)[2]
\global\Xthree=\pbackx \global\Ythree=\pbacky
\global\advance\Xthree by -2500 \global\advance\Ythree by -800
\put(\Xthree,\Ythree){$[H;\varphi_3;\varphi^-]$}
\drawline\photon[5 0](\pfrontx,\pfronty)[4]
\global\Xfour=\pbackx \global\Yfour=\pbacky
\global\advance\Xfour by -1000 \global\advance\Yfour by -800
\put(\Xfour,\Yfour){$W^{-\sigma}$}
\global\advance\Xeight by 2000
\put(\Xeight,\Yeight){$\displaystyle\frac{ig^2}{2}g^{\rho\sigma}$}
\end{picture}\\
\begin{picture}(14000,10000)
\drawline\photon[3 1](0,5000)[4]
\global\Xone=\pfrontx \global\Yone=\pfronty \global\Yeight=\pbacky
\global\advance\Yone by 500 \global\advance\Xone by -1000
\put(\Xone,\Yone){$Z^\mu$}
\drawline\scalar[1 0](\pbackx,\pbacky)[2]
\global\Xtwo=\pbackx \global\Ytwo=\pbacky \global\Xeight=\Xtwo
\global\advance\Xtwo by -1500 \global\advance\Ytwo by 500
\put(\Xtwo,\Ytwo){$[H;\varphi_3]$}
\drawline\scalar[3 0](\pfrontx,\pfronty)[2]
\global\Xthree=\pbackx \global\Ythree=\pbacky
\global\advance\Xthree by -1500 \global\advance\Ythree by -800
\put(\Xthree,\Ythree){$[H;\varphi_3]$}
\drawline\photon[5 0](\pfrontx,\pfronty)[4]
\global\Xfour=\pbackx \global\Yfour=\pbacky
\global\advance\Xfour by -1000 \global\advance\Yfour by -800
\put(\Xfour,\Yfour){$Z^\nu$}
\global\advance\Xeight by 2000
\put(\Xeight,\Yeight){$\displaystyle\frac{ig_Z^2}{2}g^{\mu\nu}$}
\end{picture}\\
\begin{picture}(30000,10000)
\drawline\scalar[3 0](0,5000)[2]
\global\Xone=\pfrontx \global\Yone=\pfronty \global\Yeight=\pbacky
\global\advance\Yone by 500 \global\advance\Xone by -1000
\put(\Xone,\Yone){$\varphi^{+}$}
\drawline\photon[1 0](\pbackx,\pbacky)[4]
\global\Xtwo=\pbackx \global\Ytwo=\pbacky \global\Xeight=\Xtwo
\global\advance\Xtwo by -2500 \global\advance\Ytwo by 500
\put(\Xtwo,\Ytwo){$[\gamma^\mu;\gamma^\mu;Z^\mu]$}
\drawline\photon[3 1](\pfrontx,\pfronty)[4]
\global\Xthree=\pbackx \global\Ythree=\pbacky
\global\advance\Xthree by -2500 \global\advance\Ythree by -800
\put(\Xthree,\Ythree){$[\gamma^\nu;Z^\nu;Z^\nu]$}
\drawline\scalar[5 0](\pfrontx,\pfronty)[2]
\global\Xfour=\pbackx \global\Yfour=\pbacky
\global\advance\Xfour by -1000 \global\advance\Yfour by -800
\put(\Xfour,\Yfour){$\varphi^{-}$}
\global\advance\Xeight by 2000
\put(\Xeight,\Yeight){$2ie^2\left[1;\displaystyle
\frac{c_W^2-s_W^2}{2 s_W c_W};
\left(\frac{c_W^2-s_W^2}{2 s_W c_W}\right)^2\right]g^{\mu\nu}$}
\end{picture}\\
\begin{picture}(5000,10000)
\drawline\scalar[3 0](0,5000)[2]
\global\Xone=\pfrontx \global\Yone=\pfronty \global\Yeight=\pbacky
\global\advance\Yone by 500 \global\advance\Xone by -3000
\put(\Xone,\Yone){$[H;\varphi_3;\varphi^\pm]$}
\drawline\scalar[1 0](\pbackx,\pbacky)[2]
\global\Xtwo=\pbackx \global\Ytwo=\pbacky \global\Xeight=\Xtwo
\global\advance\Xtwo by -2000 \global\advance\Ytwo by 500
\put(\Xtwo,\Ytwo){$[H;\varphi_3;\varphi^\pm]$}
\drawline\scalar[3 0](\pfrontx,\pfronty)[2]
\global\Xthree=\pbackx \global\Ythree=\pbacky
\global\advance\Xthree by -2000 \global\advance\Ythree by -800
\put(\Xthree,\Ythree){$[H;\varphi_3;\varphi^\mp]$}
\drawline\scalar[5 0](\pfrontx,\pfronty)[2]
\global\Xfour=\pbackx \global\Yfour=\pbacky
\global\advance\Xfour by -3000 \global\advance\Yfour by -800
\put(\Xfour,\Yfour){$[H;\varphi_3;\varphi^\mp]$}
\global\advance\Xeight by 2000
\put(\Xeight,\Yeight){$-i\displaystyle\frac{g^2 M_H^2}{2M_W^2}
\left[\frac{3}{2};\frac{3}{2};1\right]$}
\end{picture} \\
\begin{picture}(5000,10000)
\drawline\scalar[3 0](0,5000)[2]
\global\Xone=\pfrontx \global\Yone=\pfronty \global\Yeight=\pbacky
\global\advance\Yone by 500 \global\advance\Xone by -1000
\put(\Xone,\Yone){$H$}
\drawline\scalar[1 0](\pbackx,\pbacky)[2]
\global\Xtwo=\pbackx \global\Ytwo=\pbacky \global\Xeight=\Xtwo
\global\advance\Xtwo by -1000 \global\advance\Ytwo by 500
\put(\Xtwo,\Ytwo){$\varphi_3$}
\drawline\scalar[3 0](\pfrontx,\pfronty)[2]
\global\Xthree=\pbackx \global\Ythree=\pbacky
\global\advance\Xthree by -1000 \global\advance\Ythree by -800
\put(\Xthree,\Ythree){$\varphi_3$}
\drawline\scalar[5 0](\pfrontx,\pfronty)[2]
\global\Xfour=\pbackx \global\Yfour=\pbacky
\global\advance\Xfour by -1000 \global\advance\Yfour by -800
\put(\Xfour,\Yfour){$H$}
\global\advance\Xeight by 2000
\put(\Xeight,\Yeight){$-\displaystyle\frac{ig^2}{4M_W^2}
(M_H^2+2M_Z^2\tilde\varepsilon^2\xi_Z)$}
\end{picture}\\
\begin{picture}(25000,10000)
\drawline\scalar[3 0](0,5000)[2]
\global\Xone=\pfrontx \global\Yone=\pfronty \global\Yeight=\pbacky
\global\advance\Yone by 500 \global\advance\Xone by -1000
\put(\Xone,\Yone){$\varphi^+$}
\drawline\scalar[1 0](\pbackx,\pbacky)[2]
\global\Xtwo=\pbackx \global\Ytwo=\pbacky \global\Xeight=\Xtwo
\global\advance\Xtwo by -1500 \global\advance\Ytwo by 500
\put(\Xtwo,\Ytwo){$[H;\varphi_3]$}
\drawline\scalar[3 0](\pfrontx,\pfronty)[2]
\global\Xthree=\pbackx \global\Ythree=\pbacky
\global\advance\Xthree by -1500 \global\advance\Ythree by -800
\put(\Xthree,\Ythree){$[H;\varphi_3]$}
\drawline\scalar[5 0](\pfrontx,\pfronty)[2]
\global\Xfour=\pbackx \global\Yfour=\pbacky
\global\advance\Xfour by -1000 \global\advance\Yfour by -800
\put(\Xfour,\Yfour){$\varphi^-$}
\global\advance\Xeight by 2000
\put(\Xeight,\Yeight){$-\displaystyle\frac{ig^2}{4M_W^2}
[M_H^2+2M_W^2\tilde\delta^2\xi_W;
M_H^2+2M_W^2\tilde\kappa^2\xi_W]$}
\end{picture}\\

\subsection*{Ghosts vertices}
\noindent
\begin{picture}(25000,10000)
\drawline\photon[2 0](0,5000)[4]
\global\Xone=\pfrontx \global\Yone=\pfronty \global\Yeight=\Yone
\global\advance\Yone by 700
\put(\Xone,\Yone){$W^{\pm\mu}$}
\drawline\fermion[1 0](\pbackx,\pbacky)[4000]
\global\Xtwo=\pbackx \global\Ytwo=\pbacky
\global\advance\Xtwo by -2000 \global\advance\Ytwo by 500
\put(\Xtwo,\Ytwo){$[\bar\eta_\gamma;\bar\eta_Z](\bar p)$}
\drawline\fermion[3 0](\pfrontx,\pfronty)[4000]
\global\Xthree=\pbackx \global\Ythree=\pbacky
\global\advance\Xthree by -1000 \global\advance\Ythree by -800
\put(\Xthree,\Ythree){$\eta_\mp$}
\global\Xfour=\pbackx \global\advance\Xfour by 1000
\put(\Xfour,\Yeight){$\pm i\bar p^{\;\mu}[e;gc_W]$}
\end{picture}
\begin{picture}(5000,10000)
\drawline\photon[2 0](0,5000)[4]
\global\Xone=\pfrontx \global\Yone=\pfronty \global\Yeight=\Yone
\global\advance\Yone by 700
\put(\Xone,\Yone){$W^{\pm\mu}$}
\drawline\fermion[1 0](\pbackx,\pbacky)[4000]
\global\Xtwo=\pbackx \global\Ytwo=\pbacky
\global\advance\Xtwo by -1000 \global\advance\Ytwo by 500
\put(\Xtwo,\Ytwo){$\bar\eta_\pm(\bar p)$}
\drawline\fermion[3 0](\pfrontx,\pfronty)[4000]
\global\Xthree=\pbackx \global\Ythree=\pbacky
\global\advance\Xthree by -2000 \global\advance\Ythree by -800
\put(\Xthree,\Ythree){$[\eta_\gamma;\eta_Z](p)$}
\global\Xfour=\pbackx \global\advance\Xfour by 1000
\put(\Xfour,\Yeight){$\mp i[e;gc_W](\bar p+[\tilde\alpha;
\tilde\beta]p)^\mu$}
\end{picture} \\
\begin{picture}(25000,10000)
\drawline\scalar[2 0](0,5000)[2]
\global\Xone=\pfrontx \global\Yone=\pfronty \global\Yeight=\Yone
\global\advance\Yone by 700
\put(\Xone,\Yone){$[H;\varphi^\pm]$}
\drawline\fermion[1 0](\pbackx,\pbacky)[4000]
\global\Xtwo=\pbackx \global\Ytwo=\pbacky
\global\advance\Xtwo by -1000 \global\advance\Ytwo by 500
\put(\Xtwo,\Ytwo){$\bar\eta_Z$}
\drawline\fermion[3 0](\pfrontx,\pfronty)[4000]
\global\Xthree=\pbackx \global\Ythree=\pbacky
\global\advance\Xthree by -2000 \global\advance\Ythree by -800
\put(\Xthree,\Ythree){$[\eta_Z;\eta_\mp]$}
\global\Xfour=\pbackx \global\advance\Xfour by 1000
\put(\Xfour,\Yeight){$-\displaystyle\frac{im_Z\xi_Z}{2}
[g_Z(1+\tilde\varepsilon);-g]$}
\end{picture}
\begin{picture}(5000,10000)
\drawline\scalar[2 0](0,5000)[2]
\global\Xone=\pfrontx \global\Yone=\pfronty \global\Yeight=\Yone
\global\advance\Yone by 700
\put(\Xone,\Yone){$\varphi^\pm$}
\drawline\fermion[1 0](\pbackx,\pbacky)[4000]
\global\Xtwo=\pbackx \global\Ytwo=\pbacky
\global\advance\Xtwo by -1000 \global\advance\Ytwo by 500
\put(\Xtwo,\Ytwo){$\bar\eta_\pm$}
\drawline\fermion[3 0](\pfrontx,\pfronty)[4000]
\global\Xthree=\pbackx \global\Ythree=\pbacky
\global\advance\Xthree by -2000 \global\advance\Ythree by -800
\put(\Xthree,\Ythree){$[\eta_\gamma;\eta_Z]$}
\global\Xfour=\pbackx \global\advance\Xfour by 1000
\put(\Xfour,\Yeight){$-iM_W\xi_W[e;\displaystyle\frac{g_Z}{2}
(\tilde\kappa+c_W^2-s_W^2)]$}
\end{picture} \\
\begin{picture}(25000,10000)
\drawline\photon[2 0](0,5000)[4]
\global\Xone=\pfrontx \global\Yone=\pfronty \global\Yeight=\Yone
\global\advance\Yone by 700
\put(\Xone,\Yone){$[\gamma^\mu;Z^\mu]$}
\drawline\fermion[1 0](\pbackx,\pbacky)[4000]
\global\Xtwo=\pbackx \global\Ytwo=\pbacky
\global\advance\Xtwo by -1000 \global\advance\Ytwo by 500
\put(\Xtwo,\Ytwo){$\bar\eta_\pm(\bar p)$}
\drawline\fermion[3 0](\pfrontx,\pfronty)[4000]
\global\Xthree=\pbackx \global\Ythree=\pbacky
\global\advance\Xthree by -1000 \global\advance\Ythree by -800
\put(\Xthree,\Ythree){$\eta_\pm(p)$}
\global\Xfour=\pbackx \global\advance\Xfour by 1000
\put(\Xfour,\Yeight){$\pm i[e;gc_W](\bar p-
[\tilde\alpha;\tilde\beta]p)^\mu$}
\end{picture}
\begin{picture}(5000,10000)
\drawline\scalar[2 0](0,5000)[2]
\global\Xone=\pfrontx \global\Yone=\pfronty \global\Yeight=\Yone
\global\advance\Yone by 700
\put(\Xone,\Yone){$[H;\varphi_3]$}
\drawline\fermion[1 0](\pbackx,\pbacky)[4000]
\global\Xtwo=\pbackx \global\Ytwo=\pbacky
\global\advance\Xtwo by -1000 \global\advance\Ytwo by 500
\put(\Xtwo,\Ytwo){$\bar\eta_\pm$}
\drawline\fermion[3 0](\pfrontx,\pfronty)[4000]
\global\Xthree=\pbackx \global\Ythree=\pbacky
\global\advance\Xthree by -1000 \global\advance\Ythree by -800
\put(\Xthree,\Ythree){$\eta_\pm$}
\global\Xfour=\pbackx \global\advance\Xfour by 1000
\put(\Xfour,\Yeight){$[-i(1+\tilde\delta);
\pm(1-\tilde\kappa)]\displaystyle\frac{gM_W\xi_W}{2}$}
\end{picture} \\
\begin{picture}(20000,10000)
\drawline\scalar[3 0](0,5000)[2]
\global\Xone=\pfrontx \global\Yone=\pfronty \global\Yeight=\pbacky
\global\advance\Yone by 500 \global\advance\Xone by -1500
\put(\Xone,\Yone){$[H;\varphi_3]$}
\drawline\fermion[1 0](\pbackx,\pbacky)[4000]
\global\Xtwo=\pbackx \global\Ytwo=\pbacky \global\Xeight=\Xtwo
\global\advance\Xtwo by -1000 \global\advance\Ytwo by 500
\put(\Xtwo,\Ytwo){$\bar\eta_Z$}
\drawline\fermion[3 0](\pfrontx,\pfronty)[4000]
\global\Xthree=\pbackx \global\Ythree=\pbacky
\global\advance\Xthree by -1000 \global\advance\Ythree by -800
\put(\Xthree,\Ythree){$\eta_Z$}
\drawline\scalar[5 0](\pfrontx,\pfronty)[2]
\global\Xfour=\pbackx \global\Yfour=\pbacky
\global\advance\Xfour by -1500 \global\advance\Yfour by -800
\put(\Xfour,\Yfour){$[H;\varphi_3]$}
\global\advance\Xeight by 2000
\put(\Xeight,\Yeight){$[-;+]i\xi_Z\tilde\varepsilon
\displaystyle\frac{g_Z^2}{2}$}
\end{picture}
\begin{picture}(5000,10000)
\drawline\scalar[3 0](0,5000)[2]
\global\Xone=\pfrontx \global\Yone=\pfronty \global\Yeight=\pbacky
\global\advance\Yone by 500 \global\advance\Xone by -1500
\put(\Xone,\Yone){$[H;\varphi_3]$}
\drawline\fermion[1 0](\pbackx,\pbacky)[4000]
\global\Xtwo=\pbackx \global\Ytwo=\pbacky \global\Xeight=\Xtwo
\global\advance\Xtwo by -1000 \global\advance\Ytwo by 500
\put(\Xtwo,\Ytwo){$\bar\eta_Z$}
\drawline\fermion[3 0](\pfrontx,\pfronty)[4000]
\global\Xthree=\pbackx \global\Ythree=\pbacky
\global\advance\Xthree by -1000 \global\advance\Ythree by -800
\put(\Xthree,\Ythree){$\eta_\pm$}
\drawline\scalar[5 0](\pfrontx,\pfronty)[2]
\global\Xfour=\pbackx \global\Yfour=\pbacky
\global\advance\Xfour by -1000 \global\advance\Yfour by -800
\put(\Xfour,\Yfour){$\varphi^\pm$}
\global\advance\Xeight by 2000
\put(\Xeight,\Yeight){$[i;\mp 1]\tilde\varepsilon\xi_Z
\displaystyle\frac{gg_Z}{4}$}
\end{picture}\\
\begin{picture}(20000,10000)
\drawline\photon[3 1](0,5000)[4]
\global\Xone=\pfrontx \global\Yone=\pfronty \global\Yeight=\pbacky
\global\advance\Yone by 500 \global\advance\Xone by -1000
\put(\Xone,\Yone){$W^\pm_\mu$}
\drawline\fermion[1 0](\pbackx,\pbacky)[4000]
\global\Xtwo=\pbackx \global\Ytwo=\pbacky \global\Xeight=\Xtwo
\global\advance\Xtwo by -1000 \global\advance\Ytwo by 500
\put(\Xtwo,\Ytwo){$\bar\eta_\pm$}
\drawline\fermion[3 0](\pfrontx,\pfronty)[4000]
\global\Xthree=\pbackx \global\Ythree=\pbacky
\global\advance\Xthree by -1000 \global\advance\Ythree by -800
\put(\Xthree,\Ythree){$\eta_\gamma$}
\drawline\photon[5 0](\pfrontx,\pfronty)[4]
\global\Xfour=\pbackx \global\Yfour=\pbacky
\global\advance\Xfour by -2000 \global\advance\Yfour by -800
\put(\Xfour,\Yfour){$[\gamma_\nu;Z_\nu]$}
\global\advance\Xeight by 2000
\put(\Xeight,\Yeight){$-ieg_{\mu\nu}[e\tilde\alpha;
gc_W\tilde\beta]$}
\end{picture}
\begin{picture}(5000,10000)
\drawline\photon[3 1](0,5000)[4]
\global\Xone=\pfrontx \global\Yone=\pfronty \global\Yeight=\pbacky
\global\advance\Yone by 500 \global\advance\Xone by -1000
\put(\Xone,\Yone){$W^\pm_\mu$}
\drawline\fermion[1 0](\pbackx,\pbacky)[4000]
\global\Xtwo=\pbackx \global\Ytwo=\pbacky \global\Xeight=\Xtwo
\global\advance\Xtwo by -1000 \global\advance\Ytwo by 500
\put(\Xtwo,\Ytwo){$\bar\eta_\pm$}
\drawline\fermion[3 0](\pfrontx,\pfronty)[4000]
\global\Xthree=\pbackx \global\Ythree=\pbacky
\global\advance\Xthree by -1000 \global\advance\Ythree by -800
\put(\Xthree,\Ythree){$\eta_Z$}
\drawline\photon[5 0](\pfrontx,\pfronty)[4]
\global\Xfour=\pbackx \global\Yfour=\pbacky
\global\advance\Xfour by -2000 \global\advance\Yfour by -800
\put(\Xfour,\Yfour){$[\gamma_\nu;Z_\nu]$}
\global\advance\Xeight by 2000
\put(\Xeight,\Yeight){$-\displaystyle\frac{ie^2c_W}{s_W}g_{\mu\nu}
[\tilde\alpha;\tilde\beta\frac{c_W}{s_W}]$}
\end{picture}\\
\begin{picture}(18000,10000)
\drawline\scalar[3 0](0,5000)[2]
\global\Xone=\pfrontx \global\Yone=\pfronty \global\Yeight=\pbacky
\global\advance\Yone by 500 \global\advance\Xone by -1500
\put(\Xone,\Yone){$[H;\varphi_3]$}
\drawline\fermion[1 0](\pbackx,\pbacky)[4000]
\global\Xtwo=\pbackx \global\Ytwo=\pbacky \global\Xeight=\Xtwo
\global\advance\Xtwo by -1000 \global\advance\Ytwo by 500
\put(\Xtwo,\Ytwo){$\bar\eta_\pm$}
\drawline\fermion[3 0](\pfrontx,\pfronty)[4000]
\global\Xthree=\pbackx \global\Ythree=\pbacky
\global\advance\Xthree by -1000 \global\advance\Ythree by -800
\put(\Xthree,\Ythree){$\eta_\gamma$}
\drawline\scalar[5 0](\pfrontx,\pfronty)[2]
\global\Xfour=\pbackx \global\Yfour=\pbacky
\global\advance\Xfour by -1000 \global\advance\Yfour by -800
\put(\Xfour,\Yfour){$\varphi^\pm$}
\global\advance\Xeight by 1500
\put(\Xeight,\Yeight){$-\displaystyle\frac{e^2\xi_W}{2s_W}
[i\tilde\delta;\pm\tilde\kappa]$}
\end{picture}
\begin{picture}(5000,10000)
\drawline\scalar[3 0](0,5000)[2]
\global\Xone=\pfrontx \global\Yone=\pfronty \global\Yeight=\pbacky
\global\advance\Yone by 500 \global\advance\Xone by -1500
\put(\Xone,\Yone){$[H;\varphi_3]$}
\drawline\fermion[1 0](\pbackx,\pbacky)[4000]
\global\Xtwo=\pbackx \global\Ytwo=\pbacky \global\Xeight=\Xtwo
\global\advance\Xtwo by -1000 \global\advance\Ytwo by 500
\put(\Xtwo,\Ytwo){$\bar\eta_\pm$}
\drawline\fermion[3 0](\pfrontx,\pfronty)[4000]
\global\Xthree=\pbackx \global\Ythree=\pbacky
\global\advance\Xthree by -1000 \global\advance\Ythree by -800
\put(\Xthree,\Ythree){$\eta_Z$}
\drawline\scalar[5 0](\pfrontx,\pfronty)[2]
\global\Xfour=\pbackx \global\Yfour=\pbacky
\global\advance\Xfour by -1000 \global\advance\Yfour by -800
\put(\Xfour,\Yfour){$\varphi^\pm$}
\put(\Xeight,\Yeight){$= -\displaystyle\frac{gg_Z\xi_W}{4}
[i(\tilde\kappa+\tilde\delta(c_W^2-s_W^2));
\pm(\tilde\delta +\tilde\kappa(c_W^2-s_W^2))]$}
\end{picture}\\
\begin{picture}(20000,10000)
\drawline\photon[3 1](0,5000)[4]
\global\Xone=\pfrontx \global\Yone=\pfronty \global\Yeight=\pbacky
\global\advance\Yone by 500 \global\advance\Xone by -1000
\put(\Xone,\Yone){$W^+_\mu$}
\drawline\fermion[1 0](\pbackx,\pbacky)[4000]
\global\Xtwo=\pbackx \global\Ytwo=\pbacky \global\Xeight=\Xtwo
\global\advance\Xtwo by -1000 \global\advance\Ytwo by 500
\put(\Xtwo,\Ytwo){$\bar\eta_\pm$}
\drawline\fermion[3 0](\pfrontx,\pfronty)[4000]
\global\Xthree=\pbackx \global\Ythree=\pbacky
\global\advance\Xthree by -1000 \global\advance\Ythree by -800
\put(\Xthree,\Ythree){$\eta_\pm$}
\drawline\photon[5 0](\pfrontx,\pfronty)[4]
\global\Xfour=\pbackx \global\Yfour=\pbacky
\global\advance\Xfour by -1000 \global\advance\Yfour by -800
\put(\Xfour,\Yfour){$W^-_\nu$}
\global\advance\Xeight by 2000
\put(\Xeight,\Yeight){$-i(e^2\tilde\alpha+
g^2c_W^2\tilde\beta)g_{\mu\nu}$}
\end{picture}
\begin{picture}(5000,10000)
\drawline\photon[3 1](0,5000)[4]
\global\Xone=\pfrontx \global\Yone=\pfronty \global\Yeight=\pbacky
\global\advance\Yone by 500 \global\advance\Xone by -1000
\put(\Xone,\Yone){$W^\pm_\mu$}
\drawline\fermion[1 0](\pbackx,\pbacky)[4000]
\global\Xtwo=\pbackx \global\Ytwo=\pbacky \global\Xeight=\Xtwo
\global\advance\Xtwo by -1000 \global\advance\Ytwo by 500
\put(\Xtwo,\Ytwo){$\bar\eta_\pm$}
\drawline\fermion[3 0](\pfrontx,\pfronty)[4000]
\global\Xthree=\pbackx \global\Ythree=\pbacky
\global\advance\Xthree by -1000 \global\advance\Ythree by -800
\put(\Xthree,\Ythree){$\eta_\mp$}
\drawline\photon[5 0](\pfrontx,\pfronty)[4]
\global\Xfour=\pbackx \global\Yfour=\pbacky
\global\advance\Xfour by -1000 \global\advance\Yfour by -800
\put(\Xfour,\Yfour){$W^\pm_\nu$}
\global\advance\Xeight by 2000
\put(\Xeight,\Yeight){$2i(e^2\tilde\alpha+
g^2c_W^2\tilde\beta)g_{\mu\nu}$}
\end{picture}\\
\begin{picture}(20000,10000)
\drawline\scalar[3 0](0,5000)[2]
\global\Xone=\pfrontx \global\Yone=\pfronty \global\Yeight=\pbacky
\global\advance\Yone by 500 \global\advance\Xone by -1000
\put(\Xone,\Yone){$\varphi^+$}
\drawline\fermion[1 0](\pbackx,\pbacky)[4000]
\global\Xtwo=\pbackx \global\Ytwo=\pbacky \global\Xeight=\Xtwo
\global\advance\Xtwo by -1000 \global\advance\Ytwo by 500
\put(\Xtwo,\Ytwo){$\bar\eta_\pm$}
\drawline\fermion[3 0](\pfrontx,\pfronty)[4000]
\global\Xthree=\pbackx \global\Ythree=\pbacky
\global\advance\Xthree by -1000 \global\advance\Ythree by -800
\put(\Xthree,\Ythree){$\eta_\pm$}
\drawline\scalar[5 0](\pfrontx,\pfronty)[2]
\global\Xfour=\pbackx \global\Yfour=\pbacky
\global\advance\Xfour by -1000 \global\advance\Yfour by -800
\put(\Xfour,\Yfour){$\varphi^-$}
\global\advance\Xeight by 2000
\put(\Xeight,\Yeight){$i\displaystyle\frac{g^2\xi_W}{4}(\tilde\delta+
\tilde\kappa)$}
\end{picture}
\begin{picture}(5000,10000)
\drawline\scalar[3 0](0,5000)[2]
\global\Xone=\pfrontx \global\Yone=\pfronty \global\Yeight=\pbacky
\global\advance\Yone by 500 \global\advance\Xone by -2500
\put(\Xone,\Yone){$[H;\varphi_3;H]$}
\drawline\fermion[1 0](\pbackx,\pbacky)[4000]
\global\Xtwo=\pbackx \global\Ytwo=\pbacky \global\Xeight=\Xtwo
\global\advance\Xtwo by -1000 \global\advance\Ytwo by 500
\put(\Xtwo,\Ytwo){$\bar\eta_\pm$}
\drawline\fermion[3 0](\pfrontx,\pfronty)[4000]
\global\Xthree=\pbackx \global\Ythree=\pbacky
\global\advance\Xthree by -1000 \global\advance\Ythree by -800
\put(\Xthree,\Ythree){$\eta_\pm$}
\drawline\scalar[5 0](\pfrontx,\pfronty)[2]
\global\Xfour=\pbackx \global\Yfour=\pbacky
\global\advance\Xfour by -2500 \global\advance\Yfour by -800
\put(\Xfour,\Yfour){$[H;\varphi_3;\varphi_3]$}
\global\advance\Xeight by 2000
\put(\Xeight,\Yeight){$-\displaystyle\frac{g^2\xi_W}{4}[2i\tilde\delta;
2i\tilde\kappa;\pm(\tilde\kappa-\tilde\delta)]$}
\end{picture}\\
\begin{picture}(30000,10000)
\drawline\photon[3 1](0,5000)[4]
\global\Xone=\pfrontx \global\Yone=\pfronty \global\Yeight=\pbacky
\global\advance\Yone by 500 \global\advance\Xone by -2500
\put(\Xone,\Yone){$[\gamma_\mu;Z_\mu;\gamma_\mu]$}
\drawline\fermion[1 0](\pbackx,\pbacky)[4000]
\global\Xtwo=\pbackx \global\Ytwo=\pbacky \global\Xeight=\Xtwo
\global\advance\Xtwo by -1000 \global\advance\Ytwo by 500
\put(\Xtwo,\Ytwo){$\bar\eta_\pm$}
\drawline\fermion[3 0](\pfrontx,\pfronty)[4000]
\global\Xthree=\pbackx \global\Ythree=\pbacky
\global\advance\Xthree by -1000 \global\advance\Ythree by -800
\put(\Xthree,\Ythree){$\eta_\pm$}
\drawline\photon[5 0](\pfrontx,\pfronty)[4]
\global\Xfour=\pbackx \global\Yfour=\pbacky
\global\advance\Xfour by -2500 \global\advance\Yfour by -800
\put(\Xfour,\Yfour){$[\gamma_\nu;Z_\nu;Z_\nu]$}
\global\advance\Xeight by 2000
\put(\Xeight,\Yeight){$i[2e^2\tilde\alpha;2g^2c_W^2\tilde\beta;g^2c_Ws_W
(\tilde\alpha+\tilde\beta)]g_{\mu\nu}$}
\end{picture}
\begin{picture}(5000,10000)
\drawline\scalar[3 0](0,5000)[2]
\global\Xone=\pfrontx \global\Yone=\pfronty \global\Yeight=\pbacky
\global\advance\Yone by 500 \global\advance\Xone by -1000
\put(\Xone,\Yone){$\varphi^\pm$}
\drawline\fermion[1 0](\pbackx,\pbacky)[4000]
\global\Xtwo=\pbackx \global\Ytwo=\pbacky \global\Xeight=\Xtwo
\global\advance\Xtwo by -1000 \global\advance\Ytwo by 500
\put(\Xtwo,\Ytwo){$\bar\eta_\pm$}
\drawline\fermion[3 0](\pfrontx,\pfronty)[4000]
\global\Xthree=\pbackx \global\Ythree=\pbacky
\global\advance\Xthree by -1000 \global\advance\Ythree by -800
\put(\Xthree,\Ythree){$\eta_\mp$}
\drawline\scalar[5 0](\pfrontx,\pfronty)[2]
\global\Xfour=\pbackx \global\Yfour=\pbacky
\global\advance\Xfour by -1000 \global\advance\Yfour by -800
\put(\Xfour,\Yfour){$\varphi^\pm$}
\global\advance\Xeight by 2000
\put(\Xeight,\Yeight){$i\displaystyle\frac{g^2\xi_W}{2}(\tilde\kappa-
\tilde\delta)$}
\end{picture}\\



\begin{thebibliography}{10}

\bibitem{Topdiscovery}
F.~Abe {\em et al.}, \prl {\bf 73} (1994) 225.

\bibitem{Laser}
I.F.~Ginzburg {\em et al.}, \nuclinst {\bf 205} (1983) 47.\\ I.F. Ginzburg {\em
  et al.}, \nuclinst {\bf 219} (1984) 5.\\ V.I. Telnov, \nuclinst {\bf A294}
  (1990) 72. \\ V.I. Telnov, in Proceedings of {\em Physics and Experiments
  with Linear Colliders}, p.~739, edited by R.~Orava, P.~Eerola and
  M.~Nordberg, World Scientific, 1992. \\ V.I. Telnov, in {\em Proceedings of
  the IXth International Workshop on Photon-Photon Collisions.}, p.~369, edited
  by D.O. Caldwell and H.P. Paar, World Scientific, 1992.

\bibitem{Fernandhh}
G.~J.~Gounaris, D.~Schildknecht and F.~M.~Renard, \pl {\bf B83} (1979) 191.

\bibitem{Bijhh}
J.~J.~ vand der Bij, \np {\bf B267} (1986) 557.

\bibitem{Jikiahh}
G.V. Jikia and Yu.F. Pirogov, {\it Phys. Let.} B283 (1992) 135.\\ G.V. Jikia,
  {\em Nucl. Phys.} {\bf B412} (1994) 57.

\bibitem{Nonlineargauge}
K. Fujikawa, \pr {\bf D7} (1973) 393 \\ M. Base and N.D. Hari Dass, \annp {\bf
  94} (1975) 349 \\ M.B.\ Gavela, G. Girardi, C. Malleville and P. Sorba, \np
  {\bf B193} (1981) 257 \\ N.G. Deshpande and M. Nazerimonfared, \np {\bf B213}
  (1983) 390 \\ F.~Boudjema, \pl {\bf B187} (1987) 362. \\ M.~Baillargeon and
  F.~Boudjema, \pl {\bf B272} (1991) 158.

\bibitem{Backgroundgaugeold}
B.~S.~ de Witt, in {\em Dynamical Theory of Groups and Fields}, Gordon and
  Breach, N.Y. 1965.\\ S. Weinberg, {\it Phys. Rev.} D13 (1976) 974.\\ L.F.
  Abbott, {\it Nucl. Phys.} B185 (1981) 189.

\bibitem{BackgroundgaugeEW}
G.~Shore, {\em Ann. Phys.} {\bf 137} (1981) 262. \\ M.~B.~Einhorn and J.~Wudka,
  \pr {\bf D39} (1989) 2758.

\bibitem{BckgDenner}
A. Denner, S. Dittmaier, G. Weiglein {\it Nucl. Phys.} {\bf B440} (1995) 95. \\
  X.~Li and Y.~Liao, ASITP/94-50 {\em HEP-Ph/9409401}.\\ For a slight
  variation, see A.G.~Morgan and Z.~Bern, \pr {\bf D49} (1994) 6155.

\bibitem{EWA}
R.N.~Cahn and S.~Dawson, \pl {\bf B136} (1984) 196. \\ S.~Dawson, \np {\bf
  B249} (1984) 42.\\ S.~Chanowitz and M.K.~Gaillard, \pl {\bf B142} (1984) 196.
  \\ G.L. Kane, W.W.~Repko and W.B.~Rolnick, \pl {\bf B148} (1984) 367.\\
  J.~Lindfors, \zp {\bf C28} (1985) 427. \\ W.B.~Rolnick, \np {\bf B274} (1986)
  171. \\ Z.~Kunszt and D.E.~Soper, \np {\bf B296} (1988) 253.

\bibitem{egnuwh}
K.~Hagiwara, I.~Watanabe, P.M.~Zerwas, \pl {\bf B278} (1992) 187.

\bibitem{Parisgg}
M. Baillargeon, G. B\'elanger and F. Boudjema, in {\em Proceedings of
  Two-Photon Physics from DA$\Phi$NE to LEP200 and Beyond}, Paris, 1994, edited
  by F.~Kapusta and J.~Parisi (world Scientific, Singapore, 1995), p.267.

\bibitem{eeZhh}
V.~Barger, T.~Han and R.~J.~N~Phillips, \pr {\bf D38} (1988) 2766.

\bibitem{eennhh}
V.~Barger, T.~Han, {\em Mod. Phys. Lett.} {\bf A5} (1990) 667.

\bibitem{eehh}
K.J.F. Gaemers, F. Hoogeveen, {\it Z. Phys.} C26 (1984) 249.

\bibitem{Nousggvv}
For a list of references on this process, see G. B\'elanger and F. Boudjema,
  \pl {\bf B288} (1992) 210.

\bibitem{Nousgg3v}
M.~Baillargeon and F.~Boudjema, \pl {\bf B317} (1993) 371.

\bibitem{CahnnunuH}
R.N.~Cahn, \np {\bf B255} (1985) 341.

\bibitem{Jikia4w}
G.V.~Jikia, {\em Nucl. Instrum. Methods}, {\bf A335} (1995) 2907; {\it ibid}
  IHEP 94-77;{\em HEP-PH/94-07393}. This process has also been calculated in
  the unitary gauge by\\ K.~Cheung, \pr {\bf D50} (1994) 4290.

\bibitem{Bernreview}
For an excellent review, see Z.~Bern, Lectures presented at TASI 1992, Boulder,
  CO. UCLA/93/TEP/5; {\em HEP-PH/9304249}.

\bibitem{DennerRCggww}
A.~Denner, S.~Dittmaier and R.~Schuster, Bielefeld Preprint, BI-TP 95/04 {\em
  HEP/PH-9503442}.

\bibitem{Equivalencetheorem}
M.~Chanowitz and M.~K.~Gaillard, \np {\bf B261} (1985) 379.\\ B.~W.~Lee,
  C.~Quigg and H.~Thacker, \pr {\bf D16} (1977) 1519.\\
  J.~M.~Cornwall,D.~N.~Levin and G.~Tiktopoulos, \pr {\bf D10} (1974) 1145;
  {\bf D11} (1975) 972 (E).\\ C.~E.~Vayonakis, {\em Nuov. Cim. Lett.} {\bf 17}
  (1976) 383.

\bibitem{Sheffield95}
P.~Chen, Presented at the $\gamma \gamma$ Workshop, Sheffield, Apr. 7-8 1995.\\
  V.~Telnov, Plenary Talk, Photon'95 Conference, Sheffield, G.B., Apr. 8-15
  (1995) to appear in the proceedings.

\bibitem{pphh}
D.~A.~Dicus, K. Kallianpur and S.~S.~D.~ Willenbrock, \pl {\bf B200} (1988)
  187.\\ A. Abbasabadi, W.~W.~Repko, D.~A.~Dicus and R.~Vega, \pr {\bf D38}
  (1988) 2770.\\ K. Kallianpur, \pl {\bf B215} (1988) 392.

\bibitem{Chivukulascalar}
R.~S.~Chivukula and V. Koulovassilopoulos, \pl {\bf B309} (1993) 371; \pr {\bf
  D50} (1994) 3218.

\bibitem{Georginatural}
A.~Manohar and H.~Georgi, \np {\bf B234} (1984) 189.

\bibitem{Gracehh}
V.A. Ilyn et. al, June 1995, KEK CP-030 {\it HEP-PH/9506326}.

\bibitem{Chopinitalie}
Talk given at the 2nd European Workshop on {\em Physics with \epemt Linear
  Colliders}, Higgs Session, Assergi, Gran-Sasso, Italy, June $2^{nd}$, 1995.
  See Proceedings.

\bibitem{Chopinthese}
E.~Chopin, {\it Tests du secteur scalaire de la th\'eorie electro-faible},
  Rapport de Magist\`ere, ENS-Lyon, Septembre 1994.

\bibitem{Dicusnlgauge}
D.A.~Dicus and C.~Kao, \pr {\bf D49} (1994) 1265.

\end{thebibliography}
\end{document}